%% file: main.tex
\documentclass[a4paper,10pt]{llncs}

\usepackage[l2tabu,orthodox]{nag}
\usepackage[centering]{geometry}
\usepackage{ifxetex}

\usepackage{microtype}

\usepackage{xcolor}
\definecolor{oxygenorange}{HTML}{FFDD00}
\usepackage[color=oxygenorange]{todonotes}

\definecolor{darkblue}{rgb}{0,0,0.5}
\definecolor{darkgreen}{rgb}{0,0.5,0}
\usepackage[colorlinks=true,linkcolor=darkblue,urlcolor=darkblue,citecolor=darkgreen]{hyperref}

\def\isanonymous{0}

\usepackage{ifthen}
\newcommand{\anonymous}[2]{
\ifthenelse{\equal{\isanonymous}{1}}
{{#1}}
{{#2}}
}

\def\isshortversion{0}

\usepackage{ifthen}
\newcommand{\shortversion}[2]{
\ifthenelse{\equal{\isshortversion}{1}}
{{#1}}
{{#2}}
}

\usepackage{amsmath,amsfonts,amssymb}
\usepackage{xspace}
\usepackage{physics}
\usepackage{graphicx}
\usepackage[caption=false]{subfig}
\usepackage[counterclockwise, figuresleft]{rotating}
\usepackage{caption}
\usepackage{multirow}

\usepackage[vlined,linesnumbered,ruled]{algorithm2e}
\newcommand{\ccom}[1]{\tcc*{#1}}
\SetEndCharOfAlgoLine{}

\usepackage[lambda,landau,operators,probability,sets,logic,complexity,asymptotics]{cryptocode}
\newcommand{\qsharp}{Q\#\xspace}
\newcommand{\aes}{AES\xspace}
\newcommand{\sha}{SHA\xspace}
\newcommand{\lowmc}{LowMC\xspace}
\newcommand{\picnic}{Picnic\xspace}
\newcommand{\esbox}{\mbox{S-box}\xspace}
\newcommand{\nistmaxdepth}{\texttt{MAXDEPTH}\xspace}
\newcommand{\KE}{\textnormal{KE}\xspace}

\newcommand{\pr}{\mathrm{Pr}}

\newcommand{\depthunits}{cycles}

\usepackage{booktabs}  
\usepackage{comment}
\usepackage{enumitem}

\ifxetex%
\else%
\usepackage{embedfile}
\fi%

\usepackage{tikz,pgfplots}
\usetikzlibrary{calc}
\usetikzlibrary{arrows}

\definecolor{DarkPurple}{HTML}{332288}
\definecolor{DarkBlue}{HTML}{6699CC}
\definecolor{LightBlue}{HTML}{88CCEE}
\definecolor{DarkGreen}{HTML}{117733}
\definecolor{DarkRed}{HTML}{661100}
\definecolor{LightRed}{HTML}{CC6677}
\definecolor{LightPink}{HTML}{AA4466}
\definecolor{DarkPink}{HTML}{882255}
\definecolor{LightPurple}{HTML}{AA4499}

\definecolor{DarkBrown}{HTML}{604c38}
\definecolor{DarkTeal}{HTML}{23373b}
\definecolor{LightBrown}{HTML}{EB811B}
\definecolor{LightGreen}{HTML}{14B03D}

\usepackage{listings}
\lstdefinelanguage{Sage}[]{Python}{morekeywords={True,False,sage,cdef,cpdef,ctypedef,self},sensitive=true}
\makeatletter
\renewcommand*{\@fnsymbol}[1]{\ensuremath{\ifcase#1\or *\or \dagger\or \ddagger\or
		\mathsection\or \mathparagraph\or \|\or **\or \dagger\dagger
		\or \ddagger\ddagger \else\@ctrerr\fi}}
\makeatother

\newcommand*\samethanks[1][\value{footnote}]{\footnotemark[#1]}

\title{Implementing Grover oracles for quantum key search on \aes and \lowmc}
\anonymous{
	\author{}
	\institute{}
}{
	\author{ \
        Samuel Jaques  \inst{1}\thanks{Partially supported by the University of Oxford Clarendon fund.}\thanks{This work was done while Fernando and Sam were interns at Microsoft Research.} \and
		Michael Naehrig  \inst{2} \and
		Martin Roetteler \inst{3} \and
		Fernando Virdia \inst{4}\samethanks\thanks{
			Partially supported by the EPSRC and the UK government as part of the Centre for Doctoral Training in Cyber Security at Royal Holloway, University of London (EP/P009301/1).}
	}
	\institute{
		Department of Materials, University of Oxford, UK\\
		\email{samuel.jaques@materials.ox.ac.uk}
		\and
		Microsoft Research, Redmond, WA, USA \\
		\email{mnaehrig@microsoft.com}
		\and
		Microsoft Quantum, Redmond, WA, USA \\
		\email{martinro@microsoft.com}
		\and
		Information Security Group, Royal Holloway, University of London, UK\\
		\email{fernando.virdia.2016@rhul.ac.uk}
	}
}

\begin{document}

\maketitle

\begin{abstract}
	Grover's search algorithm gives a quantum attack against block ciphers by searching for a key that matches a small number of plaintext-ciphertext pairs. This attack uses $O(\sqrt{N})$ calls to the cipher to search a key space of size $N$. Previous work in the specific case of \aes derived the full gate cost by analyzing quantum circuits for the cipher, but focused on minimizing the number of qubits. 
	
	In contrast, we study the cost of quantum key search attacks under a depth restriction and introduce techniques that reduce the oracle depth, even if it requires more qubits. As cases in point, we design quantum circuits for the block ciphers \aes and \lowmc. Our circuits give a lower overall attack cost in both the gate count and depth-times-width cost models. In NIST's post-quantum cryptography standardization process, security categories are defined based on the concrete cost of quantum key search against \aes. We present new, lower cost estimates for each category, so our work has immediate implications for the security assessment of post-quantum cryptography.
	
	As part of this work, we release \qsharp implementations of the full Grover oracle for \aes-128, -192, -256 and for the three \lowmc instantiations used in \picnic, including unit tests and code to reproduce our quantum resource estimates. To the best of our knowledge, these are the first two such full implementations and automatic resource estimations. 
    
\end{abstract}

\section{Introduction}\label{sec:introduction}
The prospect of a large-scale, cryptographically relevant quantum computer has prompted increased scrutiny of the post-quantum security of our cryptographic primitives. Shor's algorithm for factoring and computing discrete logarithms introduced in \cite{shor} and \cite{Sho97} will completely break public-key schemes such as RSA, ECDSA and ECDH. But symmetric schemes such as block ciphers and hash functions are widely considered post-quantum secure. The only caveat thus far is a security reduction due to key search or pre-image attacks with Grover's algorithm~\cite{Grover}. As Grover's algorithm only provides at most a square root speedup, the rule of thumb is to simply double the cipher's key size to make it post-quantum secure. Such conventional wisdom reflects the asymptotic behavior and only gives a rough idea of the security penalties that quantum computers inflict on symmetric primitives. In particular, the cost of evaluating the Grover oracle is often ignored.

In their call for proposals to the standardization of post-quantum cryptography~\cite{NIST:PQ16c}, the National Institute of Standards and Technology (NIST) proposes security categories for post-quantum public-key schemes such as key encapsulation and digital signatures. The categories are defined by the cost of quantum algorithms for exhaustive key search on the block cipher \aes and collision search for the hash function \sha-3, and measure the attack cost in the number of quantum gates. Because the gate count of Grover's algorithm increases with parallelization, they impose a total upper bound on the depth of a quantum circuit, called \nistmaxdepth, and account for this in the gate counts. There is no bound on width. An algorithm meets the requirements of a specific security category if the best known attack uses more resources (gates) than are needed to solve the reference problem. Hence, a concrete and meaningful definition of these security categories depends on precise resource estimates of the Grover oracle for key search on \aes.

Security categories 1, 3 and 5 correspond to key recovery against \aes-128, \aes-192 and \aes-256, respectively. The NIST proposal derives gate cost estimates from the concrete, gate-level descriptions of the \aes oracle by Grassl, Langenberg, Roetteler and Steinwandt~\cite{grassl2016applying}. Grassl et al. aim to minimize the circuit width, i.e. the number of qubits needed.

\subsubsection{Prior work.}
Since the publication of~\cite{grassl2016applying}, other works have studied quantum circuits for \aes, the \aes Grover oracle and its use in Grover's algorithm.
Almazrooie, Samsudin, Abdullah and Mutter~\cite{ASAM18} improve the quantum circuit for \aes-128. 
As in~\cite{grassl2016applying}, the focus is on minimizing the number of qubits. The improvements are a slight reduction in the total number of Toffoli gates and the number of qubits by using a wider binary field inversion circuit that saves one multiplication. 
Kim, Han and Jeong~\cite{KHJ18} discuss time-space trade-offs for key search on block ciphers in general and use \aes as an example. They discuss NIST's \nistmaxdepth parameter and hence study parallelization strategies for Grover's algorithm to address the depth constraint. They take the Toffoli gate depth as the relevant metric for the \nistmaxdepth bound arguing that it is a conservative approximation.

Recently, independent and concurrent to parts of this work, Langenberg, Pham and Steinwandt~\cite{cryptoeprint:2019:854} developed quantum circuits for \aes that demonstrate significant improvements over those presented in \cite{grassl2016applying} and \cite{ASAM18}. The main source of optimization is a different \esbox design derived from work by Boyar and Peralta in \cite{boyar2010new} and \cite{BP12}, which greatly reduces the number of Toffoli gates in the \esbox as well as its Toffoli depth. Another improvement is that fewer ancilla qubits are required for the \aes key expansion. Again, this work aligns with the objectives in~\cite{grassl2016applying} to keep the number of qubits small.

Bonnetain et al.~\cite{BonnetainNS19} study the post-quantum security of \aes within a new framework for classical and quantum structured search. The work cites \cite{grassl2016applying} for deducing concrete gate counts for reduced-round attacks. 

\subsubsection{Our contributions.}
We present implementations of the full Grover oracle for key search on \aes and \lowmc in \qsharp~\cite{qsharp}, including full implementations of the block ciphers themselves. In contrast to previous work~\cite{grassl2016applying}, \cite{ASAM18} and \cite{cryptoeprint:2019:854}, having a concrete implementation allows us to get more precise, flexible and automatic estimates of the resources required to compute these operations. It also allows us to unit test our circuits, to make sure that the implementations are correct. 

The source code will be released\footnote{A public repository will be available after the internal code release process has been finalized.} under a free license to allow independent verification of our results, further investigation of different trade-offs and cost models and re-costing as the \qsharp compiler improves and as automatic optimization software becomes available. We hope that it can serve as a useful starting point for cryptanalytic work to assess the post-quantum security of other schemes.

We review the literature on the parallelization of Grover's algorithm (\cite{BBHT98}, \cite{Zal99}, \cite{QIC:GR03}, \cite{KHJ18}) to explore the cost of attacking \aes and \lowmc in the presence of a bound on the total depth, such as \nistmaxdepth proposed by NIST. We conclude that using parallelization by dividing the search space is advantageous. We also give a rigorous justification for the number of plaintext-ciphertext blocks needed in Grover's oracle in the context of parallelization. Smaller values than those proposed by Grassl et al.~\cite{grassl2016applying} are sufficient, as is also pointed out by Langenberg et al.~\cite{cryptoeprint:2019:854}.

Our quantum circuit optimization approach differs from those in the previous literature \cite{grassl2016applying}, \cite{ASAM18} and \cite{cryptoeprint:2019:854} in that our implementations do not aim for the lowest possible number of qubits. Instead, we designed them to minimize the gate-count and depth-times-width cost metrics for quantum circuits under a depth constraint. The gate-count metric is relevant for defining the NIST security categories and the depth-times-width cost metric is a more realistic measure of quantum resources when quantum error correction is deployed. Favoring lower depth at the cost of a slightly larger width in the oracle circuit leads to costs that are smaller in both metrics than for the circuits presented in \cite{grassl2016applying}, \cite{ASAM18} and \cite{cryptoeprint:2019:854}. 
Grover's algorithm does not parallelize well, meaning that minimizing depth rather than width is crucial to make the most out of the available depth. 

To the best of our knowledge, our work results in the most shallow quantum implementation of \aes proposed so far, and the first ever for \lowmc. 

We chose to also implement \lowmc to provide an example of a quantum circuit for another block cipher. It is used in the \picnic signature scheme (\cite{chase2017post}, \cite{_NISTPQC-R1:Picnic17}), a round-2 candidate in the NIST standardization process. Thus, our implementation can contribute to more precise cost estimates for attacks on \picnic and the assessment of its post-quantum security. 

We present our results for quantum key search on \aes in the context of the NIST post-quantum cryptography standardization process and derive new and lower cost estimates for the definition of the NIST security strength categories. We see a consistent gate cost reduction between 11 and 13 bits, making it easier for submitters to claim a certain quantum security category.

\section{Finding a block cipher key with Grover's algorithm}\label{sec:preliminaries}
Given plaintext-ciphertext pairs created by encrypting a small number of messages under a block cipher, Grover's quantum search algorithm~\cite{Grover} can be used to find the secret key~\cite{yamamura2000quantum}. This section provides some preliminaries on Grover's algorithm, how it can be applied to the key search problem and how it parallelizes under depth constraints.

\subsection{Grover's algorithm}\label{sec:grover}
Grover's algorithm~\cite{Grover} searches through a space of $N$ elements; for simplicity, we restrict to $N=2^k$ right away and label elements by their indices in $\{0,1\}^k$. The algorithm works with a superposition $\ket{\psi}=2^{-k/2}\sum_{x\in\{0,1\}^k}\ket{x}$ of all indices, held in a register of $k$ qubits. It makes use of an operator $U_f$ for evaluating a Boolean function $f:\{0,1\}^k\rightarrow \{0,1\}$ that marks solutions to the search problem, i.e. $f(x)=1$ if and only if the element corresponding to $x$ is a solution. When applying the Grover oracle $U_f$ to a state $\ket{x}\ket{y}$ for a single qubit $\ket{y}$, it acts as $\ket{x}\ket{y} \mapsto \ket{x}\ket{y\oplus f(x)}$ in the computational basis. When $\ket{y}$ is in the state $\ket{\varphi} = (\ket{0}- \ket{1})/\sqrt{2}$, then this action can be written as $\ket{x}\ket{\varphi} \mapsto (-1)^{f(x)}\ket{x}\ket{\varphi}$. This means that the oracle applies a phase shift to exactly the solution indices.

The algorithm first prepares the state $\ket{\psi}\ket{\varphi}$ with $\ket{\psi}$ and $\ket{\varphi}$ as above. It then repeatedly applies the so-called Grover iteration
$$G = (2\dyad{\psi}{\psi}- I)U_f,$$
an operator that consists of the oracle $U_f$ followed by the operator $2\outerproduct{\psi}-I$, which can be viewed as an inversion about the mean amplitude.
Each iteration can be visualized as a rotation of the state vector in the plane spanned by two orthogonal vectors: the superposition of all indices corresponding to solutions and non-solutions, respectively. The operator $G$ rotates the vector by a constant angle towards the superposition of solution indices. Let $1\leq M \le N$ be the number of solutions and let $0 < \theta \le \pi/2$ such that $\sin[2](\theta) = M/N$. Note that if $M \ll N$, then $\sin(\theta)$ is very small and $\theta \approx \sin(\theta) = \sqrt{M/N}$. 

When measuring the first $k$ qubits after $j > 0$ iterations of $G$, the success probability $p(j)$ for obtaining one of the solutions is  
$p(j) = \sin[2]((2j+1)\theta)$~\cite{BBHT98}, which is close to $1$ for $j\approx \frac{\pi}{4\theta}$. Hence, after $\floor{\frac{\pi}{4}\sqrt{\frac{N}{M}}}$ iterations, measurement yields a solution with overwhelming probability of at least $1-\frac{M}{N}$.

Grover's algorithm is optimal in the sense that any quantum search algorithm needs at least $\Omega(\sqrt{N})$ oracle queries to solve the problem~\cite{BBHT98}. In~\cite{Zal99}, Zalka shows that for any number of oracle queries, Grover's algorithm gives the largest probability to find a solution.

\subsection{Key search for a block cipher}\label{subsec:blockcipher}
Let $C$ be a block cipher with block length $n$ and key length $k$; for a key $K\in \{0,1\}^k$ denote by $C_K(m)\in\{0,1\}^n$ the encryption of message block $m\in\{0,1\}^n$ under the key $K$. Given $r$ plaintext-ciphertext pairs $(m_i, c_i)$ with $c_i=C_K(m_i)$, we aim to apply Grover's algorithm to find the unknown key $K$~\cite{yamamura2000quantum}. The  Boolean function $f$ for the Grover oracle takes a key $K$ as input, and is defined as
$$
f(K) = \begin{cases} 1,& \text{if }\,C_K(m_i)=c_i\text{ for all }1\leq i\leq r, \\0,&\text{otherwise.}\end{cases}
$$

Possibly, there exist other keys than $K$ that encrypt the known plaintexts to the same ciphertexts. We call such keys \emph{spurious keys}. If their number is known to be, say $M-1$, the $M$-solution version of Grover's algorithm has the same probability of measuring each spurious key as measuring the correct $K$.

\subsubsection{Spurious keys.}
We start by determining the probability that a single message encrypts to the same ciphertext under two different keys, for which we make the usual heuristic assumptions about the block cipher $C$. We assume that under a fixed key $K$, the map $\{0,1\}^n \rightarrow \{0,1\}^n, m \mapsto C_K(m)$ is a pseudo-random permutation; and under a fixed message block $m$, the map $\{0,1\}^k \rightarrow \{0,1\}^n, K \mapsto C_K(m)$ is a pseudo-random function. Now let $K$ be the correct key, i.e. the one that was used for the encryption.
It follows that for a single message block of length $n$,
$\pr_{K\neq K'}\left(C_K(m)= C_{K'}(m)\right) = 2^{-n}.$

This probability becomes smaller when the equality condition is extended to multiple blocks. Given $r$ distinct messages $m_1, \dots, m_r\in \{0,1\}^n$, we have
\begin{equation}\label{pBinomial}
p = \pr_{K\neq K'}\left((C_K(m_1),\dots,C_K(m_r)) = (C_{K'}(m_1),\dots,C_{K'}(m_r))\right) = 2^{-rn}.
\end{equation}
Since the number of keys different from $K$ is $2^k-1$, we expect the number of spurious keys for an $r$-block message to be $(2^k-1)2^{-rn}$. Choosing $r$ such that this quantity is very small ensures with high probability that there is a unique key and we can parameterize Grover's algorithm for a single solution.

\begin{remark}
Grassl et al.~\cite[\S3.1]{grassl2016applying} work with a similar argument. They take the probability over pairs $(K',K'')$ of keys with $K' \neq K''$. Since there are $2^{2k}-2^k$ such pairs, they conclude that about $(2^{2k}-2^k)2^{-rn}$ satisfy the above condition that the ciphertexts coincide on all $r$ blocks. But this also counts pairs of keys for which the ciphertexts match each other, but do not match the images under the correct $K$. Thus, using the number of pairs overestimates the number of spurious keys and hence the number $r$ of message blocks needed to ensure a unique key.
\end{remark}

Based on the above heuristic assumptions, one can determine the probability for a specific number of spurious keys. Let $X$ be the random variable whose value is the number of spurious keys for a given set of $r$ message blocks and a given key $K$. Then, $X$ is distributed according to a binomial distribution:
\shortversion{$\pr(X=t) = \binom{2^{k}-1}{t}p^t(1-p)^{2^{k}-1-t},$}{$$\pr(X=t) = \binom{2^{k}-1}{t}p^t(1-p)^{2^{k}-1-t},$$}
where $p=2^{-rn}$. We use the Poisson limit theorem to conclude that this is approximately a Poisson distribution with
\begin{equation}\label{eq:spurious}
\pr(X=t)\approx e^{-\frac{2^{k}-1}{2^{rn}}}\frac{(2^k-1)^t(2^{-rn})^t}{t!}\approx e^{-2^{k-rn}}\frac{2^{t(k-rn)}}{t!}.
\end{equation}

The probability that $K$ is the unique key consistent with the $r$ plaintext-ciphertext pairs is $\pr(X=0)\approx e^{-2^{k-rn}}$. Thus we can choose $r$ such that $rn$ is slightly larger than $k$; $rn=k+10$ gives $\pr(X=0)\approx 0.999$. 
In a block cipher where $k = b\cdot n$ is a multiple of $n$, taking $r=b+1$ will give the unique key $K$ with probability at least $1-2^{-n}$, which is negligibly close to $1$ for typical block sizes. If $rn < k$, then $K$ is almost certainly not unique. Even $rn=k-3$ gives less than a 1\% chance of a unique key. Hence, $r$ must be at least $\ceil{k/n}$.    

The case $k=rn$, when the total message length is equal to the key length, remains interesting if one aims to minimize the number of qubits. The probability for a unique $K$  is $\pr(X=0) \approx 1/e \approx 0.3679$, and the probability of exactly one spurious key is the same.  Kim et al.~\cite[Equation~(7)]{KHJ18} describe the success probability after a certain number of Grover iterations when the number of spurious keys is unknown. The optimal number of iterations gives a  maximum success probability of $0.556$, making it likely that the first attempt will not find the correct key and one must repeat the algorithm.

\subsubsection{Depth constraints for cryptanalysis.} 
In this work, we assume that any quantum adversary is bounded by a constraint on its total depth for running a quantum circuit. 
In its call for proposals to the post-quantum cryptography standardization effort~\cite{NIST:PQ16c}, NIST introduces the parameter \nistmaxdepth as such a bound and suggests that reasonable values\footnote{Suggested \nistmaxdepth values are justified by assumptions about the total available time and speed of each gate. The limit $2^{96}$ is given as ``the approximate number of gates that atomic scale qubits with speed of light propagation times could perform in a millennium''~\cite{NIST:PQ16c}. An adversary could only run a higher-depth circuit if they were able to use smaller qubits, faster propagation, or had more available time.} are between $2^{40}$ and $2^{96}$. Whenever an algorithm's overall depth exceeds this bound, parallelization becomes necessary. We do assume that \nistmaxdepth constitutes a hard upper bound on the total depth of a quantum attack, including possible repetitions of a Grover instance. 

We consider it reasonable to expect that the overall attack strategy is guaranteed to return a solution with high probability close to $1$ within the given depth bound. E.g., a success probability of $1/2$ for a Grover instance to find the correct key requires multiple runs to increase the overall probability closer to $1$. These runs, either sequentially or in parallel, need to be taken into account for determining the overall cost and must respect the depth limit. While this setting is our main focus, it can be adequate to allow and cost a quantum algorithm with a success probability noticeably smaller than $1$. Where not given in this paper, the corresponding analysis can be derived in a straightforward manner.

\shortversion{}{
\input{grover-repeated.tex}
}

\subsection{Parallelization}\label{sec:groverparallel} 
There are different ways to parallelize Grover's algorithm. Kim, Han, and Jeong~\cite{KHJ18} describe two, which they denote as \emph{inner} and \emph{outer} parallelization.
Outer parallelization runs multiple instances of the full algorithm in parallel. Only one instance must succeed, allowing us to reduce the necessary success probability, and hence number of iterations, for all.
Inner parallelization divides the search space into disjoint subsets and assigns each subset to a parallel machine. Each machine's search space is smaller, so the number of necessary iterations shrinks. 

Zalka~\cite{Zal99} concludes that in both cases, one only obtains a factor $\sqrt{S}$ gain in the number of Grover iterations when working with $S$ parallel Grover oracles, and that this is asymptotically optimal. Compared to many classical algorithms, this is an inefficient parallelization, since we must increase the width by a factor of $S$ to reduce the depth by a factor of $\sqrt{S}$.
Both methods avoid any communication, quantum or classical, during the Grover iterations. They require communication at the beginning, to distribute the plaintext-ciphertext pairs to each machine and to delegate the search space for inner parallelization, and communication at the end to collect the measured keys and decide which one, if any, is the true key. The next section discusses why our setting favours inner parallelization.

\subsubsection{Advantages of inner parallelization.}
Consider $S$ parallel machines that we run for $j$ iterations, using the notation of \S\ref{sec:grover}, and a unique key. For a single machine, the success probability is $p(j)=\sin^2\left((2j+1)\theta\right)$. Using outer parallelization, the probability that at least one machine recovers the correct key is $p_S(j)=1-(1-p(j))^S$. We hope to gain a factor $\sqrt{S}$ in the number of iterations, so instead of iterating $\floor{\frac{\pi}{4\theta}}$ times, we run each machine for $j_S=\floor{\frac{\pi}{4\theta\sqrt{S}}}$ iterations.

Considering some small values of $S$, we get $S=1:\ p_1(j_1) \approx 1$, $S=2:\ p_2(j_2) \approx 0.961$ and $S=3:\ p_3(j_3) \approx 0.945$. As $S$ gets larger, 
we use a series expansion to find that  
\begin{equation}
p_S(j_S) \approx 1-\left(1-\frac{\pi^2}{4S}+O\left(\frac{1}{S^2}\right)\right)^S \xrightarrow{S\rightarrow \infty} 1-e^{-\frac{\pi^2}{4}}\approx 0.915.
\end{equation}
This means that by simply increasing $S$, it is not possible to gain a factor $\sqrt{S}$ in the number of iterations if one aims for a success probability close to $1$. In contrast, with inner parallelization, the correct key lies in the search space of exactly one machine. With $j_S$ iterations, this machine has near certainty of measuring the correct key, while other machines are guaranteed not to measure the correct key. Overall, we have near-certainty of finding the correct key. Inner parallelization thus achieves a higher success probability with the same number $S$ of parallel instances and the same number of iterations.

Another advantage of inner parallelization is that dividing the search space separates any spurious keys into different subsets and reduces the search problem to finding a unique key. This allows us to reduce the number $r$ of message blocks in the Grover oracle and was already observed by Kim, Han, and Jeong~\cite{KHJ18} in the context of measure-and-repeat methods.
In fact, the correct key lies in exactly one subset of the search space. If the spurious keys fall into different subsets, the respective machines measure spurious keys, which can be discarded classically after measurement with access to the appropriate number of plaintext-ciphertext pair. The only relevant question is whether there is a spurious key in the correct key's subset of size $2^k/S$. The probability for this is 
\begin{equation}
\mathrm{SKP}(k, n, r, S) = \sum_{t=1}^{\infty} \pr(X=t) = 1 - e^{-\frac{2^{k-rn}}{S}},
\end{equation}
using Equation~\eqref{eq:spurious} with $2^k$ replaced by $2^k/S$. If $k=rn$, this probability is roughly $1/S$ when $S$ gets larger. In general, high parallelization makes spurious keys irrelevant, and the Grover oracle can simply use the smallest $r$ such that $\mathrm{SKP}(k, n, r, S)$ is still small enough, i.e. less than a desired bound.

\section{Quantum circuit design}\label{sec:quantum-circuit-design}
Quantum computation is usually described in the quantum circuit model. This section describes our interpretation of quantum circuits, methods and criteria for quantum circuit design, and cost models to estimate quantum resources. 

\subsection{Assumptions about the fault-tolerant gate set and architecture} 
The quantum circuits we are concerned with in this paper operate on qubits. The circuits themselves are composed of so-called Clifford+$T$ gates, which is a commonly used universal fault-tolerant gate set exposed by several families of quantum error-correcting codes. The primitive gates consist of single-qubit Clifford gates, controlled-NOT (CNOT) gates, T gates, and measurements. We make the standard assumption of \emph{full parallelism}, meaning that a quantum circuit can apply any number of gates simultaneously so long as these gates act on disjoint sets of qubits \cite{PRSA:BBGHKLSS13,QIC:GR03}. 

All quantum circuits for \aes and \lowmc described in this paper were designed, tested, and costed in the \qsharp programming language~\cite{qsharp}, which supports all assumptions discussed here. We adopt the computational model presented in \cite{jaques2019quantum}.  The \qsharp compiler allows us to compute circuit depth automatically by moving gates around through a circuit if the qubits it acts on were previously idle. In particular, this means that the depth of two circuits applied in series may be less than the sum of the individual depths of each circuit.
The \qsharp language allows the circuit to \emph{allocate} ancilla qubits as needed, which adds new qubits initialized to $\ket{0}$. If an ancilla is returned to the state $\ket{0}$ after it has been operated on, the circuit can \emph{release} it. Such a qubit is no longer entangled with the state used for computation and the circuit can now maintain or measure it.

Grover's algorithm is a far-future quantum algorithm, making it difficult to decide on the right cost for each gate. Previous work assumed that T gates constitute the main cost (\cite{grassl2016applying}, \cite{ASAM18}, \cite{cryptoeprint:2019:854}). They are exceptionally expensive for a surface code \cite{FMMC12}; however, for a future error-correcting code, T gates may be transversal and cheap while a different gate may be expensive. Thus, we present costs for both counting T gates only, and costing all gates equally. For most of the circuits, these concerns do not change the optimal design. 

We ignore all concerns of layout and communication costs for the Grover oracle circuit. Though making this assumption is unrealistic for a surface code, where qubits can only interact with neighboring ones, other codes may not have these issues. A single oracle circuit uses relatively few logical qubits ($<2^{20}$), so these costs are unlikely to dominate. This allows us to compare our work with previous proposals, which also ignore these costs. This also implies that uncontrolled swaps are free, since the classical controller can simply track such swaps and rearrange where it applies subsequent gates.

While previous work on quantum circuits for \aes such as \cite{grassl2016applying}, \cite{ASAM18} and \cite{cryptoeprint:2019:854} mainly uses Toffoli gates, we use AND gates instead. A quantum AND gate has the same functionality as a Toffoli gate, except the target qubit is assumed to be in the state $\ket{0}$, rather than an arbitrary state. We use a combination\anonymous{}{\footnote{We thank Mathias Soeken for providing the implementation of the AND gate circuit.}}of the circuit by Selinger~\cite{CCNOT} and the one by Gidney~\cite{AND2} to express the AND gate in terms of Clifford and T gates, see \S\ref{sec:and-gate}. This circuit uses 4 T gates and 11 Clifford gates in T-depth 1 and total depth 8. It uses one ancilla qubit which it immediately releases, while its adjoint circuit is slightly smaller.

\subsection{Automated resource estimation and unit tests}\label{sec:circuit-size-estimation}
One incentive for producing full implementations of the Grover oracle and its components is to obtain precise resource estimates automatically and directly from the circuit descriptions.
Another incentive is to test the circuits for correctness and to compare results on classical inputs against existing classical software implementations that are known (or believed) to be correct. Yet quantum circuits are in general not testable, since they rely on hardware yet to be constructed. To partially address this issue, the \qsharp compiler can classically simulate a subset of quantum circuits, enabling partial test coverage. We thus designed our circuits such that this tool can fully classically simulate them, by using $X$, CNOT, CCNOT, SWAP, and AND gates only, together with measurements (denoted throughout as $M$ ``gates''). This approach limits the design space since we cannot use true quantum methods within the oracle. Yet, it is worthwhile to implement components that are testable and can be fully simulated to increase confidence in the validity of resource estimates deduced from such implementations. 

As part of the development process, we first implemented \aes (resp. \lowmc) in Python3, and tested the resulting code against the \aes implementation in PyCryptodome~3.8.2~\cite{PyCryptodome} (resp. the C++ reference implementation in~\cite{LowMCRepo}). Then, we proceeded to write our \qsharp implementations (running on the Dotnet Core version 2.1.507, using the Microsoft Quantum Development Kit version 0.7.1905.3109), and tested these against our Python3 implementations, by making use of the I\qsharp interface (see \cite{IQSharp},\cite{PyQSharp}. For the \qsharp simulator to run, we are required to use the Microsoft QDK standard library's Toffoli gate for evaluating both Toffoli and AND gates, which results in deeper than necessary circuits. We also have to explicitly SWAP values across wires, which costs 3 CNOT gates, rather than simply keeping track of the necessary free rewiring. Hence, to mitigate these effects, our functions admit a Boolean flag indicating whether the code is being run as part of a unit test by the simulator, or as part of a cost estimate. In the latter case, Toffoli and AND gate designs are automatically replaced by shallower ones, and SWAP instructions are disregarded as free (after manually checking that this does not allow for incompatible circuit optimizations).
All numbers reporting the total width of a circuit include the initial number of qubits plus the maximal number of temporarily allocated auxiliary qubits within the \qsharp function. For numbers describing the total depth, all gates such as Clifford gates, CNOT and T gates as well as measurements are assigned a depth of $1$.  

The AND and Toffoli gate designs we chose use measurements, hence CNOT, 1-qubit Clifford, measurement and depth counts are probabilistic. The \qsharp Toffoli simulator does not currently support PRNG seeding for de-randomizing the measurements,\anonymous{}{\footnote{\url{https://github.com/microsoft/qsharp-runtime/issues/30}, last visited on 2019-08-24.}} which means that estimating differently sized circuits with the same or similar depth (or re-estimating the same circuit multiple times) may result in slightly different numbers.
We also note that the compiler is currently unable to optimize a given circuit by, e.g., searching through small circuit variations that may result in functionally the same operation at a smaller cost (say by allowing better use of the circuit area).

\subsection{Reversible circuits for linear maps}\label{sec:linear_maps}
Linear maps $ f \colon \FF_2^n \rightarrow \FF_2^m $ for varying dimensions $n$ and $m$ are essential building blocks of \aes and \lowmc. 
In general, such a map $f$, expressed as multiplication by a constant matrix $M_f\in \FF_2^{m\times n}$, can be implemented as a reversible circuit on $n$ input wires and $m$ additional output wires (initialized to $\ket{0}$), by using an adequate sequence of CNOT gates: if the $(i,j)$-th coefficient of $M_f$ is $1$, we set a CNOT gate targeting the $i$-th output wire, controlled on the $j$-th input wire. 

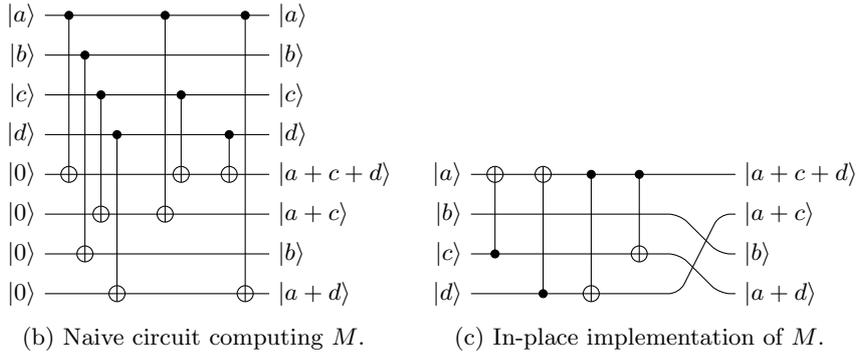
\begin{figure}
	\centering
	\subfloat[Invertible linear transformation $M$ and its PLU decomposition.]{\( M = \left(\begin{array}{rrrr}
		1 & 0 & 1 & 1 \\
		1 & 0 & 1 & 0 \\
		0 & 1 & 0 & 0 \\
		1 & 0 & 0 & 1
		\end{array}\right) =
		\left(\begin{array}{rrrr}
		1 & 0 & 0 & 0 \\
		0 & 0 & 0 & 1 \\
		0 & 1 & 0 & 0 \\
		0 & 0 & 1 & 0
		\end{array}\right)
		\cdot \left(\begin{array}{rrrr}
		1 & 0 & 0 & 0 \\
		0 & 1 & 0 & 0 \\
		1 & 0 & 1 & 0 \\
		1 & 0 & 0 & 1
		\end{array}\right)
		\cdot \left(\begin{array}{rrrr}
		1 & 0 & 1 & 1 \\
		0 & 1 & 0 & 0 \\
		0 & 0 & 1 & 0 \\
		0 & 0 & 0 & 1
		\end{array}\right) = P \cdot L \cdot U \)}\\
	\subfloat[Naive circuit computing $M$.]{\label{fig:naive-linear-map}\input{diagrams/general/linear_map_w_ancillas.tikz}}
	\subfloat[In-place implementation of $M$.]{\label{fig:plu-linear-map}\input{diagrams/general/linear_map_in_place.tikz}}
	\caption{Alternative circuits implementing the same linear transformation \mbox{$M \colon \FF_2^4 \rightarrow \FF_2^4$}, by using the two strategies described in \S\ref{sec:linear_maps}. Both are direct implementations, and could potentially be reduced in size by automatic means as in~\cite{DATE:MeuliSRBM19}, \cite{CORR:MSCRM19}, \cite{DBLP:journals/qic/GossetKMR14} and \cite{zhang2019optimizing}, or manually. Figure~\ref{fig:naive-linear-map} is wider and has a larger gate count, but is shallower, than Figure~\ref{fig:plu-linear-map}.}
	\label{fig:linear_maps}
\end{figure}

Yet, if a linear map $ g \colon \FF_2^n \rightarrow \FF_2^n $ is invertible, one can reversibly compute it in-place on the input wires via a PLU decomposition of $M_g$, $M_g=P\cdot L\cdot U$~\cite[Lecture 21]{TB97}. The lower- and upper-triangular components $L$ and $U$ of the decomposition can be implemented as described above by using the appropriate CNOT gates, while the final permutation $P$ does not require any quantum gates and instead, is realized by appropriately keeping track of the necessary rewiring. An example of a linear map decomposed in both ways is shown in Figure~\ref{fig:linear_maps}. While rewiring is not easily supported in \qsharp, the same effect can be obtained by defining a custom \texttt{REWIRE} operation that computes an in-place swap of any two wires when testing an implementation, and that can be disabled when costing it. We note that such decompositions are not generally unique, but it is not clear whether sparser decompositions can be consistently obtained with any particular technique. For our implementations, we adopt the PLU decomposition algorithm from~\cite[Algorithm 21.1]{TB97}, as implemented in SageMath~8.1~\cite{SageMath}.

\subsection{Cost metrics for quantum circuits}\label{sec:costmetrics}
For a meaningful cost analysis, we assume that an adversary has fixed constraints on its total available resources, as well as a specific cost metric they wish to minimize. Without such limits, we might conclude that \aes-128 could be broken in under a second using $2^{128}$ machines, or broken using only a few thousand qubits but a billion-year runtime. Most importantly, we assume a total depth limit $D_{\max}$ as explained in \S\ref{subsec:blockcipher}.

In this paper, we use the two cost metrics that are considered by Jaques and Schanck in \cite{jaques2019quantum}. The first is the total number of gates, the \emph{$G$-cost}. It assumes non-volatile (``passive'') quantum memory, and therefore it models circuits that incur some cost with every gate, but no cost is incurred in time units during which a qubit is not operated on.

The second cost metric is the product of circuit depth and width, the \emph{$DW$-cost}. This is a more realistic cost model when quantum error correction is necessary. It assumes a volatile (``active'') quantum memory, which incurs some cost to correct errors on every qubit in each time step, i.e. each layer of the total circuit depth. In this cost model, a released ancilla qubit would not require error correction, and the cost to correct it could be omitted. But we assume an efficient strategy for qubit allocation that avoids long idle periods for released qubits and thus choose to ignore this subtlety. Instead, we simply cost the maximum width at any point in the oracle, times its total depth. For both cost metrics, we can choose to count only $T$-gates towards gate count and depth, or count all gates equally.  

\subsubsection{The cost of Grover's algorithm.}
As in \S\ref{sec:grover}, let the search space have size $N=2^k$. Suppose we use an oracle $\mathsf{G}$ such that a single Grover iteration costs $\mathsf{G}_G$ gates, has depth $\mathsf{G}_D$, and uses $\mathsf{G}_W$ qubits. Let $S=2^s$ be the number of parallel machines that are used with the inner parallelization method by dividing the search space in $S$ disjoint parts (see \S\ref{sec:groverparallel}). In order to achieve a certain success probability $p$, the required number of iterations can be deduced from $p \leq \sin[2]((2j+1)\theta)$ which yields \mbox{$j_p = \ceil{(\arcsin(\sqrt{p})/\theta-1)/2} \approx \arcsin(\sqrt{p})/2\cdot\sqrt{N/S}$}. Let $c_p=\arcsin(\sqrt{p})/2$, then the total depth of a $j_p$-fold Grover iteration is
\begin{equation}\label{eq:total_depth}
D = j_p\mathsf{G}_D \approx c_p\sqrt{N/S}\cdot\mathsf{G}_D = c_p2^{\frac{k-s}{2}}\mathsf{G}_D\text{ \depthunits}.
\end{equation}
Note that for $p\approx 1$, we have $c_1$ such that $c_1 = \frac{\pi}{4}$. 
Each of the $S$ machines uses $j_p\mathsf{G}_G \approx c_p\sqrt{N/S}\cdot\mathsf{G}_G = c_p2^{\frac{k-s}{2}}\mathsf{G}_G$ gates, which means that the total $G$-cost over all machines is
\begin{equation}\label{eq:total_gates}
G = S\cdot j_p\mathsf{G}_G \approx c_p\sqrt{N\cdot S}\cdot\mathsf{G}_G = c_p2^{\frac{k+s}{2}}\mathsf{G}_G\text{ gates.}
\end{equation}
Finally, the total width is $W = S\cdot\mathsf{G}_W = 2^s\mathsf{G}_W\text{ qubits}$,
which leads to a $DW$-cost
\begin{equation}\label{eqn:DWcostGrover}
DW \approx c_p\sqrt{N\cdot S}\cdot\mathsf{G}_D\mathsf{G}_W = c_p2^{\frac{k+s}{2}}\mathsf{G}_D\mathsf{G}_W\text{ qubit-\depthunits}.
\end{equation}

These cost expressions show that minimizing the number $S=2^s$ of parallel machines minimizes both $G$-cost and $DW$-cost. Thus, under fixed limits on depth, width, and the number of gates, an adversary's best course of action is to use the entire depth budget and parallelize as little as possible. Under this premise, the depth limit fully determines the optimal attack strategy for a given Grover oracle. Limits on width or the number of gates simply become binary feasibility criteria and are either too tight and the adversary cannot finish the attack, or one of the limits is loose. If one resource limit is loose, we may be able to modify the oracle to use this resource to reduce depth, lowering the overall cost.

\subsubsection{Optimizing the oracle under a depth limit.}
Grover's full algorithm parallelizes so badly that it is generally preferable to parallelize \emph{within} the oracle circuit.
Reducing its depth allows more iterations within the depth limit, thus reducing the necessary parallelization. 

Let $D_{\max}$ be a fixed depth limit. Given the depth $\mathsf{G}_D$ of the oracle, we are able to run $j_{\max} = \floor{D_{\max}/\mathsf{G}_D}$ Grover iterations of the oracle $\mathsf{G}$. For a target success probability $p$, we obtain the number $S$ of parallel instances to achieve this probability in the instance whose key space partition contains the key from $p \leq \sin^2((2j_{\max}+1)\sqrt{S/N})$ as
\begin{equation}\label{eq:instances}
S = \ceil{\frac{N\cdot\arcsin[2](\sqrt{p})}{(2\cdot \floor{D_{\max}/G_D}+1)^2}}\approx c_p^22^k\frac{\mathsf{G}_D^2}{D_{\max}^2}.
\end{equation}
Using this in Equation~\eqref{eq:total_gates} gives a total gate count of
\begin{equation}\label{eq:Gcost_depthrest}
G = c_p^22^{k}\frac{\mathsf{G}_D\mathsf{G}_G}{D_{\max}}\text{ gates.}
\end{equation}
It follows that for two oracle circuits $\mathsf{G}$ and $\mathsf{F}$, the total $G$-cost is lower for $\mathsf{G}$ if and only if $\mathsf{G}_{D}\mathsf{G}_{G} < \mathsf{F}_{D}\mathsf{F}_{G}$. That is, we wish to minimize the product $\mathsf{G}_D\mathsf{G}_G$.
Similarly, the total $DW$-cost under the depth constraint is
\begin{equation}\label{eq:DWcost_depthrest}
DW = c_p^2 2^{k}\frac{\mathsf{G}^2_D\mathsf{G}_W}{D_{\max}}\text{ qubit-\depthunits}.
\end{equation}
Here, we wish to minimize $\mathsf{G}^2_D\mathsf{G}_W$ of the oracle circuit to minimize total $DW$-cost.

\shortversion{}{
	\input{grovercost_classical_communication.tex}
}

\section{A quantum circuit for \aes}\label{sec:aes}
The Advanced Encryption Standard (\aes)~\cite{AESProposal,AESSpecification} is a block cipher standardized by NIST in 2001. Using the notation from~\cite{AESProposal}, \aes is composed of an \esbox, a Round function (with subroutines ByteSub, ShiftRow, MixColumn, AddRoundKey; with the last round slightly differing from the others), and a KeyExpansion function (with subroutines SubByte, RotByte). The \esbox corresponds to inversion in the finite field $\FF_{256} \simeq \FF_2[x]/(x^8 + x^4 + x^3 + x + 1)$ (with $0 \mapsto 0$). Three different instances of \aes have been standardized, for key lengths of 128, 192 and 256 bits. Grassl et al.~\cite{grassl2016applying} describe their quantum circuit implementation of the \esbox and other components, resulting in a full description of all three instances of \aes (but no testable code has been released). Grassl et al. take care to reduce the number of ancilla qubits required, i.e. reducing the circuit  \emph{width} as much as possible. The recent improvements by Langenberg et al.~\cite{cryptoeprint:2019:854} build on the work by Grassl et al. with similar objectives.  

In this section, we describe our implementation of \aes in the quantum programming language \qsharp~\cite{qsharp}. Some of the components are taken from the description in~\cite{grassl2016applying}, while others are implemented independently, or ported from other sources. We take the circuit description from~\cite{grassl2016applying} as the basis for our work and compare to the results in~\cite{cryptoeprint:2019:854}. In general, we aim at reducing the \emph{depth} of the \aes circuit, while limitations on width are less important. Width restrictions are not explicitly considered by the NIST call for proposals~\cite[\S~4.A.5]{NIST:PQ16c}.

The internal state of \aes contains 128 bits, arranged in four 32-bit (or 4-byte) words. In the rest of this section, when referring to a `word', we intend a 4-byte word. In all tables below, we denote by \#CNOT, the number of CNOT gates, by \#1qCliff the number of 1-qubit Clifford gates, by \#T the number of T gates, by \#M the number of measurement operations and by width the number of qubits.

\subsection{\esbox, ByteSub and SubByte}
As mentioned above, the AES \esbox comprises a transformation that inverts the input as an element of $\FF_{256}$, and maps 0 to 0. The \esbox is the only source of T gates in a quantum circuit of \aes. On classical hardware, it can be implemented easily using a lookup-table. Yet, on a quantum computer, this is not efficient (see \cite{babbush2018encoding}, \cite{low2018trading} and \cite{gidney2019windowed}). Alternatively, the inversion can be computed either by using some variant of Euclid's algorithm (taking care of the special case of 0), or by applying Lagrange's theorem and raising the input to the $(|\FF_{256}^{\times}|-1)^{th}$ power (i.e. the $254^{th}$ power), which incidentally also takes care of the 0 input. Grassl et al.~\cite{grassl2016applying} suggest to use an Itoh-Tsujii inversion algorithm~\cite{IT88}, following~\cite{amento2012efficient}, and compute all required multiplications over $\FF_2[x]/(x^8 + x^4 + x^3 + x + 1)$. This idea had already been extensively explored in the vast\footnote{E.g.~see~\cite{rijmen2000efficient}, \cite{satoh2001compact}, \cite{boyar2010new}, \cite{boyar2019small}, \cite{jeon2010compact}, \cite{nogami2010mixed}, \cite{ueno2015highly}, \cite{reyhani2018smashing}, \cite{reyhani2018new}, \cite{cryptoeprint:2019:738}.}
literature on hardware design for \aes, and requires a different construction of $\FF_{256}$ to be most effective. Following this lead, we port the \esbox circuit by Boyar and Peralta from~\cite{BP12} to \qsharp. The specified linear program combining AND and XOR operations can be easily expressed as a sequence of equivalent CNOT and AND operations (we could use T-depth-1 CCNOT gates~\cite{CCNOT}, but instead opt for overall cheaper T-depth-1 AND gates~\cite{CCNOT,AND2}, see \S\ref{sec:and-gate}). Cost estimates for the \aes \esbox can be found in Table~\ref{tab:aes-sbox}. We compare to our own \qsharp implementation of the \esbox circuits from~\cite{grassl2016applying} and~\cite{cryptoeprint:2019:854}. ByteSub is a state-wide parallel application of the \esbox, requiring new output ancillas to store the result, while SubByte is a similar word-wide application of the \esbox.

\begin{table}
	\centering
	\renewcommand{\tabcolsep}{0.05in}
	\renewcommand{\arraystretch}{1.3}
		\begin{tabular}{lrrrrrrrrrr}
			\toprule
			operation & \#CNOT & \#1qCliff & \#T & \#M & T-depth & full depth & width  \\ \midrule
			\cite{grassl2016applying} \esbox & 8683 & 1028 & 3584 & 0 & 217 & 1692 & 44 \\
			\cite{boyar2010new} \esbox & 818 & 264 & 164 & 41 & 35 & 497 & 41 \\
			\cite{BP12} \esbox & 654 & 184 & 136 & 34 & 6 & 101 & 137 \\
			\midrule
	\end{tabular}
	\caption{Comparison of our reconstruction of the original \cite{grassl2016applying} \esbox circuit with the one from~\cite{boyar2010new} as used in~\cite{cryptoeprint:2019:854} and the one in this work based on~\cite{BP12}. In our implementation of~\cite{boyar2010new} from~\cite{cryptoeprint:2019:854}, we replace CCNOT gates with AND gates to allow a fairer comparison.}
	\label{tab:aes-sbox}
\end{table}

\begin{remark}
Langenberg et al.~\cite{cryptoeprint:2019:854} independently introduced a new \aes quantum circuit design using the \esbox circuit proposed in~\cite{boyar2010new}. They also present a ProjectQ~\cite{steiger2018projectq} implementation of the \esbox, albeit without unit tests. We ported their source code to \qsharp, tested and costed it. For a fairer comparison, we replaced their CCNOT gates with the AND gate design that our circuits use. Cost estimates can be found in Table~\ref{tab:aes-sbox}. Overall, the~\cite{BP12} \esbox leads to a more cost effective circuit for our purposes in both the $G$-cost and $DW$-cost metrics, and hence we did not proceed further in our analysis of costs using the~\cite{boyar2010new} design. Note that the results obtained here differ from the ones presented in~\cite[\S3.2]{cryptoeprint:2019:854}. This is due to the difference in counting gates and depth. While \cite{cryptoeprint:2019:854} counts Toffoli gates, the \qsharp resource estimator costs at a lower level of T gates and also counts all gates needed to implement a Toffoli gate. 
\end{remark}

\subsection{ShiftRow and RotByte}
ShiftRow is a permutation on the full 128-bit \aes state, happening across its four words~\cite[\S4.2.2]{AESProposal}. As a permutation of qubits, it can be entirely encoded as rewiring. Like Grassl et al.~\cite{grassl2016applying}, we consider rewiring as free and do not include it in our cost estimates. Similarly, RotByte is a circular left shift of a word by 8 bits, and can be implemented by appropriate rewiring as well.

\subsection{MixColumn}
The operation MixColumn interprets each word in the state as a polynomial in $\FF_{256}[x]/(x^4+1)$. Each word is multiplied by a fixed polynomial $c(x)$~\cite[\S~4.2.3]{AESProposal}. Since the latter is coprime to $x^4+1$, this operation can be seen as an invertible linear transformation, and hence can be implemented in place by a PLU decomposition of a matrix in $\FF_2^{32 \times 32}$. To simplify this tedious operation, we use SageMath~\cite{SageMath} code that performs the PLU decomposition, and outputs equivalent \qsharp code. Note that \cite{grassl2016applying} describes the same technique, while achieving a significantly smaller design than the one we obtain (ref. Table~\ref{tab:mixcolumn}), but we were not able to reproduce these results. However, highly optimized, shallower circuits have been proposed in the hardware design literature such as~\cite{jean2017bit}, \cite{ToSC:KLSW17}, \cite{cryptoeprint:2019:856}, \cite{cryptoeprint:2018:1143}, \cite{cryptoeprint:2019:847}. Hence, we chose to use one of those and experiment with a recent design by Maximov~\cite{cryptoeprint:2019:833}. Both circuits are costed independently in Table~\ref{tab:mixcolumn}. Maximov's circuit has a much lower depth, but it only reduces the total depth, does not reduce the T-depth (which is already 0) and comes at the cost of an increased width. Our experiments show that without a depth restriction, it seems advantageous to use the in-place version to minimize both $G$-cost and $DW$-cost metrics, while for a depth restricted setting, Maximov's circuit seems better due to the square in the depth term in Equation~\eqref{eq:DWcost_depthrest}.

\begin{table}
\centering
\renewcommand{\tabcolsep}{0.05in}
\renewcommand{\arraystretch}{1.3}
	\begin{tabular}{lrrrrrrrrrr}
		\toprule
		operation & \#CNOT & \#1qCliff & \#T & \#M & T-depth & full depth & width  \\ \midrule
		In-place MixColumn & 1108 & 0 & 0 & 0 & 0 & 111 & 128 \\
		\cite{cryptoeprint:2019:833} MixColumn & 1248 & 0 & 0 & 0 & 0 & 22 & 318 \\
		\midrule
\end{tabular}
\caption{Comparison of an in-place implementation of MixColumn (via PLU decomposition) versus the recent shallow out-of-place design in~\cite{cryptoeprint:2019:833}.}
\label{tab:mixcolumn}
\end{table}

\subsection{AddRoundKey}
AddRoundKey performs a bitwise XOR of a round key to the internal \aes state and can be realized with a parallel application of 128 CNOT gates, controlled on the round key qubits and targeted on the state qubits. Grassl et al.~\cite{grassl2016applying} and Langenberg et al.~\cite{cryptoeprint:2019:854} use the same approach.

\subsection{KeyExpansion}\label{sec:aes-key-expansion}
Key expansion is one of the two sources of T gates in the design of \aes, and hence might have a strong impact on the overall efficiency of the circuit. A simple solution to implement KeyExpansion consists of allocating enough ancilla qubits to store the full expanded key, including all round keys. This option is easy to implement, and has relatively low depth, but clearly uses more qubits than necessary. The authors of~\cite{grassl2016applying} propose an approach that caches only those key bytes that require \esbox evaluations, which amortizes the width cost. 

Instead, we propose to minimize width by not requiring ancilla qubits at all. At the same time, we are able to reduce the depth in comparison with the naive key expansion using ancilla qubits for all key bits as described above.

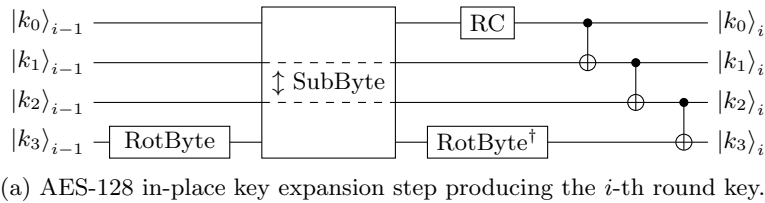
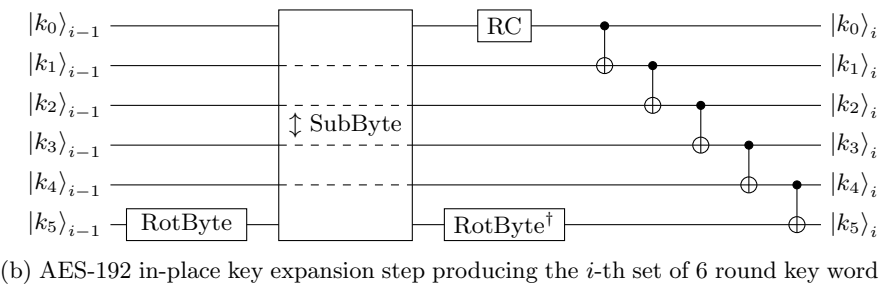
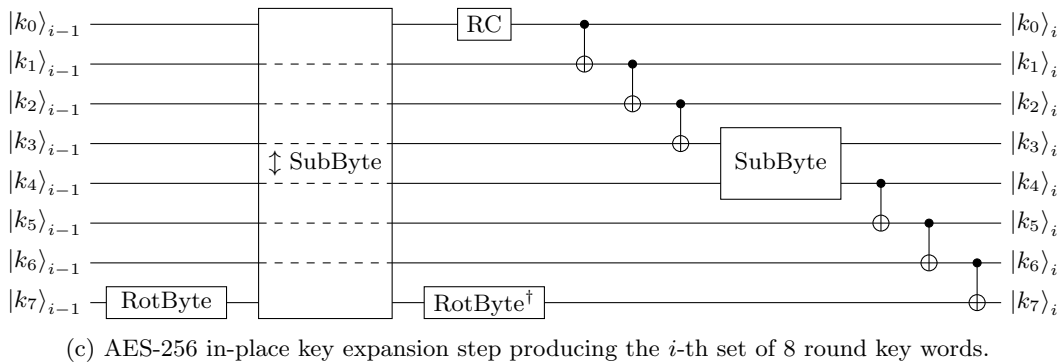
\begin{figure}
	\centering
	\subfloat[\aes-128 in-place key expansion step producing the $i$-th round key.]{\input{diagrams/aes/key_expansion_128.tikz}}\\
	\subfloat[\aes-192 in-place key expansion step producing the $i$-th set of $6$ round key words.]{\input{diagrams/aes/key_expansion_192.tikz}}\\
	\subfloat[\aes-256 in-place key expansion step producing the $i$-th set of $8$ round key words.]{\input{diagrams/aes/key_expansion_256.tikz}}\\
	\caption{In-place \aes key expansion for \aes-128, \aes-192, and \aes-256, deriving the $i^{th}$ set of $N_k$ round key words from the ${(i-1)}^{th}$ set. Each $\ket{k_j}_i$ represents the $j^{th}$ word of $\ket{k}_i$. SubByte takes the input state on the top wire, and returns the output on the bottom wire, while $\updownarrow$~SubByte takes inputs on the bottom wire, and returns outputs on the top. Dashed lines indicate that the corresponding wire is not used in the $\updownarrow$~SubByte operation. RC represents the round constant addition, which is implemented by applying X gates as appropriate.} 
	\label{fig:aes_key_expansion}
\end{figure}

Let $\ket{k}_0$ denote the \aes key consisting of $N_k \in \{4,6,8\}$ key words and $\ket{k}_i$ the $i$-th set of $N_k$ consecutive round key words. The first such block $\ket{k}_1$ can be computed in-place as shown in the appropriately sized circuit in Figure~\ref{fig:aes_key_expansion}. This circuit produces the $i$-th set of $N_k$ key words from the $(i-1)$-th set. Note that for \aes-128, these sets correspond to the actual round keys as the key size is equal to the block size, for \aes-192 and \aes-256, each round key set generates more words than needed in a single round key. 
The full operation mapping $\ket{k}_{i-1} \mapsto \ket{k}_i$ is denoted by \KE. As for the two larger key sizes, each round only needs parts of these sets of round key words, we specify $\KE_j^l$ to denote the part of the operation \KE that produces the words $j \dots l$ of the new set, disregarding other words. $\KE_j^l$ can be used as part of the round strategy from \S\ref{sec:aes-rounds} to only compute as many words of the round key as necessary, resulting in an overall narrower and shallower circuit. A comparison of this strategy and the naive KeyExpansion can be found in \S\ref{sec:aes-in-place-vs-widest}.

\begin{remark}
In addition to improving the \esbox circuit over~\cite{grassl2016applying}, Langenberg et al.~\cite[\S4]{cryptoeprint:2019:854} achieve significant savings by reducing the number of qubits and the depth of key expansion. This is achieved by an improved scheduling of key expansion during \aes encryption, namely by computing round key words only at the time they are required and un-computing them early. While their method is based on the one in~\cite{grassl2016applying} using ancilla qubits for the round keys, our approach works completely in place and reduces width and depth at the same time.     
\end{remark}

\subsection{Round, FinalRound and full \aes}\label{sec:aes-rounds}
To encrypt a message block using \aes-128 (resp. -192, -256), we initially XOR the input message with the first 4 words of the key, and then execute 10 (resp. 12, 14) rounds consisting of ByteSub, ShiftRow, MixColumn (except in the final round) and AddRoundKey. The C-like pseudo-code from~\cite[\S 4.4]{AESProposal} is reported in simplified fashion in \S\ref{sec:aes-algorithm},~Algorithm~\ref{alg:aes-encrypt}. The quantum circuits for \aes we propose follow the same blueprint with the exception that key expansion is interleaved with the algorithm in such a way that the operations $\KE_j^l$ only produce the key words that are immediately required. 

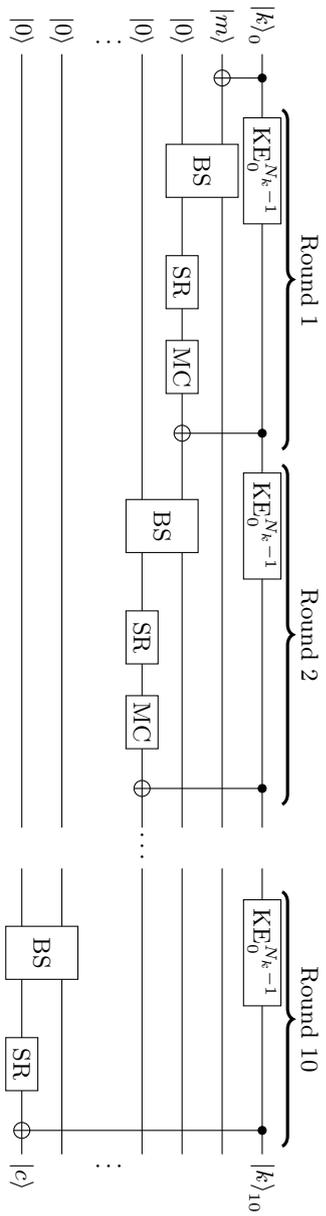
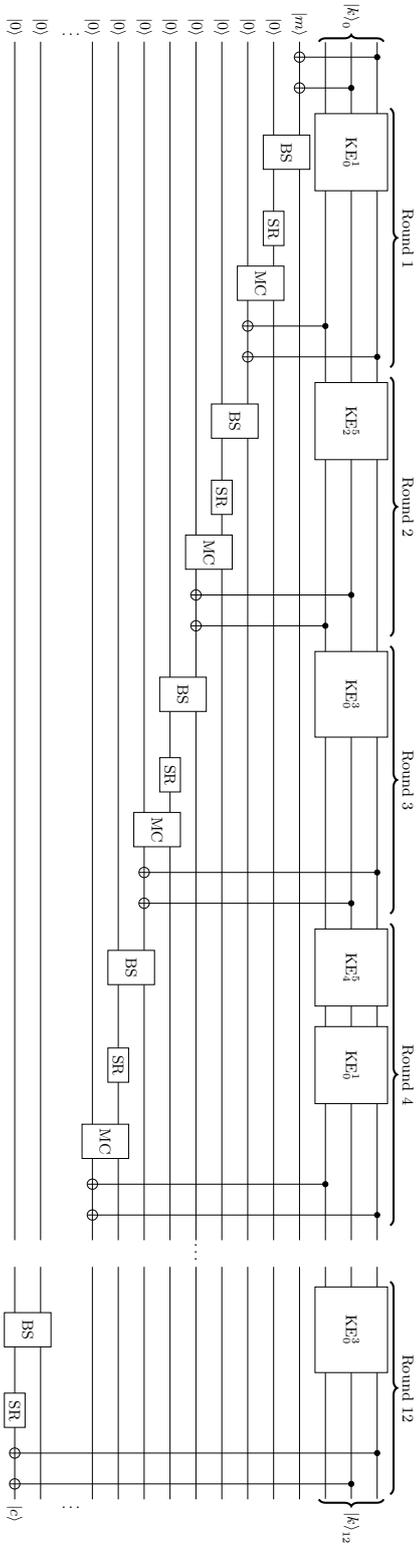
\begin{sidewaysfigure}
	\centering
	\subfloat[\aes-128 operation.]{\resizebox{!}{!}{\input{diagrams/aes/aes128.tikz}}}\\
	\subfloat[\aes-192 operation.]{\resizebox{\textwidth}{!}{\input{diagrams/aes/aes192_maximov.tikz}}}\\
	\caption{Circuit sketches for the \aes-128 and \aes-192 operation. Each wire under the $\ket{k}_0$ label represents 4 words of the key for \aes-128 and 2 words for \aes-192. Each subsequent wire (initially labeled $\ket{m}$ and $\ket{0}$) represents 4 words. CNOT gates between word-sized wires should be read as multiple parallel CNOT gates applied bitwise (e.g. at the beginning of \aes-192 the intention is of XORing 128 bits from $\ket{k}_0$ onto the state). BS stands for ByteSub, SR for ShiftRow and MC for MixColumn. For \aes-128, the circuit shows an in-place implementation of MixColumn, while for \aes-192, it uses an out-of-place version like Maximov's MixColumn linear program~\cite{cryptoeprint:2019:833}.}
	\label{fig:aes-encrypt-128-192}
\end{sidewaysfigure}

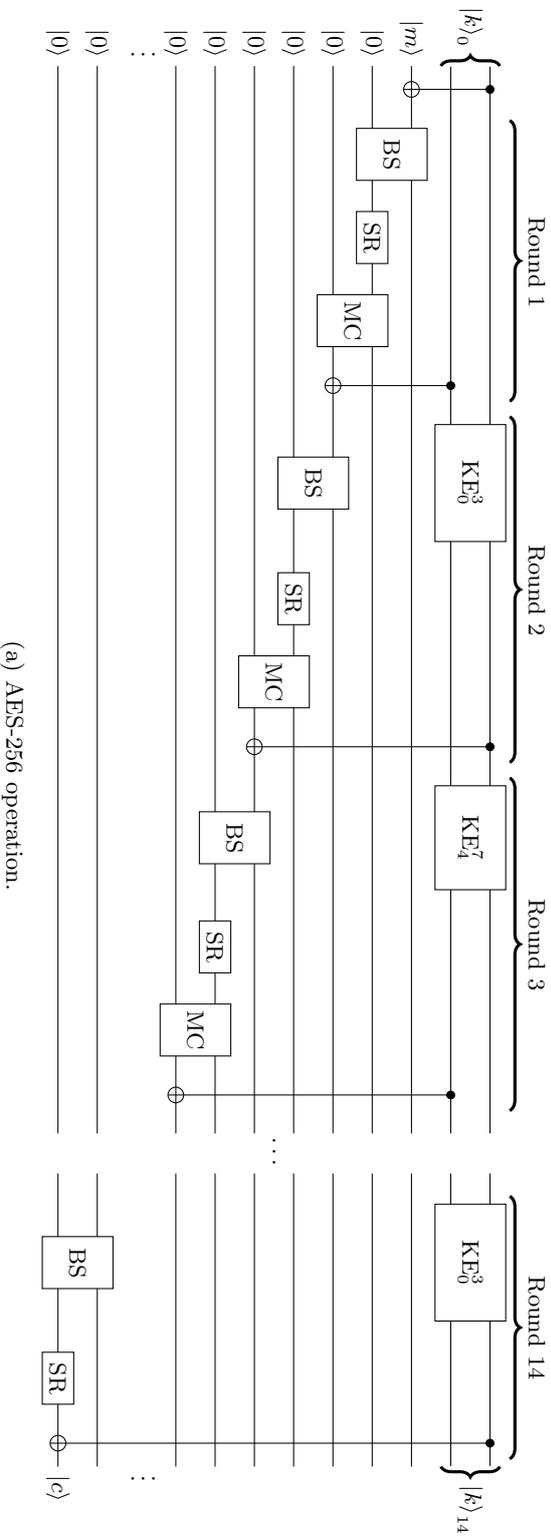
\begin{sidewaysfigure}
	\centering
	\subfloat[\aes-256 operation.]{\resizebox{\textwidth}{!}{\input{diagrams/aes/aes256_maximov.tikz}}}
	\caption{Circuit sketch for the \aes-256 operation. Each wire under the $\ket{k}_0$ label represents 4 words of the key. Each subsequent wire (initially labeled $\ket{m}$ and $\ket{0}$) represents 4 words. CNOT gates between word-sized wires should be read as multiple parallel CNOT gates applied bitwise. The value of $N_k$ is 8. BS stands for ByteSub, SR for ShiftRow and MC for MixColumn. MixColumn uses an out-of-place version like Maximov's MixColumn linear program~\cite{cryptoeprint:2019:833}.}
	\label{fig:aes-encrypt-256}
\end{sidewaysfigure}

The resulting circuits are shown in Figures~\ref{fig:aes-encrypt-128-192} and \ref{fig:aes-encrypt-256}. For formatting reasons, we omit the repeating round pattern, and only represent a subset of the full set of qubits used. In \aes-128, each round is identical until round 9. In \aes-192 rounds 5, 8 and 11 use the same \KE call and order as round 2; rounds 6 and 9 do as round 3; rounds 7 and 10 do as round 4. In \aes-256, rounds 4, 6, 8, 10, 12 (resp. 5, 7, 9, 11, 13) use the same \KE call and order as round 2 (resp. 3). Cost estimates for the resulting \aes encryption circuits can be found in Table~\ref{tab:aes-encrypt}. In contrast to \cite{grassl2016applying} and \cite{cryptoeprint:2019:854}, we aim to reduce circuit depth, hence un-computing of rounds is delayed until the output ciphertext is produced. In order to allow easier testability and modularity, the Round circuit is divided into two parts: a ForwardRound operator that computes the output state but does not clean ancilla qubits, and its adjoint. For unit-testing Round in isolation, we compose ForwardRound with its adjoint operator. For testing \aes, we first run all ForwardRound instances without ancilla cleaning, resulting in a similar Forward\aes operator, copy out the ciphertext, and then undo the Forward\aes operation.

Table~\ref{tab:aes-encrypt} presents results for the \aes circuit for both versions of MixColumn, the in-place implementation using a PLU decomposition as well as Maximov's out-of-place, but lower depth circuit. We keep both because, depending on the application one or the other is preferable. The full depth corresponds to the value $\mathsf{G}_D$ as in \S\ref{sec:costmetrics} and \S\ref{sec:groverparallel}, while width corresponds to $\mathsf{G}_W$. While for \aes-128 and \aes-192, $\mathsf{G}_D\mathsf{G}_W$ is smaller for the in-place implementation, $\mathsf{G}_D^2\mathsf{G}_W$ is smaller for Maximov's circuit. Hence, \S\ref{sec:groverparallel} indicates that when optimizing the $DW$-cost metric with depth restriction, Maximov's circuit should be preferred. If there is no depth restriction, though, the in-place design leads to a lower $DW$-cost.

\begin{table}
	\centering
	\renewcommand{\tabcolsep}{0.05in}
	\renewcommand{\arraystretch}{1.3}
		\begin{tabular}{lrrrrrrrrrr}
			\toprule
			operation & \#CNOT & \#1qCliff & \#T & \#M & T-depth & full depth & width  \\ \midrule
			\aes-128 (in-place MC) & 291150 & 83116 & 54400 & 13600 & 120 & 2827 & 1785 \\
			\aes-192 (in-place MC) & 328612 & 93160 & 60928 & 15232 & 120 & 2987 & 2105 \\
			\aes-256 (in-place MC) & 402878 & 114778 & 75072 & 18768 & 126 & 3353 & 2425 \\
			\aes-128 (Maximov's MC) & 293730 & 83236 & 54400 & 13600 & 120 & 2094 & 2937 \\
			\aes-192 (Maximov's MC) & 331752 & 93280 & 60928 & 15232 & 120 & 1879 & 3513 \\
			\aes-256 (Maximov's MC) & 406288 & 114318 & 75072 & 18768 & 126 & 1955 & 4089 \\
			\midrule
	\end{tabular}
	\caption{Circuit cost estimates for the \aes operator, using the \cite{BP12} \esbox and either an in-place or Maximov's~\cite{cryptoeprint:2019:833} MixColumn design. A discussion of the apparently inconsistent T-depth can be found in \S\ref{sec:aes-total-t-depth}.}
	\label{tab:aes-encrypt}
\end{table}

\subsection{T depth.}\label{sec:aes-total-t-depth}
Every round of \aes (as implemented in Figures~\ref{fig:aes-encrypt-128-192} and \ref{fig:aes-encrypt-256}) computes at least one layer of {\esbox}es as part of ByteSub, which must later be uncomputed. We would thus expect the T-depth of $n$ rounds of \aes to be $2n$ times the T-depth of the \esbox. Instead, Table~\ref{tab:aes-encrypt} shows smaller depths. We find this effect when using both the AND circuit and the standard library's CCNOT implementation, which passes unit tests. To test if this is a bug, we used a placeholder \esbox circuit which has an arbitrary T-depth $d$ and which the compiler cannot parallelize (see \S\ref{sef:dummy-sbox} for the design). This ``dummy`` \aes design had the expected T-depth of $2n \cdot d$. Thus we believe the \qsharp compiler found non-trivial parallelization between components of the \esbox and the surrounding circuit. This provides a strong case for full explicit implementations of quantum cryptanalytic algorithms in \qsharp or other languages that allow automatic resource estimates and optimizations; in our case the T-depth of AES-256 is 25\% less than naively expected.

\section{A quantum circuit for \lowmc}\label{sec:lowmc}
\lowmc~\cite{albrecht2015ciphers,cryptoeprint:2016:687} is a family of block ciphers aiming to result in low multiplicative complexity circuits. Originally designed to reduce the high cost of binary multiplication in the MPC and FHE scenarios, it has been adopted as a fundamental component by the \picnic  signature scheme (see \cite{chase2017post} and \cite{_NISTPQC-R1:Picnic17}) proposed for standardization as part of the NIST process for standardizing post-quantum cryptography. 

To achieve low multiplicative complexity, \lowmc proposes an S-box layer of AND-depth 1, which contains a user-defined number of parallel 3-bit \esbox computations. In general, any instantiation of \lowmc comprises a specific number of rounds, each consisting of calls to an \esbox layer, an affine transformation, and a round key addition. Key-scheduling can either be precomputed or computed on the fly. In this work, we study the original \lowmc design. This results in a sub-optimal circuit, which can clearly be improved by porting the more recent version from~\cite{_EC19:DKPRRYV} instead. Even for the original \lowmc, our work shows that the overhead imposed by the cost of the Grover oracle is very small, in particular under the T-depth metric. Since \lowmc has the potential for being standardized as a component of Picnic, we deem it appropriate to point out the differences in Grover oracle cost between different block ciphers and that generalization from \aes requires caution.

In this section we describe our \qsharp implementation of the \lowmc instances used as part of \picnic. In particular, \picnic proposes three parameter sets, with $(\text{key size}, \text{block size}) \in \{(128, 128), (192, 192), (256, 256)\}$. 

\subsection{\esbox and \mbox{S-boxLayer}}
\begin{figure}
	\centering
	\resizebox{\textwidth}{!}{
		\subfloat[\lowmc in-place \esbox.]{\label{fig:lowmc-sbox-in-place}\input{diagrams/lowmc/in_place_sbox.tikz}}
		\subfloat[\lowmc T-depth 1 \esbox.]{\label{fig:lowmc-sbox-t-depth-1}\input{diagrams/lowmc/sbox.tikz}}
	}
	\caption{Alternative quantum circuit designs for the \lowmc \esbox. We note that the in-place design does require ancilla qubits as part of the concrete CCNOT implementation.}
	\label{fig:lowmc-sbox}
\end{figure}
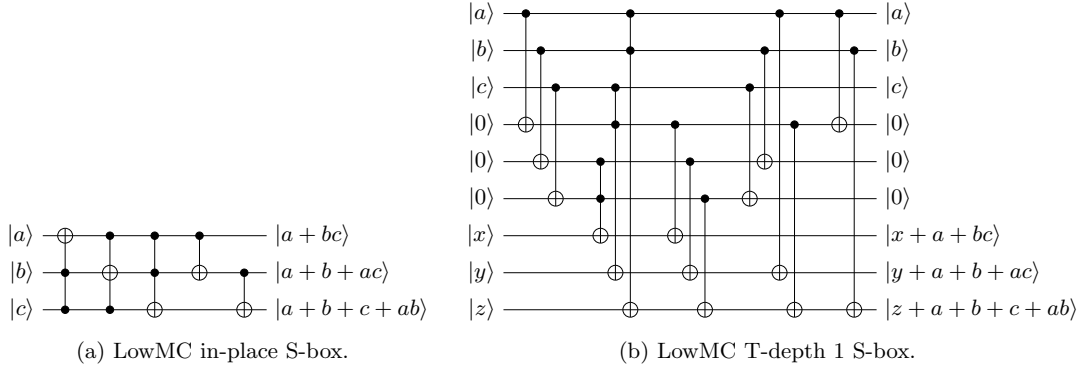

The \lowmc \esbox can be naturally implemented using Toffoli (CCNOT) gates. In particular, a simple in-place implementation with depth 5 (T-depth 3) is shown in Figure~\ref{fig:lowmc-sbox}, alongside a T-depth 1 out-of-place circuit, both of which were produced manually. Costs for both circuits can be found in Table~\ref{tab:lowmc-sbox}. We use the CCNOT implementation from~\cite{CCNOT}, which does not use measurements.
In the case of \lowmc inside of \picnic, the full \mbox{S-boxLayer} consists of 10 parallel \mbox{S-boxes} run on the 30 low order bits of the state.

\begin{table}
	\centering
	\renewcommand{\tabcolsep}{0.05in}
	\renewcommand{\arraystretch}{1.3}
	\begin{tabular}{lrrrrrrrrrr}
		\toprule
		operation & \#CNOT & \#1qCliff & \#T & \#M & T-depth & full depth & width  \\ \midrule
		In-place \esbox & 50 & 6 & 21 & 0 & 3 & 23 & 7 \\
		Shallow \esbox & 60 & 6 & 21 & 0 & 1 & 11 & 13 \\
		\midrule
	\end{tabular}
	\caption{Cost estimates for a single \lowmc \esbox circuit, following the two designs proposed in Figure~\ref{fig:lowmc-sbox}. We note that the circuit size may seem different at first sight due to Figure~\ref{fig:lowmc-sbox} not displaying the concrete CCNOT implementation.}
	\label{tab:lowmc-sbox}
\end{table}

\begin{table}
	\centering
	\renewcommand{\tabcolsep}{0.05in}
	\renewcommand{\arraystretch}{1.3}
	\begin{tabular}{lrrrrrrrrrr}
		\toprule
		operation & \#CNOT & \#1qCliff & \#T & \#M & T-depth & full depth & width  \\ \midrule
		AffineLayer L1 R1 & 8093 & 60 & 0 & 0 & 0 & 2365 & 128 \\
		AffineLayer L3 R1 & 18080 & 90 & 0 & 0 & 0 & 5301 & 192 \\
		AffineLayer L5 R1 & 32714 & 137 & 0 & 0 & 0 & 8603 & 256 \\
		\midrule
	\end{tabular}
	\caption{Costs for in-place circuits implementing the first round (R1) AffineLayer transformation for the three instantiations of \lowmc used in \picnic.}
	\label{tab:lowmc-affinelayer}
\end{table}

\begin{table}
	\centering
	\renewcommand{\tabcolsep}{0.05in}
	\renewcommand{\arraystretch}{1.3}
	\begin{tabular}{lrrrrrrrrrr}
		\toprule
		operation & \#CNOT & \#1qCliff & \#T & \#M & T-depth & full depth & width  \\ \midrule
		KeyExpansion L1 R1 & 8104 & 0 & 0 & 0 & 0 & 2438 & 128 \\
		KeyExpansion L3 R1 & 18242 & 0 & 0 & 0 & 0 & 4896 & 192 \\
		KeyExpansion L5 R1 & 32525 & 0 & 0 & 0 & 0 & 9358 & 256 \\
		\midrule
	\end{tabular}
	\caption{Costs for in-place circuits implementing the first round (R1) KeyExpansion operation for the three instantiations of \lowmc used in \picnic.}
	\label{tab:lowmc-keyexpansion}
\end{table}

\subsection{LinearLayer, ConstantAddition and AffineLayer}
\lowmc applies an affine transformation called AffineLayer to the state at every round. It consists of a matrix multiplication (LinearLayer) and the addition of a constant vector (ConstantAddition). Both matrix and vector are different for every round and are predefined constants that are populated pseudo-randomly. ConstantAddition can be implemented with the appropriate application of X gates wherever the vector's corresponding entry is a 1. In \picnic, for every round and every parameter set, all LinearLayer matrices are invertible (due to \lowmc's specification requirements), and hence the matrix multiplication can be implemented via a PLU decomposition. Cost estimates for the first round affine transformation in \lowmc as used in \picnic can be found in Table~\ref{tab:lowmc-affinelayer}.

\subsection{KeyExpansion and KeyAddition}
To generate the round keys, for each round, the \lowmc key $k$ is multiplied by a different key derivation pseudo-random matrix. In round $i$, the matrix $KM_i$ is used to compute round key $rk_i = KM_i\cdot k$. Again, for Picnic, just like the matrices used in the LinearLayer transformation, all $KM_i$ are invertible. Hence, we can use a PLU decomposition again. Furthermore, to reduce width of the circuit, rather than computing the $i^{th}$ round key $rk_i$ as $KM_i \cdot k$, we compute it in place from $rk_{i-1}$ as $rk_i = KM_i \cdot KM_{i-1}^{-1} \cdot rk_{i-1} $ by generating the PLU decomposition of each $ KM_i \cdot KM_{i-1}^{-1} $, sparing us unnecessary extra matrix multiplications or wires. We call this operation KeyExpansion. KeyAddition is equivalent to AddRoundKey in \aes, and is implemented the same way. Cost estimates for the first round key expansion in \lowmc as used in \picnic can be found in Table~\ref{tab:lowmc-keyexpansion}.

\subsection{Round and \lowmc}
A \lowmc round consists of sequentially applying \mbox{S-boxLayer}, AffineLayer and KeyAddition to the state. In our implementation, we also run KeyExpansion before AffineLayer. A full \lowmc encryption can be obtained by first adding the \lowmc key $k$ to the message, producing the initial state, and then running the specified number of rounds on the latter. Costs of the resulting encryption circuit are shown in Table~\ref{tab:lowmc-encrypt}.

\begin{table}
\centering
\renewcommand{\tabcolsep}{0.05in}
\renewcommand{\arraystretch}{1.3}
\begin{tabular}{lrrrrrrrrrr}
	\toprule
	operation & \#CNOT & \#1qCliff & \#T & \#M & T-depth & full depth & width  \\ \midrule
	\lowmc L1 & 689944 & 4932 & 8400 & 0 & 40 & 98699 & 991 \\
	\lowmc L3 & 2271870 & 9398 & 12600 & 0 & 60 & 319317 & 1483 \\
	\lowmc L5 & 5070324 & 14274 & 15960 & 0 & 76 & 693471 & 1915 \\
	\midrule
\end{tabular}
\caption{Costs for the full encryption circuit for \lowmc as used in \picnic.}
\label{tab:lowmc-encrypt}
\end{table}

\section{Grover oracles and key search resource estimates}
Equipped with \qsharp implementations of the \aes and \lowmc encryption circuits, this section describes the implementation of full Grover oracles for both block ciphers. Eventually, based on the cost estimates obtained automatically from these \qsharp Grover oracles, we provide quantum resource estimates for full key search attacks via Grover's algorithm. Beyond comparing to previous work, our emphasis is on evaluating algorithms that respect a total depth limit, for which we consider the values for \nistmaxdepth as proposed by NIST in~\cite{NIST:PQ16c}. This necessarily means that we must include parallelization, which we assume to use inner parallelization via splitting up the search space, see \S\ref{sec:groverparallel}.

\subsection{Grover oracles}\label{subsec:grovercosts}
\begin{figure}
    \centering
    \input{diagrams/aes/oracle.tikz}
    \caption{Grover oracle construction from \aes using two message-ciphertext pairs. Fw\aes represents the Forward\aes operator described in \S\ref{sec:aes-rounds}. The middle operator ``$=$'' compares the output of \aes with the provided ciphertexts and flips the target qubit if they are equal.}
    \label{fig:aes-grover-oracle}
\end{figure}
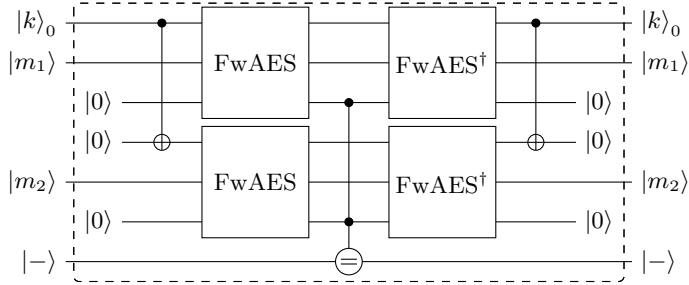

As discussed in \S\ref{subsec:blockcipher} and \S\ref{sec:groverparallel}, we must determine the parameter $r$, the number of known plaintext-ciphertext pairs that are required for a successful key-recovery attack. The Grover oracle encrypts $r$ plaintext blocks under the same candidate key and computes a Boolean value that encodes whether all $r$ resulting ciphertext blocks match the given classical results. Having a circuit for the block cipher allows us to build the oracle for an arbitrary number $r$ in a simple fashion by fanning out the key qubits to the $r$ instances and running the $r$ block cipher circuits in parallel. Then a comparison operation with the classical ciphertexts conditionally flips the result qubit and the $r$ encryptions are un-computed.
Figure~\ref{fig:aes-grover-oracle} shows the construction for \aes and $r = 2$, using the ForwardAES operation from \S\ref{sec:aes-rounds}. 

\subsubsection{The required number of plaintext-ciphertext blocks.}
The explicit computation of the probabilities in Equation~\eqref{pBinomial} shows that it using $r=2$ for \aes-$128$, $r=2$ for \aes-$192$ and $r=3$ for \aes-$256$ will guarantee a unique key with overwhelming probability. The probabilities that there are no spurious keys are $1-\epsilon$, where $\epsilon < 2^{-128}$, $\epsilon < 2^{-64}$ and $\epsilon < 2^{-128}$, respectively. Grassl et al.~\cite[\S~3.1]{grassl2016applying} used $r=3$, $r=4$ and $r=5$, respectively. Hence, these values are too large and the Grover oracle can work correctly with fewer full \aes evaluations. 

If one is content with a success probability lower than 1, it suffices to use $r=\ceil{k/n}$ blocks of plaintext-ciphertext pairs. In this case, it is enough to use $r=1$ for \aes-$128$, $r=2$ for \aes-$192$ and $r=2$ for \aes-$256$. Langenberg et al.~\cite{cryptoeprint:2019:854} also propose these values.
As an example, if we use $r=1$ for \aes-128, the probability of not having spurious keys is $1/e\approx 0.368$, which could be a high enough chance for a successful attack in certain scenarios, e.g., when there is a strict limit on the width of the attack circuit. 
Furthermore, when a large number of parallel machines are used in an instance of the attack, as discussed in \S\ref{sec:groverparallel}, even the value $r=1$ can be enough in order to guarantee with high probability that the relevant subset of the key space contains the correct key as a unique solution.

The \lowmc parameter sets we consider here all have $k=n$. Therefore, $r=2$ plaintext-ciphertext pairs are enough for all three sets ($k \in \{128,192,256\}$). Then, the probability that the key is unique is $1-\epsilon$, where $\epsilon < 2^{-k}$, i.e. this probability is negligibly close to 1. With high parallelization, $r=1$ is sufficient for a success probability very close to 1.

\subsubsection{Grover oracle cost for \aes.} Table~\ref{tab:aes-oracle} shows the resources needed for the full \aes Grover oracle for the relevant values of $r\in \{1,2,3\}$. Even without parallelization, more than 2 pairs are never required for \aes-128 and \aes-192. The same holds for 4 or more pairs for \aes-256. 

\subsubsection{Grover oracle cost for \lowmc.} The resources for our implementation of the full \lowmc Grover oracle for the relevant values of $r\in \{1,2\}$ are presented in Table~\ref{tab:lowmc-oracle}. In any setting, we never need more than $r=2$ plaintext-ciphertext pairs.

\begin{table}
	\centering
	\renewcommand{\tabcolsep}{0.05in}
	\renewcommand{\arraystretch}{1.3}
	\begin{tabular}{lrrrrrrrrrr}
		\toprule
		operation & \#CNOT & \#1qCliff & \#T & \#M & T-depth & full depth & width  \\ \midrule
		\aes-128 oracle (IP MC, $r = 1$) & 292313 & 84428 & 54908 & 13727 & 121 & 2816 & 1665 \\
		\aes-192 oracle (IP MC, $r = 1$) & 329697 & 94316 & 61436 & 15359 & 120 & 2978 & 1985 \\
		\aes-256 oracle (IP MC, $r = 1$) & 404139 & 116286 & 75580 & 18895 & 126 & 3353 & 2305 \\
		\aes-128 oracle (IP MC, $r = 2$) & 585051 & 169184 & 109820 & 27455 & 121 & 2815 & 3329 \\
		\aes-192 oracle (IP MC, $r = 2$) & 659727 & 188520 & 122876 & 30719 & 120 & 2981 & 3969 \\
		\aes-256 oracle (IP MC, $r = 2$) & 808071 & 231124 & 151164 & 37791 & 126 & 3356 & 4609 \\
		\aes-256 oracle (IP MC, $r = 3$) & 1212905 & 347766 & 226748 & 56687 & 126 & 3347 & 6913 \\
		\aes-128 oracle (M's MC, $r = 1$) & 294863 & 84488 & 54908 & 13727 & 121 & 2086 & 2817 \\
		\aes-192 oracle (M's MC, $r = 1$) & 332665 & 94092 & 61436 & 15359 & 120 & 1879 & 3393 \\
		\aes-256 oracle (M's MC, $r = 1$) & 407667 & 116062 & 75580 & 18895 & 126 & 1951 & 3969 \\
		\aes-128 oracle (M's MC, $r = 2$) & 589643 & 168288 & 109820 & 27455 & 121 & 2096 & 5633 \\
		\aes-192 oracle (M's MC, $r = 2$) & 665899 & 188544 & 122876 & 30719 & 120 & 1890 & 6785 \\
		\aes-256 oracle (M's MC, $r = 2$) & 815645 & 231712 & 151164 & 37791 & 126 & 1952 & 7937 \\
		\aes-256 oracle (M's MC, $r = 3$) & 1223087 & 346290 & 226748 & 56687 & 126 & 1956 & 11905 \\
		\midrule
	\end{tabular}
	\caption{Cost estimates for the \aes Grover oracle operator for $r =$~$1$, $2$ and $3$ plaintext-ciphertext pairs. ``IP MC'' (resp. ``M's MC'') means the oracle uses an in-place (resp. Maximov's~\cite{cryptoeprint:2019:833}) MixColumn design.} 
	\label{tab:aes-oracle}
\end{table}

\begin{table}
	\centering
	\renewcommand{\tabcolsep}{0.05in}
	\renewcommand{\arraystretch}{1.3}
	\begin{tabular}{lrrrrrrrrrr}
		\toprule
		operation & \#CNOT & \#1qCliff & \#T & \#M & T-depth & full depth & width  \\ \midrule
		\lowmc L1 oracle ($r = 1$) & 690961 & 5917 & 8908 & 191 & 41 & 98709 & 1585 \\
		\lowmc L3 oracle ($r = 1$) & 2273397 & 10881 & 13364 & 286 & 61 & 319323 & 2377 \\
		\lowmc L5 oracle ($r = 1$) & 5072343 & 16209 & 16980 & 372 & 77 & 693477 & 3049 \\
		\lowmc L1 oracle ($r = 2$) & 1382143 & 11774 & 17820 & 362 & 41 & 98707 & 3169 \\
		\lowmc L3 oracle ($r = 2$) & 4547191 & 21783 & 26732 & 576 & 61 & 319329 & 4753 \\
		\lowmc L5 oracle ($r = 2$) & 10145281 & 32567 & 33964 & 783 & 77 & 693483 & 6097 \\
		\midrule
	\end{tabular}
	\caption{Cost estimates for the \lowmc Grover oracle operator for $r =$~$1$ and $2$ plaintext-ciphertext pairs. \lowmc parameter sets are as used in Picnic.}
	\label{tab:lowmc-oracle}
\end{table}

\subsection{Cost estimates for block cipher key search}
Using the cost estimates for the \aes and \lowmc Grover oracles from \S\ref{subsec:grovercosts}, this section provides cost estimates for full key search attacks on both block ciphers. For the sake of a direct comparison to the previous results in~\cite{grassl2016applying} and \cite{cryptoeprint:2019:854}, we first ignore any limit on the depth and present the same setting as in these works. Then, we provide cost estimates with imposed depth limits and the consequential parallelization requirements.

\subsubsection{Comparison to previous work.}
Table~\ref{tab:aes-all-grover} shows cost estimates for a full run of Grover's algorithm when using $\floor{\frac{\pi}{4}2^{k/2}}$ iterations of the \aes Grover operator without parallelization. We only take into account the costs imposed by the oracle operator $U_f$ (in the notation of \S\ref{sec:grover}) and ignore the costs of the operator $2\dyad{\psi}{\psi}- I$. If the number of plaintext-ciphertext pairs ensures a unique key, this number of operations maximizes the success probability $p_{\mathrm{succ}}$ to be negligibly close to $1$. For smaller values of $r$ such as those proposed in \cite{cryptoeprint:2019:854}, the success probability is given by the probability that the key is unique.

The $G$-cost is the total number of gates, which is the sum of the first three columns in the table, corresponding to the numbers of 1-qubit Clifford and CNOT gates, T gates and measurements. Table~\ref{tab:aes-all-grover} shows that the $G$-cost is always better in our work when comparing values for the same \aes instance and the same value for $r$. The same holds for the $DW$-cost as we increase the width by factors less than $4$ and simultaneously reduce the depth by more than that.

\begin{table}
	\centering
	\renewcommand{\tabcolsep}{0.05in}
	\renewcommand{\arraystretch}{1.3}
	\resizebox{\textwidth}{!}{
		\begin{tabular}{lcccccccccc}
			\toprule
			\multicolumn{10}{c}{Grassl et al.~\cite{grassl2016applying}} \\ \midrule
			scheme & \#(1qCliff+CNOT) & \#T & \#M & T-depth & full depth & width & $G$-cost & $DW$-cost & $p_{\mathrm{succ}}$ \\ \midrule
			\aes-128 ($r = 3$) & $1.55\,\cdot\,2^{86}$ & $1.19\,\cdot\,2^{86}$ & $0$ & $1.06\,\cdot\,2^{80}$ & $1.16\,\cdot\,2^{81}$ & $2\,953$ & $1.37\,\cdot\,2^{87}$ & $1.67\,\cdot\,2^{92}$ & $\approx 1$ \\
			\aes-192 ($r = 4$) & $1.17\,\cdot\,2^{119}$ & $1.81\,\cdot\,2^{118}$ & $0$ & $1.21\,\cdot\,2^{112}$ & $1.33\,\cdot\,2^{113}$ & $4\,449$ & $1.04\,\cdot\,2^{120}$ & $1.44\,\cdot\,2^{125}$ & $\approx 1$\\
			\aes-256 ($r = 5$) & $1.83\,\cdot\,2^{151}$ & $1.41\,\cdot\,2^{151}$ & $0$ & $1.44\,\cdot\,2^{144}$ & $1.57\,\cdot\,2^{145}$ & $6\,681$ & $1.62\,\cdot\,2^{152}$ & $1.28\,\cdot\,2^{158}$ & $\approx 1$ \\
			\midrule
			\multicolumn{10}{c}{Langenberg et al.~\cite{cryptoeprint:2019:854}} \\ \midrule
			\aes-128 ($r = 1$) & $1.46\,\cdot\,2^{82}$ & $1.47\,\cdot\,2^{81}$ & $0$ & $1.44\,\cdot\,2^{77}$ & $1.39\,\cdot\,2^{79}$ & $865$ & $1.10\,\cdot\,2^{83}$ & $1.17\,\cdot\,2^{89}$ & $\approx 1/e$ \\
			\aes-192 ($r = 2$) & $1.71\,\cdot\,2^{115}$ & $1.68\,\cdot\,2^{114}$ & $0$ & $1.26\,\cdot\,2^{109}$ & $1.23\,\cdot\,2^{111}$ & $1\,793$ & $1.27\,\cdot\,2^{116}$ & $1.08\,\cdot\,2^{122}$ & $\approx 1$\\
			\aes-256 ($r = 2$) & $1.03\,\cdot\,2^{148}$ & $1.02\,\cdot\,2^{147}$ & $0$ & $1.66\,\cdot\,2^{141}$ & $1.61\,\cdot\,2^{143}$ & $2\,465$ & $1.54\,\cdot\,2^{148}$ & $1.94\,\cdot\,2^{154}$ & $\approx 1/e$ \\
			\midrule
			\multicolumn{10}{c}{this work} \\ \midrule
			\aes-128 (IP MC, $r = 1$) & $1.13\,\cdot\,2^{82}$ & $1.32\,\cdot\,2^{79}$ & $1.32\,\cdot\,2^{77}$ & $1.48\,\cdot\,2^{70}$ & $1.08\,\cdot\,2^{75}$ & $1665$ & $1.33\,\cdot\,2^{82}$ & $1.76\,\cdot\,2^{85}$ & $\approx 1/e$ \\
			\aes-128 (IP MC, $r = 2$) & $1.13\,\cdot\,2^{83}$ & $1.32\,\cdot\,2^{80}$ & $1.32\,\cdot\,2^{78}$ & $1.48\,\cdot\,2^{70}$ & $1.08\,\cdot\,2^{75}$ & $3329$ & $1.34\,\cdot\,2^{83}$ & $1.75\,\cdot\,2^{86}$ & $\approx 1$ \\
			\aes-192 (IP MC, $r = 2$) & $1.27\,\cdot\,2^{115}$ & $1.47\,\cdot\,2^{112}$ & $1.47\,\cdot\,2^{110}$ & $1.47\,\cdot\,2^{102}$ & $1.14\,\cdot\,2^{107}$ & $3969$ & $1.50\,\cdot\,2^{115}$ & $1.11\,\cdot\,2^{119}$ & $\approx 1$ \\
			\aes-256 (IP MC, $r = 2$) & $1.56\,\cdot\,2^{147}$ & $1.81\,\cdot\,2^{144}$ & $1.81\,\cdot\,2^{142}$ & $1.55\,\cdot\,2^{134}$ & $1.29\,\cdot\,2^{139}$ & $4609$ & $1.84\,\cdot\,2^{147}$ & $1.45\,\cdot\,2^{151}$ & $\approx 1/e$ \\
			\aes-256 (IP MC, $r = 3$) & $1.17\,\cdot\,2^{148}$ & $1.36\,\cdot\,2^{145}$ & $1.36\,\cdot\,2^{143}$ & $1.55\,\cdot\,2^{134}$ & $1.28\,\cdot\,2^{139}$ & $6913$ & $1.38\,\cdot\,2^{148}$ & $1.08\,\cdot\,2^{152}$ & $\approx 1$ \\
			\midrule
		\end{tabular}
	}
	\caption{Comparison of cost estimates for Grover's algorithm with $\floor{\frac{\pi}{4}2^{k/2}}$ \aes oracle iterations for attacks with high success probability, without a depth restriction. CNOT and 1-qubit Clifford gate counts are added to allow easier comparison to the previous work from~\cite{grassl2016applying} and \cite{cryptoeprint:2019:854}, who report both kinds of gates under ``Clifford''. \cite{cryptoeprint:2019:854} uses the \esbox design from \cite{boyar2010new}. ``IP MC'' means the oracle uses an in-place MixColumn design.} 	\label{tab:aes-all-grover}
\end{table}

Table~\ref{tab:lowmc-grover} shows cost estimates for \lowmc in the same setting. Despite \lowmc's lower multiplicative complexity and a relatively lower number of T gates, the large number of CNOT gates leads to overall higher $G$-cost and $DW$-cost than \aes, as we count all gates.

\begin{table}
	\centering
	\renewcommand{\tabcolsep}{0.05in}
	\renewcommand{\arraystretch}{1.3}
	\resizebox{\textwidth}{!}{
		\begin{tabular}{lrrrrrrrrrrr}
			\toprule
			scheme & \# CNOT & \#1qCliff & \#T & \#M & T-depth & full depth & width & $G$-cost & $DW$-cost & $p_{\mathrm{succ}}$ \\ \midrule
			\lowmc L1 ($r = 1$) & $1.04\,\cdot\,2^{83}$ & $1.13\,\cdot\,2^{76}$ & $1.71\,\cdot\,2^{76}$ & $1.17\,\cdot\,2^{71}$ & $1.01\,\cdot\,2^{69}$ & $1.18\,\cdot\,2^{80}$ & $1585$ & $1.06\,\cdot\,2^{83}$ & $1.83\,\cdot\,2^{90}$ & $\approx 1/e$ \\
			\lowmc L3 ($r = 1$) & $1.70\,\cdot\,2^{116}$ & $1.04\,\cdot\,2^{109}$ & $1.28\,\cdot\,2^{109}$ & $1.75\,\cdot\,2^{103}$ & $1.50\,\cdot\,2^{101}$ & $1.91\,\cdot\,2^{113}$ & $2377$ & $1.72\,\cdot\,2^{116}$ & $1.11\,\cdot\,2^{125}$ & $\approx 1/e$ \\
			\lowmc L5 ($r = 1$) & $1.90\,\cdot\,2^{149}$ & $1.55\,\cdot\,2^{141}$ & $1.63\,\cdot\,2^{141}$ & $1.14\,\cdot\,2^{136}$ & $1.89\,\cdot\,2^{133}$ & $1.04\,\cdot\,2^{147}$ & $3049$ & $1.91\,\cdot\,2^{149}$ & $1.55\,\cdot\,2^{158}$ & $\approx 1/e$ \\
			\lowmc L1 ($r = 2$) & $1.04\,\cdot\,2^{84}$ & $1.13\,\cdot\,2^{77}$ & $1.71\,\cdot\,2^{77}$ & $1.11\,\cdot\,2^{72}$ & $1.01\,\cdot\,2^{69}$ & $1.18\,\cdot\,2^{80}$ & $3169$ & $1.06\,\cdot\,2^{84}$ & $1.83\,\cdot\,2^{91}$ & $\approx 1$ \\
			\lowmc L3 ($r = 2$) & $1.70\,\cdot\,2^{117}$ & $1.04\,\cdot\,2^{110}$ & $1.28\,\cdot\,2^{110}$ & $1.77\,\cdot\,2^{104}$ & $1.50\,\cdot\,2^{101}$ & $1.91\,\cdot\,2^{113}$ & $4753$ & $1.72\,\cdot\,2^{117}$ & $1.11\,\cdot\,2^{126}$ & $\approx 1$ \\
			\lowmc L5 ($r = 2$) & $1.90\,\cdot\,2^{150}$ & $1.56\,\cdot\,2^{142}$ & $1.63\,\cdot\,2^{142}$ & $1.20\,\cdot\,2^{137}$ & $1.89\,\cdot\,2^{133}$ & $1.04\,\cdot\,2^{147}$ & $6097$ & $1.91\,\cdot\,2^{150}$ & $1.55\,\cdot\,2^{159}$ & $\approx 1$ \\
			\midrule
		\end{tabular}
	}
	\caption{Cost estimates for Grover's algorithm with $\floor{\frac{\pi}{4}2^{k/2}}$ \lowmc oracle iterations for attacks with high success probability, without a depth restriction.}
	\label{tab:lowmc-grover}
\end{table}

\subsubsection{Cost estimates under a depth limit.} 
Tables~\ref{tbl:full_grover_aes_maxdepth} and \ref{tbl:full_grover_lowmc_maxdepth} show cost estimates for running Grover's algorithm against \aes and \lowmc under a given depth limit. This restriction is proposed in the NIST call for proposals for standardization of post-quantum cryptography~\cite{NIST:PQ16c}. We use the notation and example values for \nistmaxdepth from the call. Imposing a depth limit forces the parallelization of Grover's algorithm, which we assume uses inner parallelization, see \S\ref{sec:groverparallel}.

The values in the table are determined as explained in \S\ref{sec:costmetrics}. Given cost estimates $\mathsf{G}_G$, $\mathsf{G}_D$ and $\mathsf{G}_W$ for the oracle circuit, we determine the maximal number of Grover iterations that can be carried out within the \nistmaxdepth limit. Then the required number $S$ of parallel instances is computed via Equation~\eqref{eq:instances} and the $G$-cost and $DW$-cost follow from Equations~\eqref{eq:Gcost_depthrest} and \eqref{eq:DWcost_depthrest}.
The number $r$ of plaintext-ciphertext pairs is the minimal value such that the probability $\mathrm{SKP}$ for having spurious keys in the subset of the key space that holds the target key is less than $2^{-20}$.

The impact of imposing a depth limit on the key search algorithm can directly be seen by comparing, for example Table~\ref{tbl:full_grover_aes_maxdepth} with Table~\ref{tab:aes-all-grover} in the case of \aes. Key search against \aes-128 without depth limit has a $G$-cost of $1.34\cdot 2^{83}$ gates and a $DW$-cost of $1.75\cdot 2^{86}$ qubit-\depthunits. Now, setting $\nistmaxdepth = 2^{40}$ increases both the $G$-cost and the $DW$-cost by a factor of roughly $2^{34}$ to $1.07\cdot 2^{117}$ gates and $1.76\cdot 2^{120}$ qubit-\depthunits. For $\nistmaxdepth = 2^{64}$, the increase is by a factor of roughly $2^{10}$. We note that for $\nistmaxdepth = 2^{96}$, key search on \aes-128 does not require any parallelization.

\subsubsection{Implications for post-quantum security categories.}
The security strength categories 1, 3 and 5 in the NIST call for proposals~\cite{NIST:PQ16c} are defined by the resources needed for key search on \aes-128, \aes-192 and \aes-256, respectively. For a cryptographic scheme to satisfy the security requirement at a given level, the best known attack must take at least as many resources as key search against the corresponding \aes instance. 

As guidance, NIST provides a table with gate cost estimates via a formula depending on the depth bound \nistmaxdepth. This formula is deduced as follows: assume that non-parallel Grover search requires a depth of $D = x \cdot \nistmaxdepth$ for some $x\ge 1$ and the circuit has $G$ gates. Then, about $x^2$ machines are needed that each run for a fraction $1/x$ of the time and use roughly $G/x$ gates in order for the quantum attack to fit within the depth budget given by \nistmaxdepth while attaining the same attack success probability. Hence, the total gate count for a parallelized Grover search is roughly $(G/x) \cdot x^2 = G \cdot D / \nistmaxdepth$. The cost formula reported in the NIST table (also provided in Table~\ref{tbl:nist-consequences} for reference) is deduced by using the values for $G$-cost and depth $D$ from Grassl et al.~\cite{grassl2016applying}. 

The above formula does not take into account that parallelization often allows us to reduce the number of required plaintext-ciphertext pairs, resulting in a $G$-cost reduction for search in each parallel Grover instance by a factor larger than $x$. Note also that \cite[Footnote~5]{NIST:PQ16c} mentions that using the formula for very small values of $x$ (very large values of \nistmaxdepth such that $D/\nistmaxdepth < 1$, where no parallelization is required) underestimates the quantum security of \aes. This is the case for \aes-128 with $\nistmaxdepth = 2^{96}$.

In Table~\ref{tbl:nist-consequences}, we compare NIST's numbers with our gate counts for parallel Grover search. Our results for each specific setting incorporate the reduction of plaintext-ciphertext pairs through parallelization, provide the correct cost if parallelization is not necessary and use improved circuit designs. The table shows that for most situations, \aes is less quantum secure than the NIST estimates predict. For each category, we provide a very rough approximation formula that could be used to replace NIST's formula. We observe a consistent reduction in $G$-cost for quantum key search by 11-13 bits. 

Since NIST clearly defines its security categories 1, 3 and 5 based on the computational resources required for key search on \aes, the explicit gate counts should be lowered to account for the best known attack.
This would mean that it is now easier for submitters to claim equivalent security, with the exception of category 1 with $\nistmaxdepth = 2^{96}$. A possible consequence of our work is that some of the NIST submissions might profit from slightly tweaking certain parameter sets to allow more efficient implementations, while at the same time satisfying the (now weaker) requirements for their intended security category.

\begin{remark}
	If NIST replaces its explicit gate cost estimates for \aes with the ones obtained in this work, key recovery against the instances of \lowmc we implemented requires at least as many gates as key recovery against \aes with the same key size, as can be seen from the $G$-cost results for the full depth in Table~\ref{tbl:full_grover_lowmc_maxdepth}.
	
	On the other hand, the same results show that these \lowmc instances do not meet the explicit gate count requirements for the original NIST post-quantum security categories. For example, \lowmc L1 can be broken with an attack having $G$-cost $1.25\cdot2^{123}$ in depth $\nistmaxdepth=2^{40}$, while the original NIST bound in category 1 requires a scheme to not be broken by an attack using less than $2^{130}$ gates. In all settings considered here, a \lowmc key can be found with a slightly smaller $G$-cost than NIST's original estimates for \aes, again with the exception when no parallelization is needed. The margin is relatively small.

	Yet, a final conclusion about the relative security between \lowmc and \aes should be deferred until quantum circuits for \lowmc have been optimized to a similar extent as the ones for \aes.
\end{remark}

\begin{table}
    \centering
    \renewcommand{\tabcolsep}{0.05in}
	\renewcommand{\arraystretch}{1.3}
	\begin{tabular}{l@{\hskip2em}l@{\hskip2em}r@{\hskip2em}r@{\hskip2em}r@{\hskip2em}r}
		\toprule
		NIST Security &  &  \multicolumn{3}{c}{$G$-cost for \nistmaxdepth} & \\
		 Strength Category & source & $2^{40}$ & $2^{64}$ & $2^{96}$  & approximation\\
		\midrule
		\multirow{2}{*}{1 \aes-128} & \cite{NIST:PQ16c} & $2^{130}$ & $2^{106}$ & $2^{74}$ & $2^{170}/\nistmaxdepth$\\
		& this work & $1.07 \cdot 2^{117}$ & $1.07 \cdot 2^{93}$ & $^*1.34 \cdot 2^{83}$ & $\approx 2^{157}/\nistmaxdepth$\\\midrule
		\multirow{2}{*}{3 \aes-192} & \cite{NIST:PQ16c} & $2^{193}$ & $2^{169}$ & $2^{137}$ & $2^{233}/\nistmaxdepth$ \\
		& this work & $1.09 \cdot 2^{181}$ & $1.09 \cdot 2^{157}$ & $1.09 \cdot 2^{126}$ & $\approx 2^{221}/\nistmaxdepth$\\\midrule
		\multirow{2}{*}{5 \aes-256} & \cite{NIST:PQ16c} & $2^{258}$ & $2^{234} $&  $2^{202}$ & $2^{298}/\nistmaxdepth$\\
		& this work & $1.39 \cdot 2^{245}$ & $1.39 \cdot 2^{221}$ & $1.39 \cdot 2^{190}$ & $\approx 2^{285}/\nistmaxdepth$\\
		\midrule
	\end{tabular}
	\caption{Comparison of our cost estimate results with NIST's approximations based on Grassl et al.~\cite{grassl2016applying}. The column entitled \emph{approximation} displays the formula used by NIST in \cite{NIST:PQ16c} for NIST numbers and a rough approximation that would replace the NIST formula based on our results. Note that \aes-128 under $\nistmaxdepth=2^{96}$ is a special case as the attack does not require any parallelization and its cost is underestimated by the approximation.}
	\label{tbl:nist-consequences}
\end{table}

\begin{table}
    \centering
    \renewcommand{\tabcolsep}{0.05in}
	\renewcommand{\arraystretch}{1.3}
	\subfloat[The depth cost metric is the full depth $D$. All circuits use Maximov's~\cite{cryptoeprint:2019:833} MixColumns implementation, except for \aes-128 at $\nistmaxdepth=2^{96}$ for which in-place MixColumns gives lower costs.]{
        \begin{tabular}{lccccccccc}
            \toprule
            scheme & \texttt{MD} & $r$ & $S$ & $\log_2{(\text{SKP})}$ & $D$ & $W$ & $G$-cost & $DW$-cost \\ \midrule
            \aes-128 & $2^{40}$ &  $1$ & $1.28 \cdot 2^{69}$ & $-69.36$ & $1.00 \cdot 2^{40}$ & $1.76 \cdot 2^{80}$ & $1.07 \cdot 2^{117}$ & $1.76 \cdot 2^{120}$ \\ 
            \aes-192 & $2^{40}$ &  $1$ & $1.04 \cdot 2^{133}$ & $-69.05$ & $1.00 \cdot 2^{40}$ & $1.72 \cdot 2^{144}$ & $1.09 \cdot 2^{181}$ & $1.72 \cdot 2^{184}$ \\ 
            \aes-256 & $2^{40}$ &  $1$ & $1.12 \cdot 2^{197}$ & $-69.16$ & $1.00 \cdot 2^{40}$ & $1.08 \cdot 2^{209}$ & $1.39 \cdot 2^{245}$ & $1.08 \cdot 2^{249}$ \\ 
            \midrule
            \aes-128 & $2^{64}$ &  $1$ & $1.28 \cdot 2^{21}$ & $-21.36$ & $1.00 \cdot 2^{64}$ & $1.76 \cdot 2^{32}$ & $1.07 \cdot 2^{93}$ & $1.76 \cdot 2^{96}$ \\ 
            \aes-192 & $2^{64}$ &  $1$ & $1.04 \cdot 2^{85}$ & $-21.05$ & $1.00 \cdot 2^{64}$ & $1.72 \cdot 2^{96}$ & $1.09 \cdot 2^{157}$ & $1.72 \cdot 2^{160}$ \\ 
            \aes-256 & $2^{64}$ &  $1$ & $1.12 \cdot 2^{149}$ & $-21.16$ & $1.00 \cdot 2^{64}$ & $1.08 \cdot 2^{161}$ & $1.39 \cdot 2^{221}$ & $1.08 \cdot 2^{225}$ \\ 
            \midrule
            \aes-128* & $2^{96}$ &  $2$ & $1.00 \cdot 2^{0}$ & $-\infty$ & $1.08 \cdot 2^{75}$ & $1.63 \cdot 2^{11}$ & $1.34 \cdot 2^{83}$ & $1.75 \cdot 2^{86}$ \\ 
            \aes-192 & $2^{96}$ &  $2$ & $1.05 \cdot 2^{21}$ & $-\infty$ & $1.00 \cdot 2^{96}$ & $1.74 \cdot 2^{33}$ & $1.09 \cdot 2^{126}$ & $1.74 \cdot 2^{129}$ \\ 
            \aes-256 & $2^{96}$ &  $2$ & $1.12 \cdot 2^{85}$ & $-85.16$ & $1.00 \cdot 2^{96}$ & $1.09 \cdot 2^{98}$ & $1.39 \cdot 2^{190}$ & $1.09 \cdot 2^{194}$ \\ 
            \midrule
        \end{tabular}
	} \\
	\subfloat[The depth cost metric is the T depth $T$-$D$ only. All circuits use the in-place MixColumns implementation.]{
        \begin{tabular}{lccccccccc}
            \toprule
            scheme & \texttt{MD} & $r$ & $S$ & $\log_2{(\text{SKP})}$ & $T$-$D$ & $W$ & $G$-cost & $T$-$DW$-cost \\ \midrule
            \aes-128 & $2^{40}$ &  $1$ & $1.10 \cdot 2^{61}$ & $-61.14$ & $1.00 \cdot 2^{40}$ & $1.79 \cdot 2^{71}$ & $1.98 \cdot 2^{112}$ & $1.79 \cdot 2^{111}$ \\ 
            \aes-192 & $2^{40}$ &  $1$ & $1.08 \cdot 2^{125}$ & $-61.12$ & $1.00 \cdot 2^{40}$ & $1.05 \cdot 2^{136}$ & $1.10 \cdot 2^{177}$ & $1.05 \cdot 2^{176}$ \\ 
            \aes-256 & $2^{40}$ &  $1$ & $1.20 \cdot 2^{189}$ & $-61.26$ & $1.00 \cdot 2^{40}$ & $1.35 \cdot 2^{200}$ & $1.42 \cdot 2^{241}$ & $1.35 \cdot 2^{240}$ \\ 
            \midrule
            \aes-128 & $2^{64}$ &  $2$ & $1.10 \cdot 2^{13}$ & $-\infty$ & $1.00 \cdot 2^{64}$ & $1.79 \cdot 2^{24}$ & $1.98 \cdot 2^{89}$ & $1.79 \cdot 2^{88}$ \\ 
            \aes-192 & $2^{64}$ &  $2$ & $1.08 \cdot 2^{77}$ & $-\infty$ & $1.00 \cdot 2^{64}$ & $1.05 \cdot 2^{89}$ & $1.11 \cdot 2^{154}$ & $1.05 \cdot 2^{153}$ \\ 
            \aes-256 & $2^{64}$ &  $2$ & $1.20 \cdot 2^{141}$ & $-141.26$ & $1.00 \cdot 2^{64}$ & $1.35 \cdot 2^{153}$ & $1.42 \cdot 2^{218}$ & $1.35 \cdot 2^{217}$ \\ 
            \midrule
            \aes-128 & $2^{96}$ &  $2$ & $1.00 \cdot 2^{0}$ & $-\infty$ & $1.48 \cdot 2^{70}$ & $1.63 \cdot 2^{11}$ & $1.34 \cdot 2^{83}$ & $1.21 \cdot 2^{82}$ \\ 
            \aes-192 & $2^{96}$ &  $2$ & $1.08 \cdot 2^{13}$ & $-\infty$ & $1.00 \cdot 2^{96}$ & $1.05 \cdot 2^{25}$ & $1.11 \cdot 2^{122}$ & $1.05 \cdot 2^{121}$ \\ 
            \aes-256 & $2^{96}$ &  $2$ & $1.20 \cdot 2^{77}$ & $-77.26$ & $1.00 \cdot 2^{96}$ & $1.35 \cdot 2^{89}$ & $1.42 \cdot 2^{186}$ & $1.35 \cdot 2^{185}$ \\ 
            \midrule
        \end{tabular}
	}
	\caption{Cost estimates for parallel Grover key search against \aes under a depth limit \nistmaxdepth with \emph{inner} parallelization (see \S\ref{sec:groverparallel}). \texttt{MD} is \nistmaxdepth, $r$ is the number of plaintext-ciphertext pairs used in the Grover oracle, $S$ is the number of subsets into which the key space is divided, $\mathrm{SKP}$ is the probability that spurious keys are present in the subset holding the target key, $W$ is the qubit width of the full circuit, $D$ the full depth, $T$-$D$ the T-depth, $DW$-cost uses the full depth and $T$-$DW$-cost the T-depth. After the Grover search is completed, each of the $S$ measured candidate keys is classically checked against 2 (resp. 2, 3) plaintext-ciphertext pairs for \aes-128 (resp. -192, -256).\\}
	\label{tbl:full_grover_aes_maxdepth}
\end{table}

\begin{table}
    \centering
    \renewcommand{\tabcolsep}{0.05in}
	\renewcommand{\arraystretch}{1.3}
	\subfloat[The depth cost metric is the full depth $D$.]{
        \begin{tabular}{lccccccccc}
            \toprule
            scheme & \texttt{MD} & $r$ & $S$ & $\log_2{(\text{SKP})}$ & $D$ & $W$ & $G$-cost & $DW$-cost \\ \midrule
            \lowmc L1 & $2^{40}$ &  $1$ & $1.40 \cdot 2^{80}$ & $-80.48$ & $1.00 \cdot 2^{40}$ & $1.08 \cdot 2^{91}$ & $1.25 \cdot 2^{123}$ & $1.08 \cdot 2^{131}$ \\
            \lowmc L3 & $2^{40}$ &  $1$ & $1.83 \cdot 2^{147}$ & $-147.87$ & $1.00 \cdot 2^{40}$ & $1.06 \cdot 2^{159}$ & $1.65 \cdot 2^{190}$ & $1.06 \cdot 2^{199}$ \\
            \lowmc L5 & $2^{40}$ &  $1$ & $1.08 \cdot 2^{214}$ & $-214.11$ & $1.00 \cdot 2^{40}$ & $1.61 \cdot 2^{225}$ & $1.99 \cdot 2^{256}$ & $1.61 \cdot 2^{265}$ \\
            \midrule
            \lowmc L1 & $2^{64}$ &  $1$ & $1.40 \cdot 2^{32}$ & $-32.48$ & $1.00 \cdot 2^{64}$ & $1.08 \cdot 2^{43}$ & $1.25 \cdot 2^{99}$ & $1.08 \cdot 2^{107}$ \\
            \lowmc L3 & $2^{64}$ &  $1$ & $1.83 \cdot 2^{99}$ & $-99.87$ & $1.00 \cdot 2^{64}$ & $1.06 \cdot 2^{111}$ & $1.65 \cdot 2^{166}$ & $1.06 \cdot 2^{175}$ \\
            \lowmc L5 & $2^{64}$ &  $1$ & $1.08 \cdot 2^{166}$ & $-166.11$ & $1.00 \cdot 2^{64}$ & $1.61 \cdot 2^{177}$ & $1.99 \cdot 2^{232}$ & $1.61 \cdot 2^{241}$ \\
            \midrule
            \lowmc L1 & $2^{96}$ &  $2$ & $1.00 \cdot 2^{0}$ & $-\infty$ & $1.18 \cdot 2^{80}$ & $1.55 \cdot 2^{11}$ & $1.06 \cdot 2^{84}$ & $1.83 \cdot 2^{91}$ \\
            \lowmc L3 & $2^{96}$ &  $1$ & $1.83 \cdot 2^{35}$ & $-35.87$ & $1.00 \cdot 2^{96}$ & $1.06 \cdot 2^{47}$ & $1.65 \cdot 2^{134}$ & $1.06 \cdot 2^{143}$ \\
            \lowmc L5 & $2^{96}$ &  $1$ & $1.08 \cdot 2^{102}$ & $-102.11$ & $1.00 \cdot 2^{96}$ & $1.61 \cdot 2^{113}$ & $1.99 \cdot 2^{200}$ & $1.61 \cdot 2^{209}$ \\
            \midrule
        \end{tabular}
	}\\
	\subfloat[The depth cost metric is the T depth $T$-$D$ only.]{
        \begin{tabular}{lccccccccc}
            \toprule
            scheme & \texttt{MD} & $r$ & $S$ & $\log_2{(\text{SKP})}$ & $T$-$D$ & $W$ & $G$-cost & $T$-$DW$-cost \\ \midrule
            \lowmc L1 & $2^{40}$ &  $1$ & $1.01 \cdot 2^{58}$ & $-58.02$ & $1.00 \cdot 2^{40}$ & $1.57 \cdot 2^{68}$ & $1.06 \cdot 2^{112}$ & $1.57 \cdot 2^{108}$ \\
            \lowmc L3 & $2^{40}$ &  $1$ & $1.12 \cdot 2^{123}$ & $-123.16$ & $1.00 \cdot 2^{40}$ & $1.30 \cdot 2^{134}$ & $1.29 \cdot 2^{178}$ & $1.30 \cdot 2^{174}$ \\
            \lowmc L5 & $2^{40}$ &  $1$ & $1.79 \cdot 2^{187}$ & $-187.84$ & $1.00 \cdot 2^{40}$ & $1.33 \cdot 2^{199}$ & $1.81 \cdot 2^{243}$ & $1.33 \cdot 2^{239}$ \\
            \midrule
            \lowmc L1 & $2^{64}$ &  $2$ & $1.01 \cdot 2^{10}$ & $-\infty$ & $1.00 \cdot 2^{64}$ & $1.57 \cdot 2^{21}$ & $1.06 \cdot 2^{89}$ & $1.57 \cdot 2^{85}$ \\
            \lowmc L3 & $2^{64}$ &  $1$ & $1.12 \cdot 2^{75}$ & $-75.16$ & $1.00 \cdot 2^{64}$ & $1.30 \cdot 2^{86}$ & $1.29 \cdot 2^{154}$ & $1.30 \cdot 2^{150}$ \\
            \lowmc L5 & $2^{64}$ &  $1$ & $1.79 \cdot 2^{139}$ & $-139.84$ & $1.00 \cdot 2^{64}$ & $1.33 \cdot 2^{151}$ & $1.81 \cdot 2^{219}$ & $1.33 \cdot 2^{215}$ \\
            \midrule
            \lowmc L1 & $2^{96}$ &  $2$ & $1.00 \cdot 2^{0}$ & $-\infty$ & $1.01 \cdot 2^{69}$ & $1.55 \cdot 2^{11}$ & $1.06 \cdot 2^{84}$ & $1.56 \cdot 2^{80}$ \\
            \lowmc L3 & $2^{96}$ &  $2$ & $1.12 \cdot 2^{11}$ & $-\infty$ & $1.00 \cdot 2^{96}$ & $1.30 \cdot 2^{23}$ & $1.29 \cdot 2^{123}$ & $1.30 \cdot 2^{119}$ \\
            \lowmc L5 & $2^{96}$ &  $1$ & $1.79 \cdot 2^{75}$ & $-75.84$ & $1.00 \cdot 2^{96}$ & $1.33 \cdot 2^{87}$ & $1.81 \cdot 2^{187}$ & $1.33 \cdot 2^{183}$ \\
            \midrule
		\end{tabular}
	}
	\caption{Cost estimates for parallel Grover key search against \lowmc under a depth limit \nistmaxdepth with \emph{inner} parallelization (see \S\ref{sec:groverparallel}). \texttt{MD} is \nistmaxdepth, $r$ is the number of plaintext-ciphertext pairs used in the Grover oracle, $S$ is the number of subsets into which the key space is divided, $\mathrm{SKP}$ is the probability that spurious keys are present in the subset holding the target key, $W$ is the qubit width of the full circuit, $D$ the full depth, $T$-$D$ the T-depth, $DW$-cost uses the full depth and $T$-$DW$-cost the T-depth.  After the Grover search is completed, each of the $S$ measured candidate keys is classically checked against 2 plaintext-ciphertext pairs.\\}
	\label{tbl:full_grover_lowmc_maxdepth}
\end{table}

\section{Future work}
This work's main focus is on exploring the setting proposed by NIST where quantum attacks are limited by a total bound on the depth of quantum circuits. Previous works \cite{grassl2016applying,ASAM18,cryptoeprint:2019:854} aim to minimize cost under a tradeoff between circuit depth and a limit on the total number of qubits needed, say a hypothetical bound \nistmaxdepth. Depth limits are not discussed when choosing a Grover strategy. Since it is somewhat unclear what exact characteristics and features a future scalable quantum hardware might have, quantum circuit and Grover strategy optimization with the goal of minimizing different cost metrics under different constraints than \nistmaxdepth could be an interesting avenue for future research.

We have studied key search problems for a single target. In classical cryptanalysis, multi-target attacks have to be taken into account for assessing the security of cryptographic systems. We leave the exploration of estimating the cost of quantum multi-target attacks, for example using the algorithm by Banegas and Bernstein~\cite{banegas2017low} under \nistmaxdepth (or alternative regimes), as future work.

Furthermore, the implementation of quantum circuits for cryptanalysis in \qsharp or another quantum-focused programming language for concrete cost estimation is a worthwhile exercise to increase confidence in the security of proposed post-quantum schemes. For example, quantum lattice sieving and enumeration appear to be prime candidates.

\anonymous{}{
	\subsubsection{Acknowledgements.}
	We thank Chris Granade and Bettina Heim for their help with the \qsharp language and compiler, Mathias Soeken and Thomas H{\"a}ner for general discussions on optimizing quantum circuits and \qsharp, Mathias Soeken for providing the AND gate circuit we use, and Daniel Kales and Greg Zaverucha for their input on Picnic and \lowmc.
}


\newcommand{\etalchar}[1]{$^{#1}$}

\clearpage
\appendix

\shortversion{
\section{Additional material for \S\ref{subsec:blockcipher}}\label{sec:repeated}
\input{grover-repeated.tex}
}{}


\shortversion{
	\clearpage
	\section{Additional material for \S\ref{sec:costmetrics}}
	\input{grovercost_classical_communication.tex}
}{}

\section{\aes encryption algorithm}\label{sec:aes-algorithm}

In Algorithm~\ref{alg:aes-encrypt}, we reproduce a simplified view of \aes encryption~\cite[\S~4.4]{AESProposal}.
\begin{algorithm}[htb!]
	\caption{\aes}
	\label{alg:aes-encrypt}
	\KwIn{$m$ \ccom{message}}
	\KwIn{$k$ \ccom{key}}
	$s \leftarrow m$ \ccom{state}
	$ek \leftarrow KeyExpansion(k)$ \ccom{expanded key}
	$s \leftarrow AddRoundKey(s, k)$ \;
	\For{$i=1\dots\textnormal{total rounds}-1$}{
		$s \leftarrow Round(s, ek)$ \;
	}
	$c \leftarrow FinalRound(s, ek)$ \ccom{ciphertext}
	\Return{$c$} \;
\end{algorithm}

\section{Comparison of in-place \aes KeyExpansion vs. naive unrolling}\label{sec:aes-in-place-vs-widest}
In \S\ref{sec:aes-key-expansion}, we discuss an in-place design for the KeyExpansion routine in \aes. While this clearly saves width by not requiring ancilla qubits for the expansion, it may look as going against our design choice of minimising depth. In particular, one may think that a naive design where a register of enough ancillas is allocated such that the whole key expansion can be performed before any rounds are run could save in depth, given that it does not need to handle any particular previous state on the qubits. In Table~\ref{tab:aes-wide-vs-in-place-key-expansion}, we report numbers comparing the sizes of our \aes circuits, with the only difference being the naive vs the in-place designs for KeyExpansion, showing that the latter is shallower (and of course narrower). This is due to being able to perform the gates for the KeyExpansion in parallel to the gates run during rounds that do not depend on the output of the new key material. In particular, the \esbox computations required to expand the key can be run in parallel to those executed on the state by ByteSub.

\begin{table}
	\centering
	\renewcommand{\tabcolsep}{0.05in}
	\renewcommand{\arraystretch}{1.3}
	\begin{tabular}{lrrrrrrrrrr}
		\toprule
		operation & \#CNOT & \#1qCliff & \#T & \#M & T-depth & full depth & width  \\ \midrule
		\aes-128 (in-place KE) & 291150 & 83116 & 54400 & 13600 & 120 & 2827 & 1785 \\
		\aes-192 (in-place KE) & 328612 & 93160 & 60928 & 15232 & 120 & 2987 & 2105 \\
		\aes-256 (in-place KE) & 402878 & 114778 & 75072 & 18768 & 126 & 3353 & 2425 \\
		\aes-128 (naive KE) & 293758 & 83212 & 54400 & 13600 & 132 & 2995 & 3065 \\
		\aes-192 (naive KE) & 331496 & 93040 & 60928 & 15232 & 132 & 3113 & 3577 \\
		\aes-256 (naive KE) & 406176 & 114718 & 75072 & 18768 & 138 & 3385 & 4089 \\
		\midrule
	\end{tabular}
	\caption{Size comparison for \aes quantum circuits using ``in-place'' vs ``naive'' KeyExpansion (see \S\ref{sec:aes-key-expansion}). In both cases, an ``in-place'' MixColumn circuit is used. We notice that the difference in width between equivalent circuits corresponds to $4 \cdot 32 \cdot (Nr+1) - 32 \cdot Nk$ qubits, where $Nr$ (resp. $Nk$) is the number of \aes rounds (resp. words in the \aes key), see~\cite{AESProposal}.}
	\label{tab:aes-wide-vs-in-place-key-expansion}
\end{table}

\section{AND gate}\label{sec:and-gate}
In our \aes implementation, we use a T-depth 1 circuit for an AND gate which is a combination of Selinger~\cite{CCNOT} and Gidney~\cite{AND2}, and that was designed by \anonymous{---}{Mathias Soeken}. A diagram can be found in Figure~\ref{fig:and_gate}.
\begin{figure}
	\centering
	\subfloat[AND gate.]{\resizebox{!}{!}{\input{diagrams/general/AND.tikz}}}\\
	\subfloat[AND${}^\dagger$ gate.]{\resizebox{!}{!}{\input{diagrams/general/tweaked_adjAND.tikz}}}\\
	\anonymous{
	\caption{AND gate design used in our circuit. The circuit is a combination of Selinger~\cite{CCNOT} and Gidney~\cite{AND2}, and was designed by ---.}
	\label{fig:and_gate}
	}{
	\caption{AND gate design used in our circuit. We notice that in (b), the measurement returns a classical bit $b$ and leaves the original qubit in the state $\ket{b}$.}
	\label{fig:and_gate}}
\end{figure}

\section{Placeholder \esbox}\label{sef:dummy-sbox}
\begin{figure}
	\centering
	\input{diagrams/general/dummy_sbox.tikz}
	\caption{Dummy \esbox design, that tries to forcefully avoid non-parallel calls to the \esbox to be partially executed at the same time.}
	\label{fig:dummy-sbox}
\end{figure}
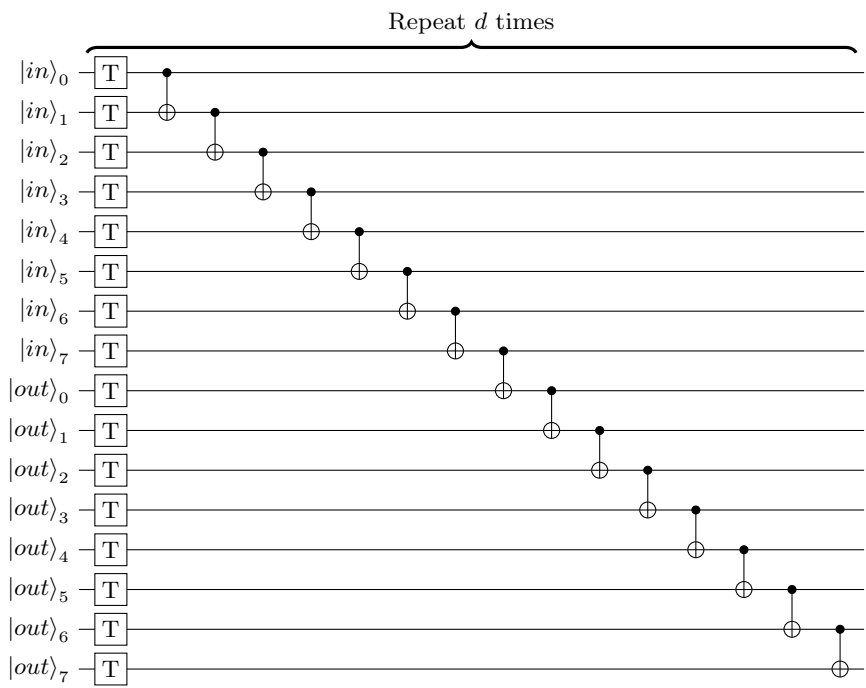

As part of our sanity checking of the \qsharp resource estimator in \S\ref{sec:aes-total-t-depth}, we replaced the \aes \esbox with the design in Figure~\ref{fig:dummy-sbox}, that tries to force all the wires to ``synchronize'' such that the T gates between two neighboring {\esbox}es cannot be partially computed in parallel. Costing the T depth of the resulting dummy \aes operation returns the expected value of $2 \times \emph{\# of rounds} \times d$, where $d$ is the depth of the dummy \esbox.

\end{document}

%% file: grover-repeated.tex
\subsubsection{Repeated measurements.}
It is shown by Boyer et al.~\cite{BBHT98} that instead of iterating $\floor{\frac{\pi}{4}\sqrt{\frac{N}{M}}}$ times, the expected number of iterations needed to find a solution is smaller if we stop early after a fixed number $j$ and repeat the algorithm until it succeeds. In this case, one expects $j/p(j)$ iterations, with a minimum of roughly $0.690\sqrt{N/M}$ expected iterations when measuring and restarting after $0.583\sqrt{N/M}$ iterations. But this observation is only useful when it is possible to run the search procedure many times. In cryptanalysis, the situation is typically different. 
In light of the above assumption of having a depth bound and the goal of achieving high success probability, a repeating strategy requiring on the average a small number of iterations is undesirable if the variance of the number of necessary iterations is high. The above optimal value of $0.583\sqrt{N/M}$ sequential quantum iterations before measuring a candidate solution identified by Boyer et al.~\cite{BBHT98} means that, if the first measurement fails and we must repeat the partial search, we end up using at least $1.166\sqrt{N/M}$ Grover iterations, exceeding $\frac{\pi}{4}\sqrt{N/M}$. 

In general, for an integer $m$ such that the depth limit allows a total of $mj$ iterations, we can decide whether to repeat a $j$-fold iteration instance of Grover's algorithm $m$ times or to use $mj$ iterations in one instance. The former succeeds with probability $1-(1-p(j))^m$, and the latter with probability $p(mj)=\sin^2((2mj+1)\theta)$. 
It can be shown by induction that, if $0<\phi<\pi/(2m)$, then $1-\sin[2](m\phi) \leq (1-\sin[2](\phi))^m$ for all $m\ge 1$;
with $\phi=(2j+1)\theta$ and the observation that $\phi\gg\theta$ it follows, that to reach a fixed probability, we are better off using more consecutive quantum iterations than measuring and repeating.

\begin{remark}\label{rmk:picnic-unicity-distance}
    While for some cryptanalytic applications, it is important to find the correct key, for others, any key that matches the plaintext-ciphertext pairs can be sufficient. For example, the \picnic signature scheme (\cite{chase2017post}, \cite{_NISTPQC-R1:Picnic17}) uses a block cipher $C$ and encrypts a message $m$ to $c$, and $(m, c)$ is the public key. The signature is a zero-knowledge proof that the signer knows a secret key $K$ such that $C_K(m)=c$. Any other key $K'$ with $C_{K'}(m)=c$ produces a valid signature for the original public key. Thus, to forge signatures, a spurious key works just as well. However, since in general, the number of spurious keys is unknown, Grover's algorithm needs to be adjusted for example as in~\cite[\S4]{BBHT98} or by running a quantum counting algorithm first~\cite[\S5]{BBHT98}. This requires repeated runs of various Grover instances. As argued above for the measure-and-repeat approach, under a total depth limitation, this reduces the success probability. 
\end{remark}

%% file: diagrams/general/linear_map_w_ancillas.tikz
\providecommand{\K}[1]{\left|#1\right\rangle}
\providecommand{\adj}[1]{#1${}^\dagger$}
\begin{tikzpicture}[scale=1.000000,x=1pt,y=1pt]
\filldraw[color=white] (0.000000, -7.500000) rectangle (84.000000, 112.500000);
\draw[color=black] (0.000000,105.000000) -- (84.000000,105.000000);
\draw[color=black] (0.000000,105.000000) node[left] {$\K{a}$};
\draw[color=black] (0.000000,90.000000) -- (84.000000,90.000000);
\draw[color=black] (0.000000,90.000000) node[left] {$\K{b}$};
\draw[color=black] (0.000000,75.000000) -- (84.000000,75.000000);
\draw[color=black] (0.000000,75.000000) node[left] {$\K{c}$};
\draw[color=black] (0.000000,60.000000) -- (84.000000,60.000000);
\draw[color=black] (0.000000,60.000000) node[left] {$\K{d}$};
\draw[color=black] (0.000000,45.000000) -- (84.000000,45.000000);
\draw[color=black] (0.000000,45.000000) node[left] {$\K{0}$};
\draw[color=black] (0.000000,30.000000) -- (84.000000,30.000000);
\draw[color=black] (0.000000,30.000000) node[left] {$\K{0}$};
\draw[color=black] (0.000000,15.000000) -- (84.000000,15.000000);
\draw[color=black] (0.000000,15.000000) node[left] {$\K{0}$};
\draw[color=black] (0.000000,0.000000) -- (84.000000,0.000000);
\draw[color=black] (0.000000,0.000000) node[left] {$\K{0}$};
\draw (9.000000,105.000000) -- (9.000000,45.000000);
\begin{scope}
\draw[fill=white] (9.000000, 45.000000) circle(3.000000pt);
\clip (9.000000, 45.000000) circle(3.000000pt);
\draw (6.000000, 45.000000) -- (12.000000, 45.000000);
\draw (9.000000, 42.000000) -- (9.000000, 48.000000);
\end{scope}
\filldraw (9.000000, 105.000000) circle(1.500000pt);
\draw (15.000000,90.000000) -- (15.000000,15.000000);
\begin{scope}
\draw[fill=white] (15.000000, 15.000000) circle(3.000000pt);
\clip (15.000000, 15.000000) circle(3.000000pt);
\draw (12.000000, 15.000000) -- (18.000000, 15.000000);
\draw (15.000000, 12.000000) -- (15.000000, 18.000000);
\end{scope}
\filldraw (15.000000, 90.000000) circle(1.500000pt);
\draw (21.000000,75.000000) -- (21.000000,30.000000);
\begin{scope}
\draw[fill=white] (21.000000, 30.000000) circle(3.000000pt);
\clip (21.000000, 30.000000) circle(3.000000pt);
\draw (18.000000, 30.000000) -- (24.000000, 30.000000);
\draw (21.000000, 27.000000) -- (21.000000, 33.000000);
\end{scope}
\filldraw (21.000000, 75.000000) circle(1.500000pt);
\draw (27.000000,60.000000) -- (27.000000,0.000000);
\begin{scope}
\draw[fill=white] (27.000000, 0.000000) circle(3.000000pt);
\clip (27.000000, 0.000000) circle(3.000000pt);
\draw (24.000000, 0.000000) -- (30.000000, 0.000000);
\draw (27.000000, -3.000000) -- (27.000000, 3.000000);
\end{scope}
\filldraw (27.000000, 60.000000) circle(1.500000pt);
\draw (45.000000,105.000000) -- (45.000000,30.000000);
\begin{scope}
\draw[fill=white] (45.000000, 30.000000) circle(3.000000pt);
\clip (45.000000, 30.000000) circle(3.000000pt);
\draw (42.000000, 30.000000) -- (48.000000, 30.000000);
\draw (45.000000, 27.000000) -- (45.000000, 33.000000);
\end{scope}
\filldraw (45.000000, 105.000000) circle(1.500000pt);
\draw (51.000000,75.000000) -- (51.000000,45.000000);
\begin{scope}
\draw[fill=white] (51.000000, 45.000000) circle(3.000000pt);
\clip (51.000000, 45.000000) circle(3.000000pt);
\draw (48.000000, 45.000000) -- (54.000000, 45.000000);
\draw (51.000000, 42.000000) -- (51.000000, 48.000000);
\end{scope}
\filldraw (51.000000, 75.000000) circle(1.500000pt);
\draw (69.000000,60.000000) -- (69.000000,45.000000);
\begin{scope}
\draw[fill=white] (69.000000, 45.000000) circle(3.000000pt);
\clip (69.000000, 45.000000) circle(3.000000pt);
\draw (66.000000, 45.000000) -- (72.000000, 45.000000);
\draw (69.000000, 42.000000) -- (69.000000, 48.000000);
\end{scope}
\filldraw (69.000000, 60.000000) circle(1.500000pt);
\draw (75.000000,105.000000) -- (75.000000,0.000000);
\begin{scope}
\draw[fill=white] (75.000000, 0.000000) circle(3.000000pt);
\clip (75.000000, 0.000000) circle(3.000000pt);
\draw (72.000000, 0.000000) -- (78.000000, 0.000000);
\draw (75.000000, -3.000000) -- (75.000000, 3.000000);
\end{scope}
\filldraw (75.000000, 105.000000) circle(1.500000pt);
\draw[color=black] (84.000000,105.000000) node[right] {$\K{a}$};
\draw[color=black] (84.000000,90.000000) node[right] {$\K{b}$};
\draw[color=black] (84.000000,75.000000) node[right] {$\K{c}$};
\draw[color=black] (84.000000,60.000000) node[right] {$\K{d}$};
\draw[color=black] (84.000000,45.000000) node[right] {$\K{a+c+d}$};
\draw[color=black] (84.000000,30.000000) node[right] {$\K{a+c}$};
\draw[color=black] (84.000000,15.000000) node[right] {$\K{b}$};
\draw[color=black] (84.000000,0.000000) node[right] {$\K{a+d}$};
\end{tikzpicture}

%% file: diagrams/general/linear_map_in_place.tikz
\providecommand{\K}[1]{\left|#1\right\rangle}
\providecommand{\adj}[1]{#1${}^\dagger$}
\begin{tikzpicture}[scale=1.000000,x=1pt,y=1pt]
\filldraw[color=white] (0.000000, -7.500000) rectangle (99.000000, 52.500000);
\draw[color=black] (0.000000,45.000000) -- (99.000000,45.000000);
\draw[color=black] (0.000000,45.000000) node[left] {$\K{a}$};
\draw[color=black,rounded corners=4.000000pt] (0.000000,30.000000) -- (78.000000,30.000000) -- (85.500000,22.500000);
\draw[color=black,rounded corners=4.000000pt] (85.500000,22.500000) -- (93.000000,15.000000) -- (99.000000,15.000000);
\draw[color=black] (0.000000,30.000000) node[left] {$\K{b}$};
\draw[color=black,rounded corners=4.000000pt] (0.000000,15.000000) -- (78.000000,15.000000) -- (85.500000,7.500000);
\draw[color=black,rounded corners=4.000000pt] (85.500000,7.500000) -- (93.000000,0.000000) -- (99.000000,0.000000);
\draw[color=black] (0.000000,15.000000) node[left] {$\K{c}$};
\draw[color=black,rounded corners=4.000000pt] (0.000000,0.000000) -- (78.000000,0.000000) -- (85.500000,15.000000);
\draw[color=black,rounded corners=4.000000pt] (85.500000,15.000000) -- (93.000000,30.000000) -- (99.000000,30.000000);
\draw[color=black] (0.000000,0.000000) node[left] {$\K{d}$};
\draw (9.000000,45.000000) -- (9.000000,15.000000);
\begin{scope}
\draw[fill=white] (9.000000, 45.000000) circle(3.000000pt);
\clip (9.000000, 45.000000) circle(3.000000pt);
\draw (6.000000, 45.000000) -- (12.000000, 45.000000);
\draw (9.000000, 42.000000) -- (9.000000, 48.000000);
\end{scope}
\filldraw (9.000000, 15.000000) circle(1.500000pt);
\draw (27.000000,45.000000) -- (27.000000,0.000000);
\begin{scope}
\draw[fill=white] (27.000000, 45.000000) circle(3.000000pt);
\clip (27.000000, 45.000000) circle(3.000000pt);
\draw (24.000000, 45.000000) -- (30.000000, 45.000000);
\draw (27.000000, 42.000000) -- (27.000000, 48.000000);
\end{scope}
\filldraw (27.000000, 0.000000) circle(1.500000pt);
\draw (45.000000,45.000000) -- (45.000000,0.000000);
\begin{scope}
\draw[fill=white] (45.000000, 0.000000) circle(3.000000pt);
\clip (45.000000, 0.000000) circle(3.000000pt);
\draw (42.000000, 0.000000) -- (48.000000, 0.000000);
\draw (45.000000, -3.000000) -- (45.000000, 3.000000);
\end{scope}
\filldraw (45.000000, 45.000000) circle(1.500000pt);
\draw (63.000000,45.000000) -- (63.000000,15.000000);
\begin{scope}
\draw[fill=white] (63.000000, 15.000000) circle(3.000000pt);
\clip (63.000000, 15.000000) circle(3.000000pt);
\draw (60.000000, 15.000000) -- (66.000000, 15.000000);
\draw (63.000000, 12.000000) -- (63.000000, 18.000000);
\end{scope}
\filldraw (63.000000, 45.000000) circle(1.500000pt);
\draw[color=black] (99.000000,45.000000) node[right] {$\K{a+c+d}$};
\draw[color=black] (99.000000,15.000000) node[right] {$\K{b}$};
\draw[color=black] (99.000000,0.000000) node[right] {$\K{a+d}$};
\draw[color=black] (99.000000,30.000000) node[right] {$\K{a+c}$};
\end{tikzpicture}

%% file: grovercost_classical_communication.tex
\subsubsection{Comparing parallel Grover search to classical search.}
In the computational model of~\cite{jaques2019quantum}, each quantum gate is interpreted as some computation done by a classical controller. For certain parameter settings, these controllers may find the key more efficiently through a classical search. Assume, this is done with a brute force algorithm, which simply iterates through all potential keys and checks if they are correct. Let $\mathsf{C}$ be the classical gate cost to test a single key. Then for a search space of size $N=2^k$, the total cost for the brute force attack to achieve success probability $p$ is $p2^k\mathsf{C}$. 
Comparing this cost to the gate cost for Grover's algorithm in Equation~\eqref{eq:total_gates}, we conclude that if we use more than $(pC/(c_p\mathsf{G}_G))^2 2^{k}$ parallel machines, Grover's algorithm is slower and more costly than a classical search on the same hardware.

Since the Grover oracle $\mathsf{G}$ includes a reversible evaluation of the block cipher and quantum computation of a function is likely more costly than its classical counterpart, we may assume that the classical gate cost $\mathsf{C}$ is smaller than the quantum gate cost $\mathsf{G}_G$ of the Grover oracle, i.e. $\mathsf{C}\leq \mathsf{G}_G$. It holds that $p/c_p < 1.45$, so $(pC/(c_p\mathsf{G}_G))^2 < 2.11$ and for $p=1$, we have $(pC/(c_p\mathsf{G}_G))^2 = 16/\pi^2\cdot C^2/\mathsf{G}_G^2 \approx 1.62\cdot C^2/\mathsf{G}_G^2 \leq 1.62$. Depending on the actual cost ratio, this bound may be in a meaningful range.  

\subsubsection{Communication cost to assemble the results in parallel Grover.}
We briefly discuss the communication cost incurred by communicating a found solution from one of the machines in a large network of parallel computers to a central processor. Each machine measures a candidate key after a specified number of Grover iterations. The classical controller then checks this key against a small number of given plaintext-ciphertext pairs in order to determine whether it is a valid solution. If the key is correct, it is communicated to a central processor. 

If the number of machines is small, the central processor simply queries each machine sequentially for the correct key. For a large number of machines, we instead assume they are connected in a binary tree structure with one machine designated as the root. The central processor queries this one for the final result. If it has measured a correct key, it is returned, otherwise it asynchronously queries two other machines which form the roots of equally-sized sub-trees, in which the same process is repeated. For $S$ machines this requires $S$ requests, but only $\lg S$ must be sequential.

We assume that the spatial arrangement of the $S$ machines is in a two-dimensional plane in form of an H tree. Furthermore, it can be assumed that communication between machines is via classical channels with very small signal propagation times. The total distance any signal must travel is proportional to the square root of the size of this tree, i.e. $\sqrt{S}$. Thus, the total time to recover the final key is $O(\lg S) + c_S \sqrt{S\mathsf{G}_W}$ \depthunits, where $c_S$ is a constant to account for the relationship between signal propagation speed and quantum gate times. For large $S$, the $O(\lg S)$ term is insignificant.

We assume that $c_S \ll 1$, meaning that these classical channels can propagate a signal across a qubit-sized distance much faster than we can apply a gate to that qubit. This means the depth of each Grover search will dwarf the communication costs so long as $S\leq 2^{k/2}\frac{c_p\mathsf{G}_D}{c_S\sqrt{\mathsf{G}_W}}$. If we use more machines than this, the communication costs dominate the depth. These costs increase with $S$ and thus $S=2^{k/2}\frac{c_p\mathsf{G}_D}{c_S\sqrt{\mathsf{G}_W}}$ gives the minimum possible depth of
\begin{equation}\label{eq:min-possible-depth}
2^{\frac{k}{4}}\sqrt{c_pc_S\mathsf{G}_D\sqrt{\mathsf{G}_W}}\text{ \depthunits.}
\end{equation}

Similar reasoning shows that a classical brute force search, which assembles its results in the same way, has a minimum depth of $2^{\frac{k}{3}}(\mathsf{G}_W\mathsf{C}c_S^2/p^2)^{1/3}$. Thus, unless we can construct a three-dimensional layout\footnote{A truly three-dimensional layout seems unlikely, though an adversary with the resources to build $2^{64}$ quantum computers may also be able to launch them into orbit and assemble them into a sphere.}, we cannot solve the search problem with quantum or classical computers in depth less than Equation \ref{eq:min-possible-depth}. For \aes-128, 192 and 256 this implies minimum depths of $2^{40.2}c_s$, $2^{56.2}c_s$ and $2^{72.3}c_s$, respectively. For \lowmc-128, 192, and 256 the minimum depths are respectively $2^{41.1}c_s$, $2^{59.8}c_s$ and $2^{76.4}c_s$.

%% file: diagrams/aes/key_expansion_128.tikz
\providecommand{\K}[1]{\left|#1\right\rangle}
\providecommand{\adj}[1]{#1${}^\dagger$}
\begin{tikzpicture}[scale=1.000000,x=1pt,y=1pt]
\filldraw[color=white] (0.000000, -7.500000) rectangle (230.000000, 52.500000);
\draw[color=black] (0.000000,45.000000) -- (230.000000,45.000000);
\draw[color=black] (0.000000,45.000000) node[left] {$\K{k_0}_{i-1}$};
\draw[color=black] (0.000000,30.000000) -- (230.000000,30.000000);
\draw[color=black] (0.000000,30.000000) node[left] {$\K{k_1}_{i-1}$};
\draw[color=black] (0.000000,15.000000) -- (230.000000,15.000000);
\draw[color=black] (0.000000,15.000000) node[left] {$\K{k_2}_{i-1}$};
\draw[color=black] (0.000000,0.000000) -- (230.000000,0.000000);
\draw[color=black] (0.000000,0.000000) node[left] {$\K{k_3}_{i-1}$};
\begin{scope}
\draw[fill=white] (28.500000, -0.000000) +(-45.000000:31.819805pt and 8.485281pt) -- +(45.000000:31.819805pt and 8.485281pt) -- +(135.000000:31.819805pt and 8.485281pt) -- +(225.000000:31.819805pt and 8.485281pt) -- cycle;
\clip (28.500000, -0.000000) +(-45.000000:31.819805pt and 8.485281pt) -- +(45.000000:31.819805pt and 8.485281pt) -- +(135.000000:31.819805pt and 8.485281pt) -- +(225.000000:31.819805pt and 8.485281pt) -- cycle;
\draw (28.500000, -0.000000) node {RotByte};
\end{scope}
\draw (88.000000,45.000000) -- (88.000000,0.000000);
\begin{scope}
\draw[fill=white] (88.000000, 22.500000) +(-45.000000:35.355339pt and 40.305087pt) -- +(45.000000:35.355339pt and 40.305087pt) -- +(135.000000:35.355339pt and 40.305087pt) -- +(225.000000:35.355339pt and 40.305087pt) -- cycle;
\clip (88.000000, 22.500000) +(-45.000000:35.355339pt and 40.305087pt) -- +(45.000000:35.355339pt and 40.305087pt) -- +(135.000000:35.355339pt and 40.305087pt) -- +(225.000000:35.355339pt and 40.305087pt) -- cycle;
\draw (88.000000, 22.500000) node {{$\updownarrow$ SubByte}};
\end{scope}
\draw[color=black,dashed] (63.000000, 30.000000) -- (113.000000, 30.000000);
\draw[color=black,dashed] (63.000000, 15.000000) -- (113.000000, 15.000000);
\begin{scope}
\draw[fill=white] (147.500000, -0.000000) +(-45.000000:31.819805pt and 8.485281pt) -- +(45.000000:31.819805pt and 8.485281pt) -- +(135.000000:31.819805pt and 8.485281pt) -- +(225.000000:31.819805pt and 8.485281pt) -- cycle;
\clip (147.500000, -0.000000) +(-45.000000:31.819805pt and 8.485281pt) -- +(45.000000:31.819805pt and 8.485281pt) -- +(135.000000:31.819805pt and 8.485281pt) -- +(225.000000:31.819805pt and 8.485281pt) -- cycle;
\draw (147.500000, -0.000000) node {\adj{RotByte}};
\end{scope}
\begin{scope}
\draw[fill=white] (147.500000, 45.000000) +(-45.000000:14.142136pt and 8.485281pt) -- +(45.000000:14.142136pt and 8.485281pt) -- +(135.000000:14.142136pt and 8.485281pt) -- +(225.000000:14.142136pt and 8.485281pt) -- cycle;
\clip (147.500000, 45.000000) +(-45.000000:14.142136pt and 8.485281pt) -- +(45.000000:14.142136pt and 8.485281pt) -- +(135.000000:14.142136pt and 8.485281pt) -- +(225.000000:14.142136pt and 8.485281pt) -- cycle;
\draw (147.500000, 45.000000) node {RC};
\end{scope}
\draw (185.000000,45.000000) -- (185.000000,30.000000);
\filldraw (185.000000, 45.000000) circle(1.500000pt);
\begin{scope}
\draw[fill=white] (185.000000, 30.000000) circle(3.000000pt);
\clip (185.000000, 30.000000) circle(3.000000pt);
\draw (182.000000, 30.000000) -- (188.000000, 30.000000);
\draw (185.000000, 27.000000) -- (185.000000, 33.000000);
\end{scope}
\draw (203.000000,30.000000) -- (203.000000,15.000000);
\filldraw (203.000000, 30.000000) circle(1.500000pt);
\begin{scope}
\draw[fill=white] (203.000000, 15.000000) circle(3.000000pt);
\clip (203.000000, 15.000000) circle(3.000000pt);
\draw (200.000000, 15.000000) -- (206.000000, 15.000000);
\draw (203.000000, 12.000000) -- (203.000000, 18.000000);
\end{scope}
\draw (221.000000,15.000000) -- (221.000000,0.000000);
\filldraw (221.000000, 15.000000) circle(1.500000pt);
\begin{scope}
\draw[fill=white] (221.000000, 0.000000) circle(3.000000pt);
\clip (221.000000, 0.000000) circle(3.000000pt);
\draw (218.000000, 0.000000) -- (224.000000, 0.000000);
\draw (221.000000, -3.000000) -- (221.000000, 3.000000);
\end{scope}
\draw[color=black] (230.000000,45.000000) node[right] {$\K{k_0}_i$};
\draw[color=black] (230.000000,30.000000) node[right] {$\K{k_1}_i$};
\draw[color=black] (230.000000,15.000000) node[right] {$\K{k_2}_i$};
\draw[color=black] (230.000000,0.000000) node[right] {$\K{k_3}_i$};
\end{tikzpicture}

%% file: diagrams/aes/key_expansion_192.tikz
\providecommand{\K}[1]{\left|#1\right\rangle}
\providecommand{\adj}[1]{#1${}^\dagger$}
\begin{tikzpicture}[scale=1.000000,x=1pt,y=1pt]
\filldraw[color=white] (0.000000, -7.500000) rectangle (266.000000, 82.500000);
\draw[color=black] (0.000000,75.000000) -- (266.000000,75.000000);
\draw[color=black] (0.000000,75.000000) node[left] {$\K{k_0}_{i-1}$};
\draw[color=black] (0.000000,60.000000) -- (266.000000,60.000000);
\draw[color=black] (0.000000,60.000000) node[left] {$\K{k_1}_{i-1}$};
\draw[color=black] (0.000000,45.000000) -- (266.000000,45.000000);
\draw[color=black] (0.000000,45.000000) node[left] {$\K{k_2}_{i-1}$};
\draw[color=black] (0.000000,30.000000) -- (266.000000,30.000000);
\draw[color=black] (0.000000,30.000000) node[left] {$\K{k_3}_{i-1}$};
\draw[color=black] (0.000000,15.000000) -- (266.000000,15.000000);
\draw[color=black] (0.000000,15.000000) node[left] {$\K{k_4}_{i-1}$};
\draw[color=black] (0.000000,0.000000) -- (266.000000,0.000000);
\draw[color=black] (0.000000,0.000000) node[left] {$\K{k_5}_{i-1}$};
\begin{scope}
\draw[fill=white] (28.500000, -0.000000) +(-45.000000:31.819805pt and 8.485281pt) -- +(45.000000:31.819805pt and 8.485281pt) -- +(135.000000:31.819805pt and 8.485281pt) -- +(225.000000:31.819805pt and 8.485281pt) -- cycle;
\clip (28.500000, -0.000000) +(-45.000000:31.819805pt and 8.485281pt) -- +(45.000000:31.819805pt and 8.485281pt) -- +(135.000000:31.819805pt and 8.485281pt) -- +(225.000000:31.819805pt and 8.485281pt) -- cycle;
\draw (28.500000, -0.000000) node {RotByte};
\end{scope}
\draw (88.000000,75.000000) -- (88.000000,0.000000);
\begin{scope}
\draw[fill=white] (88.000000, 37.500000) +(-45.000000:35.355339pt and 61.518290pt) -- +(45.000000:35.355339pt and 61.518290pt) -- +(135.000000:35.355339pt and 61.518290pt) -- +(225.000000:35.355339pt and 61.518290pt) -- cycle;
\clip (88.000000, 37.500000) +(-45.000000:35.355339pt and 61.518290pt) -- +(45.000000:35.355339pt and 61.518290pt) -- +(135.000000:35.355339pt and 61.518290pt) -- +(225.000000:35.355339pt and 61.518290pt) -- cycle;
\draw (88.000000, 37.500000) node {{$\updownarrow$ SubByte}};
\end{scope}
\draw[color=black,dashed] (63.000000, 60.000000) -- (113.000000, 60.000000);
\draw[color=black,dashed] (63.000000, 45.000000) -- (113.000000, 45.000000);
\draw[color=black,dashed] (63.000000, 30.000000) -- (113.000000, 30.000000);
\draw[color=black,dashed] (63.000000, 15.000000) -- (113.000000, 15.000000);
\begin{scope}
\draw[fill=white] (147.500000, -0.000000) +(-45.000000:31.819805pt and 8.485281pt) -- +(45.000000:31.819805pt and 8.485281pt) -- +(135.000000:31.819805pt and 8.485281pt) -- +(225.000000:31.819805pt and 8.485281pt) -- cycle;
\clip (147.500000, -0.000000) +(-45.000000:31.819805pt and 8.485281pt) -- +(45.000000:31.819805pt and 8.485281pt) -- +(135.000000:31.819805pt and 8.485281pt) -- +(225.000000:31.819805pt and 8.485281pt) -- cycle;
\draw (147.500000, -0.000000) node {\adj{RotByte}};
\end{scope}
\begin{scope}
\draw[fill=white] (147.500000, 75.000000) +(-45.000000:14.142136pt and 8.485281pt) -- +(45.000000:14.142136pt and 8.485281pt) -- +(135.000000:14.142136pt and 8.485281pt) -- +(225.000000:14.142136pt and 8.485281pt) -- cycle;
\clip (147.500000, 75.000000) +(-45.000000:14.142136pt and 8.485281pt) -- +(45.000000:14.142136pt and 8.485281pt) -- +(135.000000:14.142136pt and 8.485281pt) -- +(225.000000:14.142136pt and 8.485281pt) -- cycle;
\draw (147.500000, 75.000000) node {RC};
\end{scope}
\draw (185.000000,75.000000) -- (185.000000,60.000000);
\filldraw (185.000000, 75.000000) circle(1.500000pt);
\begin{scope}
\draw[fill=white] (185.000000, 60.000000) circle(3.000000pt);
\clip (185.000000, 60.000000) circle(3.000000pt);
\draw (182.000000, 60.000000) -- (188.000000, 60.000000);
\draw (185.000000, 57.000000) -- (185.000000, 63.000000);
\end{scope}
\draw (203.000000,60.000000) -- (203.000000,45.000000);
\filldraw (203.000000, 60.000000) circle(1.500000pt);
\begin{scope}
\draw[fill=white] (203.000000, 45.000000) circle(3.000000pt);
\clip (203.000000, 45.000000) circle(3.000000pt);
\draw (200.000000, 45.000000) -- (206.000000, 45.000000);
\draw (203.000000, 42.000000) -- (203.000000, 48.000000);
\end{scope}
\draw (221.000000,45.000000) -- (221.000000,30.000000);
\filldraw (221.000000, 45.000000) circle(1.500000pt);
\begin{scope}
\draw[fill=white] (221.000000, 30.000000) circle(3.000000pt);
\clip (221.000000, 30.000000) circle(3.000000pt);
\draw (218.000000, 30.000000) -- (224.000000, 30.000000);
\draw (221.000000, 27.000000) -- (221.000000, 33.000000);
\end{scope}
\draw (239.000000,30.000000) -- (239.000000,15.000000);
\filldraw (239.000000, 30.000000) circle(1.500000pt);
\begin{scope}
\draw[fill=white] (239.000000, 15.000000) circle(3.000000pt);
\clip (239.000000, 15.000000) circle(3.000000pt);
\draw (236.000000, 15.000000) -- (242.000000, 15.000000);
\draw (239.000000, 12.000000) -- (239.000000, 18.000000);
\end{scope}
\draw (257.000000,15.000000) -- (257.000000,0.000000);
\filldraw (257.000000, 15.000000) circle(1.500000pt);
\begin{scope}
\draw[fill=white] (257.000000, 0.000000) circle(3.000000pt);
\clip (257.000000, 0.000000) circle(3.000000pt);
\draw (254.000000, 0.000000) -- (260.000000, 0.000000);
\draw (257.000000, -3.000000) -- (257.000000, 3.000000);
\end{scope}
\draw[color=black] (266.000000,75.000000) node[right] {$\K{k_0}_{i}$};
\draw[color=black] (266.000000,60.000000) node[right] {$\K{k_1}_{i}$};
\draw[color=black] (266.000000,45.000000) node[right] {$\K{k_2}_{i}$};
\draw[color=black] (266.000000,30.000000) node[right] {$\K{k_3}_{i}$};
\draw[color=black] (266.000000,15.000000) node[right] {$\K{k_4}_{i}$};
\draw[color=black] (266.000000,0.000000) node[right] {$\K{k_5}_{i}$};
\end{tikzpicture}

%% file: diagrams/aes/key_expansion_256.tikz
\providecommand{\K}[1]{\left|#1\right\rangle}
\providecommand{\adj}[1]{#1${}^\dagger$}
\begin{tikzpicture}[scale=1.000000,x=1pt,y=1pt]
\filldraw[color=white] (0.000000, -7.500000) rectangle (341.000000, 112.500000);
\draw[color=black] (0.000000,105.000000) -- (341.000000,105.000000);
\draw[color=black] (0.000000,105.000000) node[left] {$\K{k_0}_{i-1}$};
\draw[color=black] (0.000000,90.000000) -- (341.000000,90.000000);
\draw[color=black] (0.000000,90.000000) node[left] {$\K{k_1}_{i-1}$};
\draw[color=black] (0.000000,75.000000) -- (341.000000,75.000000);
\draw[color=black] (0.000000,75.000000) node[left] {$\K{k_2}_{i-1}$};
\draw[color=black] (0.000000,60.000000) -- (341.000000,60.000000);
\draw[color=black] (0.000000,60.000000) node[left] {$\K{k_3}_{i-1}$};
\draw[color=black] (0.000000,45.000000) -- (341.000000,45.000000);
\draw[color=black] (0.000000,45.000000) node[left] {$\K{k_4}_{i-1}$};
\draw[color=black] (0.000000,30.000000) -- (341.000000,30.000000);
\draw[color=black] (0.000000,30.000000) node[left] {$\K{k_5}_{i-1}$};
\draw[color=black] (0.000000,15.000000) -- (341.000000,15.000000);
\draw[color=black] (0.000000,15.000000) node[left] {$\K{k_6}_{i-1}$};
\draw[color=black] (0.000000,0.000000) -- (341.000000,0.000000);
\draw[color=black] (0.000000,0.000000) node[left] {$\K{k_7}_{i-1}$};
\begin{scope}
\draw[fill=white] (28.500000, -0.000000) +(-45.000000:31.819805pt and 8.485281pt) -- +(45.000000:31.819805pt and 8.485281pt) -- +(135.000000:31.819805pt and 8.485281pt) -- +(225.000000:31.819805pt and 8.485281pt) -- cycle;
\clip (28.500000, -0.000000) +(-45.000000:31.819805pt and 8.485281pt) -- +(45.000000:31.819805pt and 8.485281pt) -- +(135.000000:31.819805pt and 8.485281pt) -- +(225.000000:31.819805pt and 8.485281pt) -- cycle;
\draw (28.500000, -0.000000) node {RotByte};
\end{scope}
\draw (88.000000,105.000000) -- (88.000000,0.000000);
\begin{scope}
\draw[fill=white] (88.000000, 52.500000) +(-45.000000:35.355339pt and 82.731493pt) -- +(45.000000:35.355339pt and 82.731493pt) -- +(135.000000:35.355339pt and 82.731493pt) -- +(225.000000:35.355339pt and 82.731493pt) -- cycle;
\clip (88.000000, 52.500000) +(-45.000000:35.355339pt and 82.731493pt) -- +(45.000000:35.355339pt and 82.731493pt) -- +(135.000000:35.355339pt and 82.731493pt) -- +(225.000000:35.355339pt and 82.731493pt) -- cycle;
\draw (88.000000, 52.500000) node {{$\updownarrow$ SubByte}};
\end{scope}
\draw[color=black,dashed] (63.000000, 90.000000) -- (113.000000, 90.000000);
\draw[color=black,dashed] (63.000000, 75.000000) -- (113.000000, 75.000000);
\draw[color=black,dashed] (63.000000, 60.000000) -- (113.000000, 60.000000);
\draw[color=black,dashed] (63.000000, 45.000000) -- (113.000000, 45.000000);
\draw[color=black,dashed] (63.000000, 30.000000) -- (113.000000, 30.000000);
\draw[color=black,dashed] (63.000000, 15.000000) -- (113.000000, 15.000000);
\begin{scope}
\draw[fill=white] (147.500000, -0.000000) +(-45.000000:31.819805pt and 8.485281pt) -- +(45.000000:31.819805pt and 8.485281pt) -- +(135.000000:31.819805pt and 8.485281pt) -- +(225.000000:31.819805pt and 8.485281pt) -- cycle;
\clip (147.500000, -0.000000) +(-45.000000:31.819805pt and 8.485281pt) -- +(45.000000:31.819805pt and 8.485281pt) -- +(135.000000:31.819805pt and 8.485281pt) -- +(225.000000:31.819805pt and 8.485281pt) -- cycle;
\draw (147.500000, -0.000000) node {\adj{RotByte}};
\end{scope}
\begin{scope}
\draw[fill=white] (147.500000, 105.000000) +(-45.000000:14.142136pt and 8.485281pt) -- +(45.000000:14.142136pt and 8.485281pt) -- +(135.000000:14.142136pt and 8.485281pt) -- +(225.000000:14.142136pt and 8.485281pt) -- cycle;
\clip (147.500000, 105.000000) +(-45.000000:14.142136pt and 8.485281pt) -- +(45.000000:14.142136pt and 8.485281pt) -- +(135.000000:14.142136pt and 8.485281pt) -- +(225.000000:14.142136pt and 8.485281pt) -- cycle;
\draw (147.500000, 105.000000) node {RC};
\end{scope}
\draw (185.000000,105.000000) -- (185.000000,90.000000);
\filldraw (185.000000, 105.000000) circle(1.500000pt);
\begin{scope}
\draw[fill=white] (185.000000, 90.000000) circle(3.000000pt);
\clip (185.000000, 90.000000) circle(3.000000pt);
\draw (182.000000, 90.000000) -- (188.000000, 90.000000);
\draw (185.000000, 87.000000) -- (185.000000, 93.000000);
\end{scope}
\draw (203.000000,90.000000) -- (203.000000,75.000000);
\filldraw (203.000000, 90.000000) circle(1.500000pt);
\begin{scope}
\draw[fill=white] (203.000000, 75.000000) circle(3.000000pt);
\clip (203.000000, 75.000000) circle(3.000000pt);
\draw (200.000000, 75.000000) -- (206.000000, 75.000000);
\draw (203.000000, 72.000000) -- (203.000000, 78.000000);
\end{scope}
\draw (221.000000,75.000000) -- (221.000000,60.000000);
\filldraw (221.000000, 75.000000) circle(1.500000pt);
\begin{scope}
\draw[fill=white] (221.000000, 60.000000) circle(3.000000pt);
\clip (221.000000, 60.000000) circle(3.000000pt);
\draw (218.000000, 60.000000) -- (224.000000, 60.000000);
\draw (221.000000, 57.000000) -- (221.000000, 63.000000);
\end{scope}
\draw (258.500000,60.000000) -- (258.500000,45.000000);
\begin{scope}
\draw[fill=white] (258.500000, 52.500000) +(-45.000000:31.819805pt and 19.091883pt) -- +(45.000000:31.819805pt and 19.091883pt) -- +(135.000000:31.819805pt and 19.091883pt) -- +(225.000000:31.819805pt and 19.091883pt) -- cycle;
\clip (258.500000, 52.500000) +(-45.000000:31.819805pt and 19.091883pt) -- +(45.000000:31.819805pt and 19.091883pt) -- +(135.000000:31.819805pt and 19.091883pt) -- +(225.000000:31.819805pt and 19.091883pt) -- cycle;
\draw (258.500000, 52.500000) node {SubByte};
\end{scope}
\draw (296.000000,45.000000) -- (296.000000,30.000000);
\filldraw (296.000000, 45.000000) circle(1.500000pt);
\begin{scope}
\draw[fill=white] (296.000000, 30.000000) circle(3.000000pt);
\clip (296.000000, 30.000000) circle(3.000000pt);
\draw (293.000000, 30.000000) -- (299.000000, 30.000000);
\draw (296.000000, 27.000000) -- (296.000000, 33.000000);
\end{scope}
\draw (314.000000,30.000000) -- (314.000000,15.000000);
\filldraw (314.000000, 30.000000) circle(1.500000pt);
\begin{scope}
\draw[fill=white] (314.000000, 15.000000) circle(3.000000pt);
\clip (314.000000, 15.000000) circle(3.000000pt);
\draw (311.000000, 15.000000) -- (317.000000, 15.000000);
\draw (314.000000, 12.000000) -- (314.000000, 18.000000);
\end{scope}
\draw (332.000000,15.000000) -- (332.000000,0.000000);
\filldraw (332.000000, 15.000000) circle(1.500000pt);
\begin{scope}
\draw[fill=white] (332.000000, 0.000000) circle(3.000000pt);
\clip (332.000000, 0.000000) circle(3.000000pt);
\draw (329.000000, 0.000000) -- (335.000000, 0.000000);
\draw (332.000000, -3.000000) -- (332.000000, 3.000000);
\end{scope}
\draw[color=black] (341.000000,105.000000) node[right] {$\K{k_0}_{i}$};
\draw[color=black] (341.000000,90.000000) node[right] {$\K{k_1}_{i}$};
\draw[color=black] (341.000000,75.000000) node[right] {$\K{k_2}_{i}$};
\draw[color=black] (341.000000,60.000000) node[right] {$\K{k_3}_{i}$};
\draw[color=black] (341.000000,45.000000) node[right] {$\K{k_4}_{i}$};
\draw[color=black] (341.000000,30.000000) node[right] {$\K{k_5}_{i}$};
\draw[color=black] (341.000000,15.000000) node[right] {$\K{k_6}_{i}$};
\draw[color=black] (341.000000,0.000000) node[right] {$\K{k_7}_{i}$};
\end{tikzpicture}

%% file: diagrams/aes/aes128.tikz
\providecommand{\K}[1]{\left|#1\right\rangle}
\providecommand{\adj}[1]{#1${}^\dagger$}
\begin{tikzpicture}[scale=1.000000,x=1pt,y=1pt]
\filldraw[color=white] (0.000000, -7.500000) rectangle (415.000000, 97.500000);
\draw[color=black] (0.000000,90.000000) -- (415.000000,90.000000);
\draw[color=black] (0.000000,90.000000) node[left] {$\K{k}_{0}$};
\draw[color=black] (0.000000,75.000000) -- (415.000000,75.000000);
\draw[color=black] (0.000000,75.000000) node[left] {$\K{m}$};
\draw[color=black] (0.000000,60.000000) -- (415.000000,60.000000);
\draw[color=black] (0.000000,60.000000) node[left] {$\K{0}$};
\draw[color=black] (0.000000,45.000000) -- (415.000000,45.000000);
\draw[color=black] (0.000000,45.000000) node[left] {$\K{0}$};
\draw[color=black] (0.000000,30.000000) node[anchor=mid east] {$\vdots$};
\draw[color=black] (0.000000,15.000000) -- (415.000000,15.000000);
\draw[color=black] (0.000000,15.000000) node[left] {$\K{0}$};
\draw[color=black] (0.000000,0.000000) -- (415.000000,0.000000);
\draw[color=black] (0.000000,0.000000) node[left] {$\K{0}$};
\draw (9.000000,90.000000) -- (9.000000,75.000000);
\filldraw (9.000000, 90.000000) circle(1.500000pt);
\begin{scope}
\draw[fill=white] (9.000000, 75.000000) circle(3.000000pt);
\clip (9.000000, 75.000000) circle(3.000000pt);
\draw (6.000000, 75.000000) -- (12.000000, 75.000000);
\draw (9.000000, 72.000000) -- (9.000000, 78.000000);
\end{scope}
\begin{scope}
\draw[fill=white] (44.000000, 90.000000) +(-45.000000:28.284271pt and 9.899495pt) -- +(45.000000:28.284271pt and 9.899495pt) -- +(135.000000:28.284271pt and 9.899495pt) -- +(225.000000:28.284271pt and 9.899495pt) -- cycle;
\clip (44.000000, 90.000000) +(-45.000000:28.284271pt and 9.899495pt) -- +(45.000000:28.284271pt and 9.899495pt) -- +(135.000000:28.284271pt and 9.899495pt) -- +(225.000000:28.284271pt and 9.899495pt) -- cycle;
\draw (44.000000, 90.000000) node {{KE$_0^{N_k-1}$}};
\end{scope}
\draw (44.000000,75.000000) -- (44.000000,60.000000);
\begin{scope}
\draw[fill=white] (44.000000, 67.500000) +(-45.000000:14.142136pt and 19.091883pt) -- +(45.000000:14.142136pt and 19.091883pt) -- +(135.000000:14.142136pt and 19.091883pt) -- +(225.000000:14.142136pt and 19.091883pt) -- cycle;
\clip (44.000000, 67.500000) +(-45.000000:14.142136pt and 19.091883pt) -- +(45.000000:14.142136pt and 19.091883pt) -- +(135.000000:14.142136pt and 19.091883pt) -- +(225.000000:14.142136pt and 19.091883pt) -- cycle;
\draw (44.000000, 67.500000) node {BS};
\end{scope}
\begin{scope}
\draw[fill=white] (86.000000, 60.000000) +(-45.000000:14.142136pt and 8.485281pt) -- +(45.000000:14.142136pt and 8.485281pt) -- +(135.000000:14.142136pt and 8.485281pt) -- +(225.000000:14.142136pt and 8.485281pt) -- cycle;
\clip (86.000000, 60.000000) +(-45.000000:14.142136pt and 8.485281pt) -- +(45.000000:14.142136pt and 8.485281pt) -- +(135.000000:14.142136pt and 8.485281pt) -- +(225.000000:14.142136pt and 8.485281pt) -- cycle;
\draw (86.000000, 60.000000) node {SR};
\end{scope}
\begin{scope}
\draw[fill=white] (118.000000, 60.000000) +(-45.000000:14.142136pt and 8.485281pt) -- +(45.000000:14.142136pt and 8.485281pt) -- +(135.000000:14.142136pt and 8.485281pt) -- +(225.000000:14.142136pt and 8.485281pt) -- cycle;
\clip (118.000000, 60.000000) +(-45.000000:14.142136pt and 8.485281pt) -- +(45.000000:14.142136pt and 8.485281pt) -- +(135.000000:14.142136pt and 8.485281pt) -- +(225.000000:14.142136pt and 8.485281pt) -- cycle;
\draw (118.000000, 60.000000) node {MC};
\end{scope}
\draw (143.000000,90.000000) -- (143.000000,60.000000);
\filldraw (143.000000, 90.000000) circle(1.500000pt);
\begin{scope}
\draw[fill=white] (143.000000, 60.000000) circle(3.000000pt);
\clip (143.000000, 60.000000) circle(3.000000pt);
\draw (140.000000, 60.000000) -- (146.000000, 60.000000);
\draw (143.000000, 57.000000) -- (143.000000, 63.000000);
\end{scope}
\begin{scope}
\draw[fill=white] (178.000000, 90.000000) +(-45.000000:28.284271pt and 9.899495pt) -- +(45.000000:28.284271pt and 9.899495pt) -- +(135.000000:28.284271pt and 9.899495pt) -- +(225.000000:28.284271pt and 9.899495pt) -- cycle;
\clip (178.000000, 90.000000) +(-45.000000:28.284271pt and 9.899495pt) -- +(45.000000:28.284271pt and 9.899495pt) -- +(135.000000:28.284271pt and 9.899495pt) -- +(225.000000:28.284271pt and 9.899495pt) -- cycle;
\draw (178.000000, 90.000000) node {{KE$_0^{N_k-1}$}};
\end{scope}
\draw (178.000000,60.000000) -- (178.000000,45.000000);
\begin{scope}
\draw[fill=white] (178.000000, 52.500000) +(-45.000000:14.142136pt and 19.091883pt) -- +(45.000000:14.142136pt and 19.091883pt) -- +(135.000000:14.142136pt and 19.091883pt) -- +(225.000000:14.142136pt and 19.091883pt) -- cycle;
\clip (178.000000, 52.500000) +(-45.000000:14.142136pt and 19.091883pt) -- +(45.000000:14.142136pt and 19.091883pt) -- +(135.000000:14.142136pt and 19.091883pt) -- +(225.000000:14.142136pt and 19.091883pt) -- cycle;
\draw (178.000000, 52.500000) node {BS};
\end{scope}
\begin{scope}
\draw[fill=white] (220.000000, 45.000000) +(-45.000000:14.142136pt and 8.485281pt) -- +(45.000000:14.142136pt and 8.485281pt) -- +(135.000000:14.142136pt and 8.485281pt) -- +(225.000000:14.142136pt and 8.485281pt) -- cycle;
\clip (220.000000, 45.000000) +(-45.000000:14.142136pt and 8.485281pt) -- +(45.000000:14.142136pt and 8.485281pt) -- +(135.000000:14.142136pt and 8.485281pt) -- +(225.000000:14.142136pt and 8.485281pt) -- cycle;
\draw (220.000000, 45.000000) node {SR};
\end{scope}
\begin{scope}
\draw[fill=white] (252.000000, 45.000000) +(-45.000000:14.142136pt and 8.485281pt) -- +(45.000000:14.142136pt and 8.485281pt) -- +(135.000000:14.142136pt and 8.485281pt) -- +(225.000000:14.142136pt and 8.485281pt) -- cycle;
\clip (252.000000, 45.000000) +(-45.000000:14.142136pt and 8.485281pt) -- +(45.000000:14.142136pt and 8.485281pt) -- +(135.000000:14.142136pt and 8.485281pt) -- +(225.000000:14.142136pt and 8.485281pt) -- cycle;
\draw (252.000000, 45.000000) node {MC};
\end{scope}
\draw (277.000000,90.000000) -- (277.000000,45.000000);
\filldraw (277.000000, 90.000000) circle(1.500000pt);
\begin{scope}
\draw[fill=white] (277.000000, 45.000000) circle(3.000000pt);
\clip (277.000000, 45.000000) circle(3.000000pt);
\draw (274.000000, 45.000000) -- (280.000000, 45.000000);
\draw (277.000000, 42.000000) -- (277.000000, 48.000000);
\end{scope}
\draw[fill=white,color=white] (292.000000, -6.000000) rectangle (307.000000, 96.000000);
\draw (299.500000, 45.000000) node {\dots};
\begin{scope}
\draw[fill=white] (339.000000, 90.000000) +(-45.000000:28.284271pt and 9.899495pt) -- +(45.000000:28.284271pt and 9.899495pt) -- +(135.000000:28.284271pt and 9.899495pt) -- +(225.000000:28.284271pt and 9.899495pt) -- cycle;
\clip (339.000000, 90.000000) +(-45.000000:28.284271pt and 9.899495pt) -- +(45.000000:28.284271pt and 9.899495pt) -- +(135.000000:28.284271pt and 9.899495pt) -- +(225.000000:28.284271pt and 9.899495pt) -- cycle;
\draw (339.000000, 90.000000) node {{KE$_0^{N_k-1}$}};
\end{scope}
\draw (339.000000,15.000000) -- (339.000000,0.000000);
\begin{scope}
\draw[fill=white] (339.000000, 7.500000) +(-45.000000:14.142136pt and 19.091883pt) -- +(45.000000:14.142136pt and 19.091883pt) -- +(135.000000:14.142136pt and 19.091883pt) -- +(225.000000:14.142136pt and 19.091883pt) -- cycle;
\clip (339.000000, 7.500000) +(-45.000000:14.142136pt and 19.091883pt) -- +(45.000000:14.142136pt and 19.091883pt) -- +(135.000000:14.142136pt and 19.091883pt) -- +(225.000000:14.142136pt and 19.091883pt) -- cycle;
\draw (339.000000, 7.500000) node {BS};
\end{scope}
\begin{scope}
\draw[fill=white] (381.000000, -0.000000) +(-45.000000:14.142136pt and 8.485281pt) -- +(45.000000:14.142136pt and 8.485281pt) -- +(135.000000:14.142136pt and 8.485281pt) -- +(225.000000:14.142136pt and 8.485281pt) -- cycle;
\clip (381.000000, -0.000000) +(-45.000000:14.142136pt and 8.485281pt) -- +(45.000000:14.142136pt and 8.485281pt) -- +(135.000000:14.142136pt and 8.485281pt) -- +(225.000000:14.142136pt and 8.485281pt) -- cycle;
\draw (381.000000, -0.000000) node {SR};
\end{scope}
\draw (406.000000,90.000000) -- (406.000000,0.000000);
\filldraw (406.000000, 90.000000) circle(1.500000pt);
\begin{scope}
\draw[fill=white] (406.000000, 0.000000) circle(3.000000pt);
\clip (406.000000, 0.000000) circle(3.000000pt);
\draw (403.000000, 0.000000) -- (409.000000, 0.000000);
\draw (406.000000, -3.000000) -- (406.000000, 3.000000);
\end{scope}
\draw[color=black] (415.000000,90.000000) node[right] {$\K{k}_{10}$};
\draw[color=black] (415.000000,30.000000) node[anchor=mid west] {$\vdots$};
\draw[color=black] (415.000000,0.000000) node[right] {$\K{c}$};
\draw[decorate,decoration={brace,amplitude = 4.000000pt},very thick] (21.000000,97.500000) -- (149.000000,97.500000);
\draw (85.000000, 101.500000) node[text width=144pt,above,text centered] {Round 1};
\draw[decorate,decoration={brace,amplitude = 4.000000pt},very thick] (155.000000,97.500000) -- (283.000000,97.500000);
\draw (219.000000, 101.500000) node[text width=144pt,above,text centered] {Round 2};
\draw[decorate,decoration={brace,amplitude = 4.000000pt},very thick] (316.000000,97.500000) -- (412.000000,97.500000);
\draw (364.000000, 101.500000) node[text width=144pt,above,text centered] {Round 10};
\end{tikzpicture}

%% file: diagrams/aes/aes192_maximov.tikz
\providecommand{\K}[1]{\left|#1\right\rangle}
\providecommand{\adj}[1]{#1${}^\dagger$}
\begin{tikzpicture}[scale=1.000000,x=1pt,y=1pt]
\filldraw[color=white] (0.000000, -7.500000) rectangle (851.000000, 217.500000);
\draw[color=black] (0.000000,210.000000) -- (851.000000,210.000000);
\draw[color=black] (0.000000,195.000000) -- (851.000000,195.000000);
\draw[color=black] (0.000000,180.000000) -- (851.000000,180.000000);
\filldraw[color=white,fill=white] (0.000000,176.250000) rectangle (-4.000000,213.750000);
\draw[decorate,decoration={brace,amplitude = 4.000000pt},very thick] (0.000000,176.250000) -- (0.000000,213.750000);
\draw[color=black] (-4.000000,195.000000) node[left] {$\K{k}_{0}$};
\draw[color=black] (0.000000,165.000000) -- (851.000000,165.000000);
\draw[color=black] (0.000000,165.000000) node[left] {$\K{m}$};
\draw[color=black] (0.000000,150.000000) -- (851.000000,150.000000);
\draw[color=black] (0.000000,150.000000) node[left] {$\K{0}$};
\draw[color=black] (0.000000,135.000000) -- (851.000000,135.000000);
\draw[color=black] (0.000000,135.000000) node[left] {$\K{0}$};
\draw[color=black] (0.000000,120.000000) -- (851.000000,120.000000);
\draw[color=black] (0.000000,120.000000) node[left] {$\K{0}$};
\draw[color=black] (0.000000,105.000000) -- (851.000000,105.000000);
\draw[color=black] (0.000000,105.000000) node[left] {$\K{0}$};
\draw[color=black] (0.000000,90.000000) -- (851.000000,90.000000);
\draw[color=black] (0.000000,90.000000) node[left] {$\K{0}$};
\draw[color=black] (0.000000,75.000000) -- (851.000000,75.000000);
\draw[color=black] (0.000000,75.000000) node[left] {$\K{0}$};
\draw[color=black] (0.000000,60.000000) -- (851.000000,60.000000);
\draw[color=black] (0.000000,60.000000) node[left] {$\K{0}$};
\draw[color=black] (0.000000,45.000000) -- (851.000000,45.000000);
\draw[color=black] (0.000000,45.000000) node[left] {$\K{0}$};
\draw[color=black] (0.000000,30.000000) node[anchor=mid east] {$\vdots$};
\draw[color=black] (0.000000,15.000000) -- (851.000000,15.000000);
\draw[color=black] (0.000000,15.000000) node[left] {$\K{0}$};
\draw[color=black] (0.000000,0.000000) -- (851.000000,0.000000);
\draw[color=black] (0.000000,0.000000) node[left] {$\K{0}$};
\draw (9.000000,210.000000) -- (9.000000,165.000000);
\filldraw (9.000000, 210.000000) circle(1.500000pt);
\begin{scope}
\draw[fill=white] (9.000000, 165.000000) circle(3.000000pt);
\clip (9.000000, 165.000000) circle(3.000000pt);
\draw (6.000000, 165.000000) -- (12.000000, 165.000000);
\draw (9.000000, 162.000000) -- (9.000000, 168.000000);
\end{scope}
\draw (27.000000,195.000000) -- (27.000000,165.000000);
\filldraw (27.000000, 195.000000) circle(1.500000pt);
\begin{scope}
\draw[fill=white] (27.000000, 165.000000) circle(3.000000pt);
\clip (27.000000, 165.000000) circle(3.000000pt);
\draw (24.000000, 165.000000) -- (30.000000, 165.000000);
\draw (27.000000, 162.000000) -- (27.000000, 168.000000);
\end{scope}
\draw (64.500000,210.000000) -- (64.500000,180.000000);
\begin{scope}
\draw[fill=white] (64.500000, 195.000000) +(-45.000000:31.819805pt and 29.698485pt) -- +(45.000000:31.819805pt and 29.698485pt) -- +(135.000000:31.819805pt and 29.698485pt) -- +(225.000000:31.819805pt and 29.698485pt) -- cycle;
\clip (64.500000, 195.000000) +(-45.000000:31.819805pt and 29.698485pt) -- +(45.000000:31.819805pt and 29.698485pt) -- +(135.000000:31.819805pt and 29.698485pt) -- +(225.000000:31.819805pt and 29.698485pt) -- cycle;
\draw (64.500000, 195.000000) node {{KE$_0^{1}$}};
\end{scope}
\draw (64.500000,165.000000) -- (64.500000,150.000000);
\begin{scope}
\draw[fill=white] (64.500000, 157.500000) +(-45.000000:14.142136pt and 19.091883pt) -- +(45.000000:14.142136pt and 19.091883pt) -- +(135.000000:14.142136pt and 19.091883pt) -- +(225.000000:14.142136pt and 19.091883pt) -- cycle;
\clip (64.500000, 157.500000) +(-45.000000:14.142136pt and 19.091883pt) -- +(45.000000:14.142136pt and 19.091883pt) -- +(135.000000:14.142136pt and 19.091883pt) -- +(225.000000:14.142136pt and 19.091883pt) -- cycle;
\draw (64.500000, 157.500000) node {BS};
\end{scope}
\begin{scope}
\draw[fill=white] (109.000000, 150.000000) +(-45.000000:14.142136pt and 8.485281pt) -- +(45.000000:14.142136pt and 8.485281pt) -- +(135.000000:14.142136pt and 8.485281pt) -- +(225.000000:14.142136pt and 8.485281pt) -- cycle;
\clip (109.000000, 150.000000) +(-45.000000:14.142136pt and 8.485281pt) -- +(45.000000:14.142136pt and 8.485281pt) -- +(135.000000:14.142136pt and 8.485281pt) -- +(225.000000:14.142136pt and 8.485281pt) -- cycle;
\draw (109.000000, 150.000000) node {SR};
\end{scope}
\draw (141.000000,150.000000) -- (141.000000,135.000000);
\begin{scope}
\draw[fill=white] (141.000000, 142.500000) +(-45.000000:14.142136pt and 19.091883pt) -- +(45.000000:14.142136pt and 19.091883pt) -- +(135.000000:14.142136pt and 19.091883pt) -- +(225.000000:14.142136pt and 19.091883pt) -- cycle;
\clip (141.000000, 142.500000) +(-45.000000:14.142136pt and 19.091883pt) -- +(45.000000:14.142136pt and 19.091883pt) -- +(135.000000:14.142136pt and 19.091883pt) -- +(225.000000:14.142136pt and 19.091883pt) -- cycle;
\draw (141.000000, 142.500000) node {MC};
\end{scope}
\draw (166.000000,180.000000) -- (166.000000,135.000000);
\filldraw (166.000000, 180.000000) circle(1.500000pt);
\begin{scope}
\draw[fill=white] (166.000000, 135.000000) circle(3.000000pt);
\clip (166.000000, 135.000000) circle(3.000000pt);
\draw (163.000000, 135.000000) -- (169.000000, 135.000000);
\draw (166.000000, 132.000000) -- (166.000000, 138.000000);
\end{scope}
\draw (184.000000,210.000000) -- (184.000000,135.000000);
\filldraw (184.000000, 210.000000) circle(1.500000pt);
\begin{scope}
\draw[fill=white] (184.000000, 135.000000) circle(3.000000pt);
\clip (184.000000, 135.000000) circle(3.000000pt);
\draw (181.000000, 135.000000) -- (187.000000, 135.000000);
\draw (184.000000, 132.000000) -- (184.000000, 138.000000);
\end{scope}
\draw (221.500000,210.000000) -- (221.500000,180.000000);
\begin{scope}
\draw[fill=white] (221.500000, 195.000000) +(-45.000000:31.819805pt and 29.698485pt) -- +(45.000000:31.819805pt and 29.698485pt) -- +(135.000000:31.819805pt and 29.698485pt) -- +(225.000000:31.819805pt and 29.698485pt) -- cycle;
\clip (221.500000, 195.000000) +(-45.000000:31.819805pt and 29.698485pt) -- +(45.000000:31.819805pt and 29.698485pt) -- +(135.000000:31.819805pt and 29.698485pt) -- +(225.000000:31.819805pt and 29.698485pt) -- cycle;
\draw (221.500000, 195.000000) node {{KE$_{2}^{5}$}};
\end{scope}
\draw (221.500000,135.000000) -- (221.500000,120.000000);
\begin{scope}
\draw[fill=white] (221.500000, 127.500000) +(-45.000000:14.142136pt and 19.091883pt) -- +(45.000000:14.142136pt and 19.091883pt) -- +(135.000000:14.142136pt and 19.091883pt) -- +(225.000000:14.142136pt and 19.091883pt) -- cycle;
\clip (221.500000, 127.500000) +(-45.000000:14.142136pt and 19.091883pt) -- +(45.000000:14.142136pt and 19.091883pt) -- +(135.000000:14.142136pt and 19.091883pt) -- +(225.000000:14.142136pt and 19.091883pt) -- cycle;
\draw (221.500000, 127.500000) node {BS};
\end{scope}
\begin{scope}
\draw[fill=white] (266.000000, 120.000000) +(-45.000000:14.142136pt and 8.485281pt) -- +(45.000000:14.142136pt and 8.485281pt) -- +(135.000000:14.142136pt and 8.485281pt) -- +(225.000000:14.142136pt and 8.485281pt) -- cycle;
\clip (266.000000, 120.000000) +(-45.000000:14.142136pt and 8.485281pt) -- +(45.000000:14.142136pt and 8.485281pt) -- +(135.000000:14.142136pt and 8.485281pt) -- +(225.000000:14.142136pt and 8.485281pt) -- cycle;
\draw (266.000000, 120.000000) node {SR};
\end{scope}
\draw (298.000000,120.000000) -- (298.000000,105.000000);
\begin{scope}
\draw[fill=white] (298.000000, 112.500000) +(-45.000000:14.142136pt and 19.091883pt) -- +(45.000000:14.142136pt and 19.091883pt) -- +(135.000000:14.142136pt and 19.091883pt) -- +(225.000000:14.142136pt and 19.091883pt) -- cycle;
\clip (298.000000, 112.500000) +(-45.000000:14.142136pt and 19.091883pt) -- +(45.000000:14.142136pt and 19.091883pt) -- +(135.000000:14.142136pt and 19.091883pt) -- +(225.000000:14.142136pt and 19.091883pt) -- cycle;
\draw (298.000000, 112.500000) node {MC};
\end{scope}
\draw (323.000000,195.000000) -- (323.000000,105.000000);
\filldraw (323.000000, 195.000000) circle(1.500000pt);
\begin{scope}
\draw[fill=white] (323.000000, 105.000000) circle(3.000000pt);
\clip (323.000000, 105.000000) circle(3.000000pt);
\draw (320.000000, 105.000000) -- (326.000000, 105.000000);
\draw (323.000000, 102.000000) -- (323.000000, 108.000000);
\end{scope}
\draw (341.000000,180.000000) -- (341.000000,105.000000);
\filldraw (341.000000, 180.000000) circle(1.500000pt);
\begin{scope}
\draw[fill=white] (341.000000, 105.000000) circle(3.000000pt);
\clip (341.000000, 105.000000) circle(3.000000pt);
\draw (338.000000, 105.000000) -- (344.000000, 105.000000);
\draw (341.000000, 102.000000) -- (341.000000, 108.000000);
\end{scope}
\draw (381.000000,210.000000) -- (381.000000,180.000000);
\begin{scope}
\draw[fill=white] (381.000000, 195.000000) +(-45.000000:35.355339pt and 29.698485pt) -- +(45.000000:35.355339pt and 29.698485pt) -- +(135.000000:35.355339pt and 29.698485pt) -- +(225.000000:35.355339pt and 29.698485pt) -- cycle;
\clip (381.000000, 195.000000) +(-45.000000:35.355339pt and 29.698485pt) -- +(45.000000:35.355339pt and 29.698485pt) -- +(135.000000:35.355339pt and 29.698485pt) -- +(225.000000:35.355339pt and 29.698485pt) -- cycle;
\draw (381.000000, 195.000000) node {{KE$_{0}^{3}$}};
\end{scope}
\draw (381.000000,105.000000) -- (381.000000,90.000000);
\begin{scope}
\draw[fill=white] (381.000000, 97.500000) +(-45.000000:14.142136pt and 19.091883pt) -- +(45.000000:14.142136pt and 19.091883pt) -- +(135.000000:14.142136pt and 19.091883pt) -- +(225.000000:14.142136pt and 19.091883pt) -- cycle;
\clip (381.000000, 97.500000) +(-45.000000:14.142136pt and 19.091883pt) -- +(45.000000:14.142136pt and 19.091883pt) -- +(135.000000:14.142136pt and 19.091883pt) -- +(225.000000:14.142136pt and 19.091883pt) -- cycle;
\draw (381.000000, 97.500000) node {BS};
\end{scope}
\begin{scope}
\draw[fill=white] (428.000000, 90.000000) +(-45.000000:14.142136pt and 8.485281pt) -- +(45.000000:14.142136pt and 8.485281pt) -- +(135.000000:14.142136pt and 8.485281pt) -- +(225.000000:14.142136pt and 8.485281pt) -- cycle;
\clip (428.000000, 90.000000) +(-45.000000:14.142136pt and 8.485281pt) -- +(45.000000:14.142136pt and 8.485281pt) -- +(135.000000:14.142136pt and 8.485281pt) -- +(225.000000:14.142136pt and 8.485281pt) -- cycle;
\draw (428.000000, 90.000000) node {SR};
\end{scope}
\draw (460.000000,90.000000) -- (460.000000,75.000000);
\begin{scope}
\draw[fill=white] (460.000000, 82.500000) +(-45.000000:14.142136pt and 19.091883pt) -- +(45.000000:14.142136pt and 19.091883pt) -- +(135.000000:14.142136pt and 19.091883pt) -- +(225.000000:14.142136pt and 19.091883pt) -- cycle;
\clip (460.000000, 82.500000) +(-45.000000:14.142136pt and 19.091883pt) -- +(45.000000:14.142136pt and 19.091883pt) -- +(135.000000:14.142136pt and 19.091883pt) -- +(225.000000:14.142136pt and 19.091883pt) -- cycle;
\draw (460.000000, 82.500000) node {MC};
\end{scope}
\draw (485.000000,210.000000) -- (485.000000,75.000000);
\filldraw (485.000000, 210.000000) circle(1.500000pt);
\begin{scope}
\draw[fill=white] (485.000000, 75.000000) circle(3.000000pt);
\clip (485.000000, 75.000000) circle(3.000000pt);
\draw (482.000000, 75.000000) -- (488.000000, 75.000000);
\draw (485.000000, 72.000000) -- (485.000000, 78.000000);
\end{scope}
\draw (503.000000,195.000000) -- (503.000000,75.000000);
\filldraw (503.000000, 195.000000) circle(1.500000pt);
\begin{scope}
\draw[fill=white] (503.000000, 75.000000) circle(3.000000pt);
\clip (503.000000, 75.000000) circle(3.000000pt);
\draw (500.000000, 75.000000) -- (506.000000, 75.000000);
\draw (503.000000, 72.000000) -- (503.000000, 78.000000);
\end{scope}
\draw (540.500000,210.000000) -- (540.500000,180.000000);
\begin{scope}
\draw[fill=white] (540.500000, 195.000000) +(-45.000000:31.819805pt and 29.698485pt) -- +(45.000000:31.819805pt and 29.698485pt) -- +(135.000000:31.819805pt and 29.698485pt) -- +(225.000000:31.819805pt and 29.698485pt) -- cycle;
\clip (540.500000, 195.000000) +(-45.000000:31.819805pt and 29.698485pt) -- +(45.000000:31.819805pt and 29.698485pt) -- +(135.000000:31.819805pt and 29.698485pt) -- +(225.000000:31.819805pt and 29.698485pt) -- cycle;
\draw (540.500000, 195.000000) node {{KE$_{4}^{5}$}};
\end{scope}
\draw (540.500000,75.000000) -- (540.500000,60.000000);
\begin{scope}
\draw[fill=white] (540.500000, 67.500000) +(-45.000000:14.142136pt and 19.091883pt) -- +(45.000000:14.142136pt and 19.091883pt) -- +(135.000000:14.142136pt and 19.091883pt) -- +(225.000000:14.142136pt and 19.091883pt) -- cycle;
\clip (540.500000, 67.500000) +(-45.000000:14.142136pt and 19.091883pt) -- +(45.000000:14.142136pt and 19.091883pt) -- +(135.000000:14.142136pt and 19.091883pt) -- +(225.000000:14.142136pt and 19.091883pt) -- cycle;
\draw (540.500000, 67.500000) node {BS};
\end{scope}
\draw (597.500000,210.000000) -- (597.500000,180.000000);
\begin{scope}
\draw[fill=white] (597.500000, 195.000000) +(-45.000000:31.819805pt and 29.698485pt) -- +(45.000000:31.819805pt and 29.698485pt) -- +(135.000000:31.819805pt and 29.698485pt) -- +(225.000000:31.819805pt and 29.698485pt) -- cycle;
\clip (597.500000, 195.000000) +(-45.000000:31.819805pt and 29.698485pt) -- +(45.000000:31.819805pt and 29.698485pt) -- +(135.000000:31.819805pt and 29.698485pt) -- +(225.000000:31.819805pt and 29.698485pt) -- cycle;
\draw (597.500000, 195.000000) node {{KE$_0^{1}$}};
\end{scope}
\begin{scope}
\draw[fill=white] (597.500000, 60.000000) +(-45.000000:14.142136pt and 8.485281pt) -- +(45.000000:14.142136pt and 8.485281pt) -- +(135.000000:14.142136pt and 8.485281pt) -- +(225.000000:14.142136pt and 8.485281pt) -- cycle;
\clip (597.500000, 60.000000) +(-45.000000:14.142136pt and 8.485281pt) -- +(45.000000:14.142136pt and 8.485281pt) -- +(135.000000:14.142136pt and 8.485281pt) -- +(225.000000:14.142136pt and 8.485281pt) -- cycle;
\draw (597.500000, 60.000000) node {SR};
\end{scope}
\draw (642.000000,60.000000) -- (642.000000,45.000000);
\begin{scope}
\draw[fill=white] (642.000000, 52.500000) +(-45.000000:14.142136pt and 19.091883pt) -- +(45.000000:14.142136pt and 19.091883pt) -- +(135.000000:14.142136pt and 19.091883pt) -- +(225.000000:14.142136pt and 19.091883pt) -- cycle;
\clip (642.000000, 52.500000) +(-45.000000:14.142136pt and 19.091883pt) -- +(45.000000:14.142136pt and 19.091883pt) -- +(135.000000:14.142136pt and 19.091883pt) -- +(225.000000:14.142136pt and 19.091883pt) -- cycle;
\draw (642.000000, 52.500000) node {MC};
\end{scope}
\draw (667.000000,180.000000) -- (667.000000,45.000000);
\filldraw (667.000000, 180.000000) circle(1.500000pt);
\begin{scope}
\draw[fill=white] (667.000000, 45.000000) circle(3.000000pt);
\clip (667.000000, 45.000000) circle(3.000000pt);
\draw (664.000000, 45.000000) -- (670.000000, 45.000000);
\draw (667.000000, 42.000000) -- (667.000000, 48.000000);
\end{scope}
\draw (685.000000,210.000000) -- (685.000000,45.000000);
\filldraw (685.000000, 210.000000) circle(1.500000pt);
\begin{scope}
\draw[fill=white] (685.000000, 45.000000) circle(3.000000pt);
\clip (685.000000, 45.000000) circle(3.000000pt);
\draw (682.000000, 45.000000) -- (688.000000, 45.000000);
\draw (685.000000, 42.000000) -- (685.000000, 48.000000);
\end{scope}
\draw[fill=white,color=white] (700.000000, -6.000000) rectangle (715.000000, 216.000000);
\draw (707.500000, 105.000000) node {\dots};
\draw (752.000000,210.000000) -- (752.000000,180.000000);
\begin{scope}
\draw[fill=white] (752.000000, 195.000000) +(-45.000000:35.355339pt and 29.698485pt) -- +(45.000000:35.355339pt and 29.698485pt) -- +(135.000000:35.355339pt and 29.698485pt) -- +(225.000000:35.355339pt and 29.698485pt) -- cycle;
\clip (752.000000, 195.000000) +(-45.000000:35.355339pt and 29.698485pt) -- +(45.000000:35.355339pt and 29.698485pt) -- +(135.000000:35.355339pt and 29.698485pt) -- +(225.000000:35.355339pt and 29.698485pt) -- cycle;
\draw (752.000000, 195.000000) node {{KE$_0^{3}$}};
\end{scope}
\draw (752.000000,15.000000) -- (752.000000,0.000000);
\begin{scope}
\draw[fill=white] (752.000000, 7.500000) +(-45.000000:14.142136pt and 19.091883pt) -- +(45.000000:14.142136pt and 19.091883pt) -- +(135.000000:14.142136pt and 19.091883pt) -- +(225.000000:14.142136pt and 19.091883pt) -- cycle;
\clip (752.000000, 7.500000) +(-45.000000:14.142136pt and 19.091883pt) -- +(45.000000:14.142136pt and 19.091883pt) -- +(135.000000:14.142136pt and 19.091883pt) -- +(225.000000:14.142136pt and 19.091883pt) -- cycle;
\draw (752.000000, 7.500000) node {BS};
\end{scope}
\begin{scope}
\draw[fill=white] (799.000000, -0.000000) +(-45.000000:14.142136pt and 8.485281pt) -- +(45.000000:14.142136pt and 8.485281pt) -- +(135.000000:14.142136pt and 8.485281pt) -- +(225.000000:14.142136pt and 8.485281pt) -- cycle;
\clip (799.000000, -0.000000) +(-45.000000:14.142136pt and 8.485281pt) -- +(45.000000:14.142136pt and 8.485281pt) -- +(135.000000:14.142136pt and 8.485281pt) -- +(225.000000:14.142136pt and 8.485281pt) -- cycle;
\draw (799.000000, -0.000000) node {SR};
\end{scope}
\draw (824.000000,210.000000) -- (824.000000,0.000000);
\filldraw (824.000000, 210.000000) circle(1.500000pt);
\begin{scope}
\draw[fill=white] (824.000000, 0.000000) circle(3.000000pt);
\clip (824.000000, 0.000000) circle(3.000000pt);
\draw (821.000000, 0.000000) -- (827.000000, 0.000000);
\draw (824.000000, -3.000000) -- (824.000000, 3.000000);
\end{scope}
\draw (842.000000,195.000000) -- (842.000000,0.000000);
\filldraw (842.000000, 195.000000) circle(1.500000pt);
\begin{scope}
\draw[fill=white] (842.000000, 0.000000) circle(3.000000pt);
\clip (842.000000, 0.000000) circle(3.000000pt);
\draw (839.000000, 0.000000) -- (845.000000, 0.000000);
\draw (842.000000, -3.000000) -- (842.000000, 3.000000);
\end{scope}
\filldraw[color=white,fill=white] (851.000000,176.250000) rectangle (855.000000,213.750000);
\draw[decorate,decoration={brace,mirror,amplitude = 4.000000pt},very thick] (851.000000,176.250000) -- (851.000000,213.750000);
\draw[color=black] (855.000000,195.000000) node[right] {$\K{k}_{12}$};
\draw[color=black] (851.000000,30.000000) node[anchor=mid west] {$\vdots$};
\draw[color=black] (851.000000,0.000000) node[right] {$\K{c}$};
\draw[decorate,decoration={brace,amplitude = 4.000000pt},very thick] (39.000000,217.500000) -- (190.000000,217.500000);
\draw (114.500000, 221.500000) node[text width=144pt,above,text centered] {Round 1};
\draw[decorate,decoration={brace,amplitude = 4.000000pt},very thick] (196.000000,217.500000) -- (347.000000,217.500000);
\draw (271.500000, 221.500000) node[text width=144pt,above,text centered] {Round 2};
\draw[decorate,decoration={brace,amplitude = 4.000000pt},very thick] (353.000000,217.500000) -- (509.000000,217.500000);
\draw (431.000000, 221.500000) node[text width=144pt,above,text centered] {Round 3};
\draw[decorate,decoration={brace,amplitude = 4.000000pt},very thick] (515.000000,217.500000) -- (691.000000,217.500000);
\draw (603.000000, 221.500000) node[text width=144pt,above,text centered] {Round 4};
\draw[decorate,decoration={brace,amplitude = 4.000000pt},very thick] (724.000000,217.500000) -- (848.000000,217.500000);
\draw (786.000000, 221.500000) node[text width=144pt,above,text centered] {Round 12};
\end{tikzpicture}

%% file: diagrams/aes/aes256_maximov.tikz
\providecommand{\K}[1]{\left|#1\right\rangle}
\providecommand{\adj}[1]{#1${}^\dagger$}
\begin{tikzpicture}[scale=1.000000,x=1pt,y=1pt]
\filldraw[color=white] (0.000000, -7.500000) rectangle (539.000000, 172.500000);
\draw[color=black] (0.000000,165.000000) -- (539.000000,165.000000);
\draw[color=black] (0.000000,150.000000) -- (539.000000,150.000000);
\filldraw[color=white,fill=white] (0.000000,146.250000) rectangle (-4.000000,168.750000);
\draw[decorate,decoration={brace,amplitude = 4.000000pt},very thick] (0.000000,146.250000) -- (0.000000,168.750000);
\draw[color=black] (-4.000000,157.500000) node[left] {$\K{k}_{0}$};
\draw[color=black] (0.000000,135.000000) -- (539.000000,135.000000);
\draw[color=black] (0.000000,135.000000) node[left] {$\K{m}$};
\draw[color=black] (0.000000,120.000000) -- (539.000000,120.000000);
\draw[color=black] (0.000000,120.000000) node[left] {$\K{0}$};
\draw[color=black] (0.000000,105.000000) -- (539.000000,105.000000);
\draw[color=black] (0.000000,105.000000) node[left] {$\K{0}$};
\draw[color=black] (0.000000,90.000000) -- (539.000000,90.000000);
\draw[color=black] (0.000000,90.000000) node[left] {$\K{0}$};
\draw[color=black] (0.000000,75.000000) -- (539.000000,75.000000);
\draw[color=black] (0.000000,75.000000) node[left] {$\K{0}$};
\draw[color=black] (0.000000,60.000000) -- (539.000000,60.000000);
\draw[color=black] (0.000000,60.000000) node[left] {$\K{0}$};
\draw[color=black] (0.000000,45.000000) -- (539.000000,45.000000);
\draw[color=black] (0.000000,45.000000) node[left] {$\K{0}$};
\draw[color=black] (0.000000,30.000000) node[anchor=mid east] {$\vdots$};
\draw[color=black] (0.000000,15.000000) -- (539.000000,15.000000);
\draw[color=black] (0.000000,15.000000) node[left] {$\K{0}$};
\draw[color=black] (0.000000,0.000000) -- (539.000000,0.000000);
\draw[color=black] (0.000000,0.000000) node[left] {$\K{0}$};
\draw (9.000000,165.000000) -- (9.000000,135.000000);
\filldraw (9.000000, 165.000000) circle(1.500000pt);
\begin{scope}
\draw[fill=white] (9.000000, 135.000000) circle(3.000000pt);
\clip (9.000000, 135.000000) circle(3.000000pt);
\draw (6.000000, 135.000000) -- (12.000000, 135.000000);
\draw (9.000000, 132.000000) -- (9.000000, 138.000000);
\end{scope}
\draw (34.000000,135.000000) -- (34.000000,120.000000);
\begin{scope}
\draw[fill=white] (34.000000, 127.500000) +(-45.000000:14.142136pt and 19.091883pt) -- +(45.000000:14.142136pt and 19.091883pt) -- +(135.000000:14.142136pt and 19.091883pt) -- +(225.000000:14.142136pt and 19.091883pt) -- cycle;
\clip (34.000000, 127.500000) +(-45.000000:14.142136pt and 19.091883pt) -- +(45.000000:14.142136pt and 19.091883pt) -- +(135.000000:14.142136pt and 19.091883pt) -- +(225.000000:14.142136pt and 19.091883pt) -- cycle;
\draw (34.000000, 127.500000) node {BS};
\end{scope}
\begin{scope}
\draw[fill=white] (66.000000, 120.000000) +(-45.000000:14.142136pt and 8.485281pt) -- +(45.000000:14.142136pt and 8.485281pt) -- +(135.000000:14.142136pt and 8.485281pt) -- +(225.000000:14.142136pt and 8.485281pt) -- cycle;
\clip (66.000000, 120.000000) +(-45.000000:14.142136pt and 8.485281pt) -- +(45.000000:14.142136pt and 8.485281pt) -- +(135.000000:14.142136pt and 8.485281pt) -- +(225.000000:14.142136pt and 8.485281pt) -- cycle;
\draw (66.000000, 120.000000) node {SR};
\end{scope}
\draw (98.000000,120.000000) -- (98.000000,105.000000);
\begin{scope}
\draw[fill=white] (98.000000, 112.500000) +(-45.000000:14.142136pt and 19.091883pt) -- +(45.000000:14.142136pt and 19.091883pt) -- +(135.000000:14.142136pt and 19.091883pt) -- +(225.000000:14.142136pt and 19.091883pt) -- cycle;
\clip (98.000000, 112.500000) +(-45.000000:14.142136pt and 19.091883pt) -- +(45.000000:14.142136pt and 19.091883pt) -- +(135.000000:14.142136pt and 19.091883pt) -- +(225.000000:14.142136pt and 19.091883pt) -- cycle;
\draw (98.000000, 112.500000) node {MC};
\end{scope}
\draw (123.000000,150.000000) -- (123.000000,105.000000);
\filldraw (123.000000, 150.000000) circle(1.500000pt);
\begin{scope}
\draw[fill=white] (123.000000, 105.000000) circle(3.000000pt);
\clip (123.000000, 105.000000) circle(3.000000pt);
\draw (120.000000, 105.000000) -- (126.000000, 105.000000);
\draw (123.000000, 102.000000) -- (123.000000, 108.000000);
\end{scope}
\draw (160.500000,165.000000) -- (160.500000,150.000000);
\begin{scope}
\draw[fill=white] (160.500000, 157.500000) +(-45.000000:31.819805pt and 19.091883pt) -- +(45.000000:31.819805pt and 19.091883pt) -- +(135.000000:31.819805pt and 19.091883pt) -- +(225.000000:31.819805pt and 19.091883pt) -- cycle;
\clip (160.500000, 157.500000) +(-45.000000:31.819805pt and 19.091883pt) -- +(45.000000:31.819805pt and 19.091883pt) -- +(135.000000:31.819805pt and 19.091883pt) -- +(225.000000:31.819805pt and 19.091883pt) -- cycle;
\draw (160.500000, 157.500000) node {{KE$_0^{3}$}};
\end{scope}
\draw (160.500000,105.000000) -- (160.500000,90.000000);
\begin{scope}
\draw[fill=white] (160.500000, 97.500000) +(-45.000000:14.142136pt and 19.091883pt) -- +(45.000000:14.142136pt and 19.091883pt) -- +(135.000000:14.142136pt and 19.091883pt) -- +(225.000000:14.142136pt and 19.091883pt) -- cycle;
\clip (160.500000, 97.500000) +(-45.000000:14.142136pt and 19.091883pt) -- +(45.000000:14.142136pt and 19.091883pt) -- +(135.000000:14.142136pt and 19.091883pt) -- +(225.000000:14.142136pt and 19.091883pt) -- cycle;
\draw (160.500000, 97.500000) node {BS};
\end{scope}
\begin{scope}
\draw[fill=white] (205.000000, 90.000000) +(-45.000000:14.142136pt and 8.485281pt) -- +(45.000000:14.142136pt and 8.485281pt) -- +(135.000000:14.142136pt and 8.485281pt) -- +(225.000000:14.142136pt and 8.485281pt) -- cycle;
\clip (205.000000, 90.000000) +(-45.000000:14.142136pt and 8.485281pt) -- +(45.000000:14.142136pt and 8.485281pt) -- +(135.000000:14.142136pt and 8.485281pt) -- +(225.000000:14.142136pt and 8.485281pt) -- cycle;
\draw (205.000000, 90.000000) node {SR};
\end{scope}
\draw (237.000000,90.000000) -- (237.000000,75.000000);
\begin{scope}
\draw[fill=white] (237.000000, 82.500000) +(-45.000000:14.142136pt and 19.091883pt) -- +(45.000000:14.142136pt and 19.091883pt) -- +(135.000000:14.142136pt and 19.091883pt) -- +(225.000000:14.142136pt and 19.091883pt) -- cycle;
\clip (237.000000, 82.500000) +(-45.000000:14.142136pt and 19.091883pt) -- +(45.000000:14.142136pt and 19.091883pt) -- +(135.000000:14.142136pt and 19.091883pt) -- +(225.000000:14.142136pt and 19.091883pt) -- cycle;
\draw (237.000000, 82.500000) node {MC};
\end{scope}
\draw (262.000000,165.000000) -- (262.000000,75.000000);
\filldraw (262.000000, 165.000000) circle(1.500000pt);
\begin{scope}
\draw[fill=white] (262.000000, 75.000000) circle(3.000000pt);
\clip (262.000000, 75.000000) circle(3.000000pt);
\draw (259.000000, 75.000000) -- (265.000000, 75.000000);
\draw (262.000000, 72.000000) -- (262.000000, 78.000000);
\end{scope}
\draw (297.000000,165.000000) -- (297.000000,150.000000);
\begin{scope}
\draw[fill=white] (297.000000, 157.500000) +(-45.000000:28.284271pt and 19.091883pt) -- +(45.000000:28.284271pt and 19.091883pt) -- +(135.000000:28.284271pt and 19.091883pt) -- +(225.000000:28.284271pt and 19.091883pt) -- cycle;
\clip (297.000000, 157.500000) +(-45.000000:28.284271pt and 19.091883pt) -- +(45.000000:28.284271pt and 19.091883pt) -- +(135.000000:28.284271pt and 19.091883pt) -- +(225.000000:28.284271pt and 19.091883pt) -- cycle;
\draw (297.000000, 157.500000) node {{KE$_{4}^{7}$}};
\end{scope}
\draw (297.000000,75.000000) -- (297.000000,60.000000);
\begin{scope}
\draw[fill=white] (297.000000, 67.500000) +(-45.000000:14.142136pt and 19.091883pt) -- +(45.000000:14.142136pt and 19.091883pt) -- +(135.000000:14.142136pt and 19.091883pt) -- +(225.000000:14.142136pt and 19.091883pt) -- cycle;
\clip (297.000000, 67.500000) +(-45.000000:14.142136pt and 19.091883pt) -- +(45.000000:14.142136pt and 19.091883pt) -- +(135.000000:14.142136pt and 19.091883pt) -- +(225.000000:14.142136pt and 19.091883pt) -- cycle;
\draw (297.000000, 67.500000) node {BS};
\end{scope}
\begin{scope}
\draw[fill=white] (339.000000, 60.000000) +(-45.000000:14.142136pt and 8.485281pt) -- +(45.000000:14.142136pt and 8.485281pt) -- +(135.000000:14.142136pt and 8.485281pt) -- +(225.000000:14.142136pt and 8.485281pt) -- cycle;
\clip (339.000000, 60.000000) +(-45.000000:14.142136pt and 8.485281pt) -- +(45.000000:14.142136pt and 8.485281pt) -- +(135.000000:14.142136pt and 8.485281pt) -- +(225.000000:14.142136pt and 8.485281pt) -- cycle;
\draw (339.000000, 60.000000) node {SR};
\end{scope}
\draw (371.000000,60.000000) -- (371.000000,45.000000);
\begin{scope}
\draw[fill=white] (371.000000, 52.500000) +(-45.000000:14.142136pt and 19.091883pt) -- +(45.000000:14.142136pt and 19.091883pt) -- +(135.000000:14.142136pt and 19.091883pt) -- +(225.000000:14.142136pt and 19.091883pt) -- cycle;
\clip (371.000000, 52.500000) +(-45.000000:14.142136pt and 19.091883pt) -- +(45.000000:14.142136pt and 19.091883pt) -- +(135.000000:14.142136pt and 19.091883pt) -- +(225.000000:14.142136pt and 19.091883pt) -- cycle;
\draw (371.000000, 52.500000) node {MC};
\end{scope}
\draw (396.000000,150.000000) -- (396.000000,45.000000);
\filldraw (396.000000, 150.000000) circle(1.500000pt);
\begin{scope}
\draw[fill=white] (396.000000, 45.000000) circle(3.000000pt);
\clip (396.000000, 45.000000) circle(3.000000pt);
\draw (393.000000, 45.000000) -- (399.000000, 45.000000);
\draw (396.000000, 42.000000) -- (396.000000, 48.000000);
\end{scope}
\draw[fill=white,color=white] (411.000000, -6.000000) rectangle (426.000000, 171.000000);
\draw (418.500000, 82.500000) node {\dots};
\draw (460.500000,165.000000) -- (460.500000,150.000000);
\begin{scope}
\draw[fill=white] (460.500000, 157.500000) +(-45.000000:31.819805pt and 19.091883pt) -- +(45.000000:31.819805pt and 19.091883pt) -- +(135.000000:31.819805pt and 19.091883pt) -- +(225.000000:31.819805pt and 19.091883pt) -- cycle;
\clip (460.500000, 157.500000) +(-45.000000:31.819805pt and 19.091883pt) -- +(45.000000:31.819805pt and 19.091883pt) -- +(135.000000:31.819805pt and 19.091883pt) -- +(225.000000:31.819805pt and 19.091883pt) -- cycle;
\draw (460.500000, 157.500000) node {{KE$_0^{3}$}};
\end{scope}
\draw (460.500000,15.000000) -- (460.500000,0.000000);
\begin{scope}
\draw[fill=white] (460.500000, 7.500000) +(-45.000000:14.142136pt and 19.091883pt) -- +(45.000000:14.142136pt and 19.091883pt) -- +(135.000000:14.142136pt and 19.091883pt) -- +(225.000000:14.142136pt and 19.091883pt) -- cycle;
\clip (460.500000, 7.500000) +(-45.000000:14.142136pt and 19.091883pt) -- +(45.000000:14.142136pt and 19.091883pt) -- +(135.000000:14.142136pt and 19.091883pt) -- +(225.000000:14.142136pt and 19.091883pt) -- cycle;
\draw (460.500000, 7.500000) node {BS};
\end{scope}
\begin{scope}
\draw[fill=white] (505.000000, -0.000000) +(-45.000000:14.142136pt and 8.485281pt) -- +(45.000000:14.142136pt and 8.485281pt) -- +(135.000000:14.142136pt and 8.485281pt) -- +(225.000000:14.142136pt and 8.485281pt) -- cycle;
\clip (505.000000, -0.000000) +(-45.000000:14.142136pt and 8.485281pt) -- +(45.000000:14.142136pt and 8.485281pt) -- +(135.000000:14.142136pt and 8.485281pt) -- +(225.000000:14.142136pt and 8.485281pt) -- cycle;
\draw (505.000000, -0.000000) node {SR};
\end{scope}
\draw (530.000000,165.000000) -- (530.000000,0.000000);
\filldraw (530.000000, 165.000000) circle(1.500000pt);
\begin{scope}
\draw[fill=white] (530.000000, 0.000000) circle(3.000000pt);
\clip (530.000000, 0.000000) circle(3.000000pt);
\draw (527.000000, 0.000000) -- (533.000000, 0.000000);
\draw (530.000000, -3.000000) -- (530.000000, 3.000000);
\end{scope}
\filldraw[color=white,fill=white] (539.000000,146.250000) rectangle (543.000000,168.750000);
\draw[decorate,decoration={brace,mirror,amplitude = 4.000000pt},very thick] (539.000000,146.250000) -- (539.000000,168.750000);
\draw[color=black] (543.000000,157.500000) node[right] {$\K{k}_{14}$};
\draw[color=black] (539.000000,30.000000) node[anchor=mid west] {$\vdots$};
\draw[color=black] (539.000000,0.000000) node[right] {$\K{c}$};
\draw[decorate,decoration={brace,amplitude = 4.000000pt},very thick] (21.000000,172.500000) -- (129.000000,172.500000);
\draw (75.000000, 176.500000) node[text width=144pt,above,text centered] {Round 1};
\draw[decorate,decoration={brace,amplitude = 4.000000pt},very thick] (135.000000,172.500000) -- (268.000000,172.500000);
\draw (201.500000, 176.500000) node[text width=144pt,above,text centered] {Round 2};
\draw[decorate,decoration={brace,amplitude = 4.000000pt},very thick] (274.000000,172.500000) -- (402.000000,172.500000);
\draw (338.000000, 176.500000) node[text width=144pt,above,text centered] {Round 3};
\draw[decorate,decoration={brace,amplitude = 4.000000pt},very thick] (435.000000,172.500000) -- (536.000000,172.500000);
\draw (485.500000, 176.500000) node[text width=144pt,above,text centered] {Round 14};
\end{tikzpicture}

%% file: diagrams/lowmc/in_place_sbox.tikz
\providecommand{\K}[1]{\left|#1\right\rangle}
\providecommand{\adj}[1]{#1${}^\dagger$}
\begin{tikzpicture}[scale=1.000000,x=1pt,y=1pt]
\filldraw[color=white] (0.000000, -7.500000) rectangle (90.000000, 37.500000);
\draw[color=black] (0.000000,30.000000) -- (90.000000,30.000000);
\draw[color=black] (0.000000,30.000000) node[left] {$\K{a}$};
\draw[color=black] (0.000000,15.000000) -- (90.000000,15.000000);
\draw[color=black] (0.000000,15.000000) node[left] {$\K{b}$};
\draw[color=black] (0.000000,0.000000) -- (90.000000,0.000000);
\draw[color=black] (0.000000,0.000000) node[left] {$\K{c}$};
\draw (9.000000,30.000000) -- (9.000000,0.000000);
\filldraw (9.000000, 15.000000) circle(1.500000pt);
\filldraw (9.000000, 0.000000) circle(1.500000pt);
\begin{scope}
\draw[fill=white] (9.000000, 30.000000) circle(3.000000pt);
\clip (9.000000, 30.000000) circle(3.000000pt);
\draw (6.000000, 30.000000) -- (12.000000, 30.000000);
\draw (9.000000, 27.000000) -- (9.000000, 33.000000);
\end{scope}
\draw (27.000000,30.000000) -- (27.000000,0.000000);
\filldraw (27.000000, 30.000000) circle(1.500000pt);
\filldraw (27.000000, 0.000000) circle(1.500000pt);
\begin{scope}
\draw[fill=white] (27.000000, 15.000000) circle(3.000000pt);
\clip (27.000000, 15.000000) circle(3.000000pt);
\draw (24.000000, 15.000000) -- (30.000000, 15.000000);
\draw (27.000000, 12.000000) -- (27.000000, 18.000000);
\end{scope}
\draw (45.000000,30.000000) -- (45.000000,0.000000);
\filldraw (45.000000, 30.000000) circle(1.500000pt);
\filldraw (45.000000, 15.000000) circle(1.500000pt);
\begin{scope}
\draw[fill=white] (45.000000, 0.000000) circle(3.000000pt);
\clip (45.000000, 0.000000) circle(3.000000pt);
\draw (42.000000, 0.000000) -- (48.000000, 0.000000);
\draw (45.000000, -3.000000) -- (45.000000, 3.000000);
\end{scope}
\draw (63.000000,30.000000) -- (63.000000,15.000000);
\begin{scope}
\draw[fill=white] (63.000000, 15.000000) circle(3.000000pt);
\clip (63.000000, 15.000000) circle(3.000000pt);
\draw (60.000000, 15.000000) -- (66.000000, 15.000000);
\draw (63.000000, 12.000000) -- (63.000000, 18.000000);
\end{scope}
\filldraw (63.000000, 30.000000) circle(1.500000pt);
\draw (81.000000,15.000000) -- (81.000000,0.000000);
\begin{scope}
\draw[fill=white] (81.000000, 0.000000) circle(3.000000pt);
\clip (81.000000, 0.000000) circle(3.000000pt);
\draw (78.000000, 0.000000) -- (84.000000, 0.000000);
\draw (81.000000, -3.000000) -- (81.000000, 3.000000);
\end{scope}
\filldraw (81.000000, 15.000000) circle(1.500000pt);
\draw[color=black] (90.000000,30.000000) node[right] {$\K{a + bc}$};
\draw[color=black] (90.000000,15.000000) node[right] {$\K{a + b + ac}$};
\draw[color=black] (90.000000,0.000000) node[right] {$\K{a + b + c + ab}$};
\end{tikzpicture}

%% file: diagrams/lowmc/sbox.tikz
\providecommand{\K}[1]{\left|#1\right\rangle}
\providecommand{\adj}[1]{#1${}^\dagger$}
\begin{tikzpicture}[scale=1.000000,x=1pt,y=1pt]
\filldraw[color=white] (0.000000, -7.500000) rectangle (150.000000, 127.500000);
\draw[color=black] (0.000000,120.000000) -- (150.000000,120.000000);
\draw[color=black] (0.000000,120.000000) node[left] {$\K{a}$};
\draw[color=black] (0.000000,105.000000) -- (150.000000,105.000000);
\draw[color=black] (0.000000,105.000000) node[left] {$\K{b}$};
\draw[color=black] (0.000000,90.000000) -- (150.000000,90.000000);
\draw[color=black] (0.000000,90.000000) node[left] {$\K{c}$};
\draw[color=black] (0.000000,75.000000) -- (150.000000,75.000000);
\draw[color=black] (0.000000,75.000000) node[left] {$\K{0}$};
\draw[color=black] (0.000000,60.000000) -- (150.000000,60.000000);
\draw[color=black] (0.000000,60.000000) node[left] {$\K{0}$};
\draw[color=black] (0.000000,45.000000) -- (150.000000,45.000000);
\draw[color=black] (0.000000,45.000000) node[left] {$\K{0}$};
\draw[color=black] (0.000000,30.000000) -- (150.000000,30.000000);
\draw[color=black] (0.000000,30.000000) node[left] {$\K{x}$};
\draw[color=black] (0.000000,15.000000) -- (150.000000,15.000000);
\draw[color=black] (0.000000,15.000000) node[left] {$\K{y}$};
\draw[color=black] (0.000000,0.000000) -- (150.000000,0.000000);
\draw[color=black] (0.000000,0.000000) node[left] {$\K{z}$};
\draw (9.000000,120.000000) -- (9.000000,75.000000);
\begin{scope}
\draw[fill=white] (9.000000, 75.000000) circle(3.000000pt);
\clip (9.000000, 75.000000) circle(3.000000pt);
\draw (6.000000, 75.000000) -- (12.000000, 75.000000);
\draw (9.000000, 72.000000) -- (9.000000, 78.000000);
\end{scope}
\filldraw (9.000000, 120.000000) circle(1.500000pt);
\draw (15.000000,105.000000) -- (15.000000,60.000000);
\begin{scope}
\draw[fill=white] (15.000000, 60.000000) circle(3.000000pt);
\clip (15.000000, 60.000000) circle(3.000000pt);
\draw (12.000000, 60.000000) -- (18.000000, 60.000000);
\draw (15.000000, 57.000000) -- (15.000000, 63.000000);
\end{scope}
\filldraw (15.000000, 105.000000) circle(1.500000pt);
\draw (21.000000,90.000000) -- (21.000000,45.000000);
\begin{scope}
\draw[fill=white] (21.000000, 45.000000) circle(3.000000pt);
\clip (21.000000, 45.000000) circle(3.000000pt);
\draw (18.000000, 45.000000) -- (24.000000, 45.000000);
\draw (21.000000, 42.000000) -- (21.000000, 48.000000);
\end{scope}
\filldraw (21.000000, 90.000000) circle(1.500000pt);
\draw (39.000000,60.000000) -- (39.000000,30.000000);
\filldraw (39.000000, 60.000000) circle(1.500000pt);
\filldraw (39.000000, 45.000000) circle(1.500000pt);
\begin{scope}
\draw[fill=white] (39.000000, 30.000000) circle(3.000000pt);
\clip (39.000000, 30.000000) circle(3.000000pt);
\draw (36.000000, 30.000000) -- (42.000000, 30.000000);
\draw (39.000000, 27.000000) -- (39.000000, 33.000000);
\end{scope}
\draw (45.000000,90.000000) -- (45.000000,15.000000);
\filldraw (45.000000, 90.000000) circle(1.500000pt);
\filldraw (45.000000, 75.000000) circle(1.500000pt);
\begin{scope}
\draw[fill=white] (45.000000, 15.000000) circle(3.000000pt);
\clip (45.000000, 15.000000) circle(3.000000pt);
\draw (42.000000, 15.000000) -- (48.000000, 15.000000);
\draw (45.000000, 12.000000) -- (45.000000, 18.000000);
\end{scope}
\draw (51.000000,120.000000) -- (51.000000,0.000000);
\filldraw (51.000000, 120.000000) circle(1.500000pt);
\filldraw (51.000000, 105.000000) circle(1.500000pt);
\begin{scope}
\draw[fill=white] (51.000000, 0.000000) circle(3.000000pt);
\clip (51.000000, 0.000000) circle(3.000000pt);
\draw (48.000000, 0.000000) -- (54.000000, 0.000000);
\draw (51.000000, -3.000000) -- (51.000000, 3.000000);
\end{scope}
\draw (69.000000,75.000000) -- (69.000000,30.000000);
\begin{scope}
\draw[fill=white] (69.000000, 30.000000) circle(3.000000pt);
\clip (69.000000, 30.000000) circle(3.000000pt);
\draw (66.000000, 30.000000) -- (72.000000, 30.000000);
\draw (69.000000, 27.000000) -- (69.000000, 33.000000);
\end{scope}
\filldraw (69.000000, 75.000000) circle(1.500000pt);
\draw (75.000000,60.000000) -- (75.000000,15.000000);
\begin{scope}
\draw[fill=white] (75.000000, 15.000000) circle(3.000000pt);
\clip (75.000000, 15.000000) circle(3.000000pt);
\draw (72.000000, 15.000000) -- (78.000000, 15.000000);
\draw (75.000000, 12.000000) -- (75.000000, 18.000000);
\end{scope}
\filldraw (75.000000, 60.000000) circle(1.500000pt);
\draw (81.000000,45.000000) -- (81.000000,0.000000);
\begin{scope}
\draw[fill=white] (81.000000, 0.000000) circle(3.000000pt);
\clip (81.000000, 0.000000) circle(3.000000pt);
\draw (78.000000, 0.000000) -- (84.000000, 0.000000);
\draw (81.000000, -3.000000) -- (81.000000, 3.000000);
\end{scope}
\filldraw (81.000000, 45.000000) circle(1.500000pt);
\draw (99.000000,90.000000) -- (99.000000,45.000000);
\begin{scope}
\draw[fill=white] (99.000000, 45.000000) circle(3.000000pt);
\clip (99.000000, 45.000000) circle(3.000000pt);
\draw (96.000000, 45.000000) -- (102.000000, 45.000000);
\draw (99.000000, 42.000000) -- (99.000000, 48.000000);
\end{scope}
\filldraw (99.000000, 90.000000) circle(1.500000pt);
\draw (105.000000,105.000000) -- (105.000000,60.000000);
\begin{scope}
\draw[fill=white] (105.000000, 60.000000) circle(3.000000pt);
\clip (105.000000, 60.000000) circle(3.000000pt);
\draw (102.000000, 60.000000) -- (108.000000, 60.000000);
\draw (105.000000, 57.000000) -- (105.000000, 63.000000);
\end{scope}
\filldraw (105.000000, 105.000000) circle(1.500000pt);
\draw (111.000000,120.000000) -- (111.000000,15.000000);
\begin{scope}
\draw[fill=white] (111.000000, 15.000000) circle(3.000000pt);
\clip (111.000000, 15.000000) circle(3.000000pt);
\draw (108.000000, 15.000000) -- (114.000000, 15.000000);
\draw (111.000000, 12.000000) -- (111.000000, 18.000000);
\end{scope}
\filldraw (111.000000, 120.000000) circle(1.500000pt);
\draw (117.000000,75.000000) -- (117.000000,0.000000);
\begin{scope}
\draw[fill=white] (117.000000, 0.000000) circle(3.000000pt);
\clip (117.000000, 0.000000) circle(3.000000pt);
\draw (114.000000, 0.000000) -- (120.000000, 0.000000);
\draw (117.000000, -3.000000) -- (117.000000, 3.000000);
\end{scope}
\filldraw (117.000000, 75.000000) circle(1.500000pt);
\draw (135.000000,120.000000) -- (135.000000,75.000000);
\begin{scope}
\draw[fill=white] (135.000000, 75.000000) circle(3.000000pt);
\clip (135.000000, 75.000000) circle(3.000000pt);
\draw (132.000000, 75.000000) -- (138.000000, 75.000000);
\draw (135.000000, 72.000000) -- (135.000000, 78.000000);
\end{scope}
\filldraw (135.000000, 120.000000) circle(1.500000pt);
\draw (141.000000,105.000000) -- (141.000000,0.000000);
\begin{scope}
\draw[fill=white] (141.000000, 0.000000) circle(3.000000pt);
\clip (141.000000, 0.000000) circle(3.000000pt);
\draw (138.000000, 0.000000) -- (144.000000, 0.000000);
\draw (141.000000, -3.000000) -- (141.000000, 3.000000);
\end{scope}
\filldraw (141.000000, 105.000000) circle(1.500000pt);
\draw[color=black] (150.000000,120.000000) node[right] {$\K{a}$};
\draw[color=black] (150.000000,105.000000) node[right] {$\K{b}$};
\draw[color=black] (150.000000,90.000000) node[right] {$\K{c}$};
\draw[color=black] (150.000000,75.000000) node[right] {$\K{0}$};
\draw[color=black] (150.000000,60.000000) node[right] {$\K{0}$};
\draw[color=black] (150.000000,45.000000) node[right] {$\K{0}$};
\draw[color=black] (150.000000,30.000000) node[right] {$\K{x + a + bc}$};
\draw[color=black] (150.000000,15.000000) node[right] {$\K{y + a + b + ac}$};
\draw[color=black] (150.000000,0.000000) node[right] {$\K{z + a + b + c + ab}$};
\end{tikzpicture}

%% file: diagrams/aes/oracle.tikz
\providecommand{\K}[1]{\left|#1\right\rangle}
\providecommand{\adj}[1]{#1${}^\dagger$}
\begin{tikzpicture}[scale=1.000000,x=1pt,y=1pt]
\filldraw[color=white] (0.000000, -7.500000) rectangle (212.000000, 97.500000);
\draw[color=black] (0.000000,90.000000) -- (212.000000,90.000000);
\draw[color=black] (0.000000,90.000000) node[left] {$\K{k}_{0}$};
\draw[color=black] (0.000000,75.000000) -- (212.000000,75.000000);
\draw[color=black] (0.000000,75.000000) node[left] {$\K{m_1}$};
\draw[color=black] (13.500000,60.000000) -- (198.500000,60.000000);
\draw[color=black] (13.500000,45.000000) -- (198.500000,45.000000);
\draw[color=black] (0.000000,30.000000) -- (212.000000,30.000000);
\draw[color=black] (0.000000,30.000000) node[left] {$\K{m_2}$};
\draw[color=black] (13.500000,15.000000) -- (198.500000,15.000000);
\draw[color=black] (0.000000,0.000000) -- (212.000000,0.000000);
\draw[color=black] (0.000000,0.000000) node[left] {$\K{-}$};
\draw[color=black] (21.000000,45.000000) node[fill=white,left,minimum height=15.000000pt,minimum width=15.000000pt,inner sep=0pt] {\phantom{$\K{0}$}};
\draw[color=black] (21.000000,45.000000) node[left] {$\K{0}$};
\draw[color=black] (21.000000,60.000000) node[fill=white,left,minimum height=15.000000pt,minimum width=15.000000pt,inner sep=0pt] {\phantom{$\K{0}$}};
\draw[color=black] (21.000000,60.000000) node[left] {$\K{0}$};
\draw[color=black] (21.000000,15.000000) node[fill=white,left,minimum height=15.000000pt,minimum width=15.000000pt,inner sep=0pt] {\phantom{$\K{0}$}};
\draw[color=black] (21.000000,15.000000) node[left] {$\K{0}$};
\draw (36.000000,90.000000) -- (36.000000,45.000000);
\filldraw (36.000000, 90.000000) circle(1.500000pt);
\begin{scope}
\draw[fill=white] (36.000000, 45.000000) circle(3.000000pt);
\clip (36.000000, 45.000000) circle(3.000000pt);
\draw (33.000000, 45.000000) -- (39.000000, 45.000000);
\draw (36.000000, 42.000000) -- (36.000000, 48.000000);
\end{scope}
\draw (71.000000,90.000000) -- (71.000000,60.000000);
\begin{scope}
\draw[fill=white] (71.000000, 75.000000) +(-45.000000:28.284271pt and 29.698485pt) -- +(45.000000:28.284271pt and 29.698485pt) -- +(135.000000:28.284271pt and 29.698485pt) -- +(225.000000:28.284271pt and 29.698485pt) -- cycle;
\clip (71.000000, 75.000000) +(-45.000000:28.284271pt and 29.698485pt) -- +(45.000000:28.284271pt and 29.698485pt) -- +(135.000000:28.284271pt and 29.698485pt) -- +(225.000000:28.284271pt and 29.698485pt) -- cycle;
\draw (71.000000, 75.000000) node {FwAES};
\end{scope}
\draw (71.000000,45.000000) -- (71.000000,15.000000);
\begin{scope}
\draw[fill=white] (71.000000, 30.000000) +(-45.000000:28.284271pt and 29.698485pt) -- +(45.000000:28.284271pt and 29.698485pt) -- +(135.000000:28.284271pt and 29.698485pt) -- +(225.000000:28.284271pt and 29.698485pt) -- cycle;
\clip (71.000000, 30.000000) +(-45.000000:28.284271pt and 29.698485pt) -- +(45.000000:28.284271pt and 29.698485pt) -- +(135.000000:28.284271pt and 29.698485pt) -- +(225.000000:28.284271pt and 29.698485pt) -- cycle;
\draw (71.000000, 30.000000) node {FwAES};
\end{scope}
\draw (106.000000,60.000000) -- (106.000000,0.000000);
\filldraw (106.000000, 60.000000) circle(1.500000pt);
\filldraw (106.000000, 15.000000) circle(1.500000pt);
\begin{scope}
\draw[fill=white] (106.000000, 0.000000) circle(5.000000pt);
\clip (106.000000, 0.000000) circle(5.000000pt);
\begin{scope}[shift={(106.000000,0.000000)}]
\draw (-3,-0.9) -- (3,-0.9); \draw (-3,0.9) -- (3,0.9);
\end{scope}
\end{scope}
\draw (141.000000,90.000000) -- (141.000000,60.000000);
\begin{scope}
\draw[fill=white] (141.000000, 75.000000) +(-45.000000:28.284271pt and 29.698485pt) -- +(45.000000:28.284271pt and 29.698485pt) -- +(135.000000:28.284271pt and 29.698485pt) -- +(225.000000:28.284271pt and 29.698485pt) -- cycle;
\clip (141.000000, 75.000000) +(-45.000000:28.284271pt and 29.698485pt) -- +(45.000000:28.284271pt and 29.698485pt) -- +(135.000000:28.284271pt and 29.698485pt) -- +(225.000000:28.284271pt and 29.698485pt) -- cycle;
\draw (141.000000, 75.000000) node {\adj{FwAES}};
\end{scope}
\draw (141.000000,45.000000) -- (141.000000,15.000000);
\begin{scope}
\draw[fill=white] (141.000000, 30.000000) +(-45.000000:28.284271pt and 29.698485pt) -- +(45.000000:28.284271pt and 29.698485pt) -- +(135.000000:28.284271pt and 29.698485pt) -- +(225.000000:28.284271pt and 29.698485pt) -- cycle;
\clip (141.000000, 30.000000) +(-45.000000:28.284271pt and 29.698485pt) -- +(45.000000:28.284271pt and 29.698485pt) -- +(135.000000:28.284271pt and 29.698485pt) -- +(225.000000:28.284271pt and 29.698485pt) -- cycle;
\draw (141.000000, 30.000000) node {\adj{FwAES}};
\end{scope}
\draw (176.000000,90.000000) -- (176.000000,45.000000);
\filldraw (176.000000, 90.000000) circle(1.500000pt);
\begin{scope}
\draw[fill=white] (176.000000, 45.000000) circle(3.000000pt);
\clip (176.000000, 45.000000) circle(3.000000pt);
\draw (173.000000, 45.000000) -- (179.000000, 45.000000);
\draw (176.000000, 42.000000) -- (176.000000, 48.000000);
\end{scope}
\draw[color=black] (191.000000,45.000000) node[fill=white,right,minimum height=15.000000pt,minimum width=15.000000pt,inner sep=0pt] {\phantom{$\K{0}$}};
\draw[color=black] (191.000000,45.000000) node[right] {$\K{0}$};
\draw[color=black] (191.000000,60.000000) node[fill=white,right,minimum height=15.000000pt,minimum width=15.000000pt,inner sep=0pt] {\phantom{$\K{0}$}};
\draw[color=black] (191.000000,60.000000) node[right] {$\K{0}$};
\draw[color=black] (191.000000,15.000000) node[fill=white,right,minimum height=15.000000pt,minimum width=15.000000pt,inner sep=0pt] {\phantom{$\K{0}$}};
\draw[color=black] (191.000000,15.000000) node[right] {$\K{0}$};
\draw[color=black] (212.000000,90.000000) node[right] {$\K{k}_{0}$};
\draw[color=black] (212.000000,75.000000) node[right] {$\K{m_1}$};
\draw[color=black] (212.000000,30.000000) node[right] {$\K{m_2}$};
\draw[color=black] (212.000000,0.000000) node[right] {$\K{-}$};
\draw[draw opacity=1.000000,fill opacity=0.200000,color=black,dashed,rounded corners=2pt] (3.000000,97.500000) rectangle (209.000000,-7.500000);
\draw[draw opacity=1.000000,fill opacity=0.200000,color=black,dashed,rounded corners=2pt] (3.000000,97.500000) rectangle (209.000000,-7.500000);
\end{tikzpicture}

%% file: diagrams/general/AND.tikz
\providecommand{\K}[1]{\left|#1\right\rangle}
\providecommand{\adj}[1]{#1${}^\dagger$}
\begin{tikzpicture}[scale=1.000000,x=1pt,y=1pt]
\filldraw[color=white] (0.000000, -7.500000) rectangle (199.500000, 52.500000);
\draw[color=black] (0.000000,45.000000) -- (199.500000,45.000000);
\draw[color=black] (0.000000,45.000000) node[left] {$\K{a}$};
\draw[color=black] (0.000000,30.000000) -- (199.500000,30.000000);
\draw[color=black] (0.000000,30.000000) node[left] {$\K{b}$};
\draw[color=black] (0.000000,15.000000) -- (199.500000,15.000000);
\draw[color=black] (0.000000,15.000000) node[left] {$\K{0}$};
\draw[color=black] (13.500000,0.000000) -- (186.000000,0.000000);
\draw[color=black] (21.000000,0.000000) node[fill=white,left,minimum height=15.000000pt,minimum width=15.000000pt,inner sep=0pt] {\phantom{$\K{0}$}};
\draw[color=black] (21.000000,0.000000) node[left] {$\K{0}$};
\begin{scope}
\draw[fill=white] (13.500000, 15.000000) +(-45.000000:8.485281pt and 8.485281pt) -- +(45.000000:8.485281pt and 8.485281pt) -- +(135.000000:8.485281pt and 8.485281pt) -- +(225.000000:8.485281pt and 8.485281pt) -- cycle;
\clip (13.500000, 15.000000) +(-45.000000:8.485281pt and 8.485281pt) -- +(45.000000:8.485281pt and 8.485281pt) -- +(135.000000:8.485281pt and 8.485281pt) -- +(225.000000:8.485281pt and 8.485281pt) -- cycle;
\draw (13.500000, 15.000000) node {$H$};
\end{scope}
\draw (36.000000,30.000000) -- (36.000000,0.000000);
\filldraw (36.000000, 30.000000) circle(1.500000pt);
\begin{scope}
\draw[fill=white] (36.000000, 0.000000) circle(3.000000pt);
\clip (36.000000, 0.000000) circle(3.000000pt);
\draw (33.000000, 0.000000) -- (39.000000, 0.000000);
\draw (36.000000, -3.000000) -- (36.000000, 3.000000);
\end{scope}
\draw (42.000000,45.000000) -- (42.000000,15.000000);
\filldraw (42.000000, 15.000000) circle(1.500000pt);
\begin{scope}
\draw[fill=white] (42.000000, 45.000000) circle(3.000000pt);
\clip (42.000000, 45.000000) circle(3.000000pt);
\draw (39.000000, 45.000000) -- (45.000000, 45.000000);
\draw (42.000000, 42.000000) -- (42.000000, 48.000000);
\end{scope}
\draw (60.000000,30.000000) -- (60.000000,15.000000);
\filldraw (60.000000, 15.000000) circle(1.500000pt);
\begin{scope}
\draw[fill=white] (60.000000, 30.000000) circle(3.000000pt);
\clip (60.000000, 30.000000) circle(3.000000pt);
\draw (57.000000, 30.000000) -- (63.000000, 30.000000);
\draw (60.000000, 27.000000) -- (60.000000, 33.000000);
\end{scope}
\draw (66.000000,45.000000) -- (66.000000,0.000000);
\filldraw (66.000000, 45.000000) circle(1.500000pt);
\begin{scope}
\draw[fill=white] (66.000000, 0.000000) circle(3.000000pt);
\clip (66.000000, 0.000000) circle(3.000000pt);
\draw (63.000000, 0.000000) -- (69.000000, 0.000000);
\draw (66.000000, -3.000000) -- (66.000000, 3.000000);
\end{scope}
\begin{scope}
\draw[fill=white] (87.750000, 45.000000) +(-45.000000:9.545942pt and 8.485281pt) -- +(45.000000:9.545942pt and 8.485281pt) -- +(135.000000:9.545942pt and 8.485281pt) -- +(225.000000:9.545942pt and 8.485281pt) -- cycle;
\clip (87.750000, 45.000000) +(-45.000000:9.545942pt and 8.485281pt) -- +(45.000000:9.545942pt and 8.485281pt) -- +(135.000000:9.545942pt and 8.485281pt) -- +(225.000000:9.545942pt and 8.485281pt) -- cycle;
\draw (87.750000, 45.000000) node {\adj{T}};
\end{scope}
\begin{scope}
\draw[fill=white] (87.750000, 30.000000) +(-45.000000:9.545942pt and 8.485281pt) -- +(45.000000:9.545942pt and 8.485281pt) -- +(135.000000:9.545942pt and 8.485281pt) -- +(225.000000:9.545942pt and 8.485281pt) -- cycle;
\clip (87.750000, 30.000000) +(-45.000000:9.545942pt and 8.485281pt) -- +(45.000000:9.545942pt and 8.485281pt) -- +(135.000000:9.545942pt and 8.485281pt) -- +(225.000000:9.545942pt and 8.485281pt) -- cycle;
\draw (87.750000, 30.000000) node {\adj{T}};
\end{scope}
\begin{scope}
\draw[fill=white] (87.750000, 15.000000) +(-45.000000:9.545942pt and 8.485281pt) -- +(45.000000:9.545942pt and 8.485281pt) -- +(135.000000:9.545942pt and 8.485281pt) -- +(225.000000:9.545942pt and 8.485281pt) -- cycle;
\clip (87.750000, 15.000000) +(-45.000000:9.545942pt and 8.485281pt) -- +(45.000000:9.545942pt and 8.485281pt) -- +(135.000000:9.545942pt and 8.485281pt) -- +(225.000000:9.545942pt and 8.485281pt) -- cycle;
\draw (87.750000, 15.000000) node {T};
\end{scope}
\begin{scope}
\draw[fill=white] (87.750000, -0.000000) +(-45.000000:9.545942pt and 8.485281pt) -- +(45.000000:9.545942pt and 8.485281pt) -- +(135.000000:9.545942pt and 8.485281pt) -- +(225.000000:9.545942pt and 8.485281pt) -- cycle;
\clip (87.750000, -0.000000) +(-45.000000:9.545942pt and 8.485281pt) -- +(45.000000:9.545942pt and 8.485281pt) -- +(135.000000:9.545942pt and 8.485281pt) -- +(225.000000:9.545942pt and 8.485281pt) -- cycle;
\draw (87.750000, -0.000000) node {T};
\end{scope}
\draw (109.500000,45.000000) -- (109.500000,0.000000);
\filldraw (109.500000, 45.000000) circle(1.500000pt);
\begin{scope}
\draw[fill=white] (109.500000, 0.000000) circle(3.000000pt);
\clip (109.500000, 0.000000) circle(3.000000pt);
\draw (106.500000, 0.000000) -- (112.500000, 0.000000);
\draw (109.500000, -3.000000) -- (109.500000, 3.000000);
\end{scope}
\draw (115.500000,30.000000) -- (115.500000,15.000000);
\filldraw (115.500000, 15.000000) circle(1.500000pt);
\begin{scope}
\draw[fill=white] (115.500000, 30.000000) circle(3.000000pt);
\clip (115.500000, 30.000000) circle(3.000000pt);
\draw (112.500000, 30.000000) -- (118.500000, 30.000000);
\draw (115.500000, 27.000000) -- (115.500000, 33.000000);
\end{scope}
\draw (133.500000,45.000000) -- (133.500000,15.000000);
\filldraw (133.500000, 15.000000) circle(1.500000pt);
\begin{scope}
\draw[fill=white] (133.500000, 45.000000) circle(3.000000pt);
\clip (133.500000, 45.000000) circle(3.000000pt);
\draw (130.500000, 45.000000) -- (136.500000, 45.000000);
\draw (133.500000, 42.000000) -- (133.500000, 48.000000);
\end{scope}
\draw (139.500000,30.000000) -- (139.500000,0.000000);
\filldraw (139.500000, 30.000000) circle(1.500000pt);
\begin{scope}
\draw[fill=white] (139.500000, 0.000000) circle(3.000000pt);
\clip (139.500000, 0.000000) circle(3.000000pt);
\draw (136.500000, 0.000000) -- (142.500000, 0.000000);
\draw (139.500000, -3.000000) -- (139.500000, 3.000000);
\end{scope}
\begin{scope}
\draw[fill=white] (160.500000, 15.000000) +(-45.000000:8.485281pt and 8.485281pt) -- +(45.000000:8.485281pt and 8.485281pt) -- +(135.000000:8.485281pt and 8.485281pt) -- +(225.000000:8.485281pt and 8.485281pt) -- cycle;
\clip (160.500000, 15.000000) +(-45.000000:8.485281pt and 8.485281pt) -- +(45.000000:8.485281pt and 8.485281pt) -- +(135.000000:8.485281pt and 8.485281pt) -- +(225.000000:8.485281pt and 8.485281pt) -- cycle;
\draw (160.500000, 15.000000) node {$H$};
\end{scope}
\begin{scope}
\draw[fill=white] (186.000000, 15.000000) +(-45.000000:8.485281pt and 8.485281pt) -- +(45.000000:8.485281pt and 8.485281pt) -- +(135.000000:8.485281pt and 8.485281pt) -- +(225.000000:8.485281pt and 8.485281pt) -- cycle;
\clip (186.000000, 15.000000) +(-45.000000:8.485281pt and 8.485281pt) -- +(45.000000:8.485281pt and 8.485281pt) -- +(135.000000:8.485281pt and 8.485281pt) -- +(225.000000:8.485281pt and 8.485281pt) -- cycle;
\draw (186.000000, 15.000000) node {S};
\end{scope}
\draw[color=black] (178.500000,0.000000) node[fill=white,right,minimum height=15.000000pt,minimum width=15.000000pt,inner sep=0pt] {\phantom{$\K{0}$}};
\draw[color=black] (178.500000,0.000000) node[right] {$\K{0}$};
\draw[color=black] (199.500000,45.000000) node[right] {$\K{a}$};
\draw[color=black] (199.500000,30.000000) node[right] {$\K{b}$};
\draw[color=black] (199.500000,15.000000) node[right] {$\K{a \cdot b}$};
\end{tikzpicture}

%% file: diagrams/general/tweaked_adjAND.tikz
\providecommand{\K}[1]{\left|#1\right\rangle}
\providecommand{\adj}[1]{#1${}^\dagger$}
\begin{tikzpicture}[scale=1.000000,x=1pt,y=1pt]
\filldraw[color=white] (0.000000, -7.500000) rectangle (147.000000, 52.500000);
\draw[color=black] (0.000000,45.000000) -- (147.000000,45.000000);
\draw[color=black] (0.000000,45.000000) node[left] {$\K{a}$};
\draw[color=black] (0.000000,30.000000) -- (147.000000,30.000000);
\draw[color=black] (0.000000,30.000000) node[left] {$\K{b}$};
\draw[color=black] (0.000000,15.000000) -- (97.500000,15.000000);
\draw[color=black] (97.500000,15.000000) -- (147.000000,15.000000);
\draw[color=black] (0.000000,15.000000) node[left] {$\K{a \cdot b}$};
\draw[color=black] (35.500000,-0.500000) -- (97.500000,-0.500000);
\draw[color=black] (36.500000,0.500000) -- (97.500000,0.500000);
\begin{scope}
\draw[fill=white] (12.000000, 15.000000) +(-45.000000:8.485281pt and 8.485281pt) -- +(45.000000:8.485281pt and 8.485281pt) -- +(135.000000:8.485281pt and 8.485281pt) -- +(225.000000:8.485281pt and 8.485281pt) -- cycle;
\clip (12.000000, 15.000000) +(-45.000000:8.485281pt and 8.485281pt) -- +(45.000000:8.485281pt and 8.485281pt) -- +(135.000000:8.485281pt and 8.485281pt) -- +(225.000000:8.485281pt and 8.485281pt) -- cycle;
\draw (12.000000, 15.000000) node {$H$};
\end{scope}

\draw (35.500000,15.000000) -- (35.500000,-0.700000);
\draw (36.500000,15.000000) -- (36.500000,0.300000);
\begin{scope}[shift={(0,15)}]
\draw[fill=white] (30.000000, -6.000000) rectangle (42.000000, 6.000000);
\draw[very thin] (36.000000, 0.600000) arc (90:150:6.000000pt);
\draw[very thin] (36.000000, 0.600000) arc (90:30:6.000000pt);
\draw[->,>=stealth] (36.000000, -5.400000) -- +(80:10.392305pt);
\end{scope}
\draw (97.500000,45.000000) -- (97.500000,15.000000);
\draw (97.000000,15.000000) -- (97.000000,0.000000);
\draw (98.000000,15.000000) -- (98.000000,0.000000);
\begin{scope}
\draw[fill=white] (97.500000, 30.000000) +(-45.000000:61.518290pt and 35.355339pt) -- +(45.000000:61.518290pt and 35.355339pt) -- +(135.000000:61.518290pt and 35.355339pt) -- +(225.000000:61.518290pt and 35.355339pt) -- cycle;
\clip (97.500000, 30.000000) +(-45.000000:61.518290pt and 35.355339pt) -- +(45.000000:61.518290pt and 35.355339pt) -- +(135.000000:61.518290pt and 35.355339pt) -- +(225.000000:61.518290pt and 35.355339pt) -- cycle;
\begin{scope}[shift={(97.500000,30.000000)}]
\begin{scope}[shift={(-43.5,-15)}] \draw[color=black] (0.000000,30.000000) -- (87.000000,30.000000); \draw[color=black] (0.000000,15.000000) -- (87.000000,15.000000); \draw[color=black] (0.000000,0.000000) -- (12.750000,0.000000); \draw[color=black] (12.750000,0.000000) -- (87.000000,0.000000); \begin{scope} \draw[fill=white] (12.750000, 30.000000) +(-45.000000:9.545942pt and 8.485281pt) -- +(45.000000:9.545942pt and 8.485281pt) -- +(135.000000:9.545942pt and 8.485281pt) -- +(225.000000:9.545942pt and 8.485281pt) -- cycle; \clip (12.750000, 30.000000) +(-45.000000:9.545942pt and 8.485281pt) -- +(45.000000:9.545942pt and 8.485281pt) -- +(135.000000:9.545942pt and 8.485281pt) -- +(225.000000:9.545942pt and 8.485281pt) -- cycle; \draw (12.750000, 30.000000) node {S}; \end{scope} \begin{scope} \draw[fill=white] (12.750000, 15.000000) +(-45.000000:9.545942pt and 8.485281pt) -- +(45.000000:9.545942pt and 8.485281pt) -- +(135.000000:9.545942pt and 8.485281pt) -- +(225.000000:9.545942pt and 8.485281pt) -- cycle; \clip (12.750000, 15.000000) +(-45.000000:9.545942pt and 8.485281pt) -- +(45.000000:9.545942pt and 8.485281pt) -- +(135.000000:9.545942pt and 8.485281pt) -- +(225.000000:9.545942pt and 8.485281pt) -- cycle; \draw (12.750000, 15.000000) node {S}; \end{scope} \begin{scope} \draw[fill=white] (12.750000, -0.000000) +(-45.000000:8.485281pt and 8.485281pt) -- +(45.000000:8.485281pt and 8.485281pt) -- +(135.000000:8.485281pt and 8.485281pt) -- +(225.000000:8.485281pt and 8.485281pt) -- cycle; \clip (12.750000, -0.000000) +(-45.000000:8.485281pt and 8.485281pt) -- +(45.000000:8.485281pt and 8.485281pt) -- +(135.000000:8.485281pt and 8.485281pt) -- +(225.000000:8.485281pt and 8.485281pt) -- cycle; \draw (12.750000, -0.000000) node {$X$}; \end{scope} \draw (34.500000,30.000000) -- (34.500000,15.000000); \filldraw (34.500000, 30.000000) circle(1.500000pt); \begin{scope} \draw[fill=white] (34.500000, 15.000000) circle(3.000000pt); \clip (34.500000, 15.000000) circle(3.000000pt); \draw (31.500000, 15.000000) -- (37.500000, 15.000000); \draw (34.500000, 12.000000) -- (34.500000, 18.000000); \end{scope} \begin{scope} \draw[fill=white] (56.250000, 15.000000) +(-45.000000:9.545942pt and 8.485281pt) -- +(45.000000:9.545942pt and 8.485281pt) -- +(135.000000:9.545942pt and 8.485281pt) -- +(225.000000:9.545942pt and 8.485281pt) -- cycle; \clip (56.250000, 15.000000) +(-45.000000:9.545942pt and 8.485281pt) -- +(45.000000:9.545942pt and 8.485281pt) -- +(135.000000:9.545942pt and 8.485281pt) -- +(225.000000:9.545942pt and 8.485281pt) -- cycle; \draw (56.250000, 15.000000) node {\adj{S}}; \end{scope} \draw (78.000000,30.000000) -- (78.000000,15.000000); \filldraw (78.000000, 30.000000) circle(1.500000pt); \begin{scope} \draw[fill=white] (78.000000, 15.000000) circle(3.000000pt); \clip (78.000000, 15.000000) circle(3.000000pt); \draw (75.000000, 15.000000) -- (81.000000, 15.000000); \draw (78.000000, 12.000000) -- (78.000000, 18.000000); \end{scope} \end{scope}
\end{scope}
\end{scope}
\filldraw (97.500000, 0.000000) circle(1.500000pt);
\draw[color=black] (147.000000,45.000000) node[right] {$\K{a}$};
\draw[color=black] (147.000000,30.000000) node[right] {$\K{b}$};
\draw[color=black] (147.000000,15.000000) node[right] {$\K{0}$};
\end{tikzpicture}

%% file: diagrams/general/dummy_sbox.tikz
\providecommand{\K}[1]{\left|#1\right\rangle}
\providecommand{\adj}[1]{#1${}^\dagger$}
\begin{tikzpicture}[scale=1.000000,x=1pt,y=1pt]
\filldraw[color=white] (0.000000, -7.500000) rectangle (294.000000, 232.500000);
\draw[color=black] (0.000000,225.000000) -- (294.000000,225.000000);
\draw[color=black] (0.000000,225.000000) node[left] {$\K{in}_0$};
\draw[color=black] (0.000000,210.000000) -- (294.000000,210.000000);
\draw[color=black] (0.000000,210.000000) node[left] {$\K{in}_1$};
\draw[color=black] (0.000000,195.000000) -- (294.000000,195.000000);
\draw[color=black] (0.000000,195.000000) node[left] {$\K{in}_2$};
\draw[color=black] (0.000000,180.000000) -- (294.000000,180.000000);
\draw[color=black] (0.000000,180.000000) node[left] {$\K{in}_3$};
\draw[color=black] (0.000000,165.000000) -- (294.000000,165.000000);
\draw[color=black] (0.000000,165.000000) node[left] {$\K{in}_4$};
\draw[color=black] (0.000000,150.000000) -- (294.000000,150.000000);
\draw[color=black] (0.000000,150.000000) node[left] {$\K{in}_5$};
\draw[color=black] (0.000000,135.000000) -- (294.000000,135.000000);
\draw[color=black] (0.000000,135.000000) node[left] {$\K{in}_6$};
\draw[color=black] (0.000000,120.000000) -- (294.000000,120.000000);
\draw[color=black] (0.000000,120.000000) node[left] {$\K{in}_7$};
\draw[color=black] (0.000000,105.000000) -- (294.000000,105.000000);
\draw[color=black] (0.000000,105.000000) node[left] {$\K{out}_0$};
\draw[color=black] (0.000000,90.000000) -- (294.000000,90.000000);
\draw[color=black] (0.000000,90.000000) node[left] {$\K{out}_1$};
\draw[color=black] (0.000000,75.000000) -- (294.000000,75.000000);
\draw[color=black] (0.000000,75.000000) node[left] {$\K{out}_2$};
\draw[color=black] (0.000000,60.000000) -- (294.000000,60.000000);
\draw[color=black] (0.000000,60.000000) node[left] {$\K{out}_3$};
\draw[color=black] (0.000000,45.000000) -- (294.000000,45.000000);
\draw[color=black] (0.000000,45.000000) node[left] {$\K{out}_4$};
\draw[color=black] (0.000000,30.000000) -- (294.000000,30.000000);
\draw[color=black] (0.000000,30.000000) node[left] {$\K{out}_5$};
\draw[color=black] (0.000000,15.000000) -- (294.000000,15.000000);
\draw[color=black] (0.000000,15.000000) node[left] {$\K{out}_6$};
\draw[color=black] (0.000000,0.000000) -- (294.000000,0.000000);
\draw[color=black] (0.000000,0.000000) node[left] {$\K{out}_7$};
\begin{scope}
\draw[fill=white] (12.000000, 225.000000) +(-45.000000:8.485281pt and 8.485281pt) -- +(45.000000:8.485281pt and 8.485281pt) -- +(135.000000:8.485281pt and 8.485281pt) -- +(225.000000:8.485281pt and 8.485281pt) -- cycle;
\clip (12.000000, 225.000000) +(-45.000000:8.485281pt and 8.485281pt) -- +(45.000000:8.485281pt and 8.485281pt) -- +(135.000000:8.485281pt and 8.485281pt) -- +(225.000000:8.485281pt and 8.485281pt) -- cycle;
\draw (12.000000, 225.000000) node {T};
\end{scope}
\begin{scope}
\draw[fill=white] (12.000000, 210.000000) +(-45.000000:8.485281pt and 8.485281pt) -- +(45.000000:8.485281pt and 8.485281pt) -- +(135.000000:8.485281pt and 8.485281pt) -- +(225.000000:8.485281pt and 8.485281pt) -- cycle;
\clip (12.000000, 210.000000) +(-45.000000:8.485281pt and 8.485281pt) -- +(45.000000:8.485281pt and 8.485281pt) -- +(135.000000:8.485281pt and 8.485281pt) -- +(225.000000:8.485281pt and 8.485281pt) -- cycle;
\draw (12.000000, 210.000000) node {T};
\end{scope}
\begin{scope}
\draw[fill=white] (12.000000, 195.000000) +(-45.000000:8.485281pt and 8.485281pt) -- +(45.000000:8.485281pt and 8.485281pt) -- +(135.000000:8.485281pt and 8.485281pt) -- +(225.000000:8.485281pt and 8.485281pt) -- cycle;
\clip (12.000000, 195.000000) +(-45.000000:8.485281pt and 8.485281pt) -- +(45.000000:8.485281pt and 8.485281pt) -- +(135.000000:8.485281pt and 8.485281pt) -- +(225.000000:8.485281pt and 8.485281pt) -- cycle;
\draw (12.000000, 195.000000) node {T};
\end{scope}
\begin{scope}
\draw[fill=white] (12.000000, 180.000000) +(-45.000000:8.485281pt and 8.485281pt) -- +(45.000000:8.485281pt and 8.485281pt) -- +(135.000000:8.485281pt and 8.485281pt) -- +(225.000000:8.485281pt and 8.485281pt) -- cycle;
\clip (12.000000, 180.000000) +(-45.000000:8.485281pt and 8.485281pt) -- +(45.000000:8.485281pt and 8.485281pt) -- +(135.000000:8.485281pt and 8.485281pt) -- +(225.000000:8.485281pt and 8.485281pt) -- cycle;
\draw (12.000000, 180.000000) node {T};
\end{scope}
\begin{scope}
\draw[fill=white] (12.000000, 165.000000) +(-45.000000:8.485281pt and 8.485281pt) -- +(45.000000:8.485281pt and 8.485281pt) -- +(135.000000:8.485281pt and 8.485281pt) -- +(225.000000:8.485281pt and 8.485281pt) -- cycle;
\clip (12.000000, 165.000000) +(-45.000000:8.485281pt and 8.485281pt) -- +(45.000000:8.485281pt and 8.485281pt) -- +(135.000000:8.485281pt and 8.485281pt) -- +(225.000000:8.485281pt and 8.485281pt) -- cycle;
\draw (12.000000, 165.000000) node {T};
\end{scope}
\begin{scope}
\draw[fill=white] (12.000000, 150.000000) +(-45.000000:8.485281pt and 8.485281pt) -- +(45.000000:8.485281pt and 8.485281pt) -- +(135.000000:8.485281pt and 8.485281pt) -- +(225.000000:8.485281pt and 8.485281pt) -- cycle;
\clip (12.000000, 150.000000) +(-45.000000:8.485281pt and 8.485281pt) -- +(45.000000:8.485281pt and 8.485281pt) -- +(135.000000:8.485281pt and 8.485281pt) -- +(225.000000:8.485281pt and 8.485281pt) -- cycle;
\draw (12.000000, 150.000000) node {T};
\end{scope}
\begin{scope}
\draw[fill=white] (12.000000, 135.000000) +(-45.000000:8.485281pt and 8.485281pt) -- +(45.000000:8.485281pt and 8.485281pt) -- +(135.000000:8.485281pt and 8.485281pt) -- +(225.000000:8.485281pt and 8.485281pt) -- cycle;
\clip (12.000000, 135.000000) +(-45.000000:8.485281pt and 8.485281pt) -- +(45.000000:8.485281pt and 8.485281pt) -- +(135.000000:8.485281pt and 8.485281pt) -- +(225.000000:8.485281pt and 8.485281pt) -- cycle;
\draw (12.000000, 135.000000) node {T};
\end{scope}
\begin{scope}
\draw[fill=white] (12.000000, 120.000000) +(-45.000000:8.485281pt and 8.485281pt) -- +(45.000000:8.485281pt and 8.485281pt) -- +(135.000000:8.485281pt and 8.485281pt) -- +(225.000000:8.485281pt and 8.485281pt) -- cycle;
\clip (12.000000, 120.000000) +(-45.000000:8.485281pt and 8.485281pt) -- +(45.000000:8.485281pt and 8.485281pt) -- +(135.000000:8.485281pt and 8.485281pt) -- +(225.000000:8.485281pt and 8.485281pt) -- cycle;
\draw (12.000000, 120.000000) node {T};
\end{scope}
\begin{scope}
\draw[fill=white] (12.000000, 105.000000) +(-45.000000:8.485281pt and 8.485281pt) -- +(45.000000:8.485281pt and 8.485281pt) -- +(135.000000:8.485281pt and 8.485281pt) -- +(225.000000:8.485281pt and 8.485281pt) -- cycle;
\clip (12.000000, 105.000000) +(-45.000000:8.485281pt and 8.485281pt) -- +(45.000000:8.485281pt and 8.485281pt) -- +(135.000000:8.485281pt and 8.485281pt) -- +(225.000000:8.485281pt and 8.485281pt) -- cycle;
\draw (12.000000, 105.000000) node {T};
\end{scope}
\begin{scope}
\draw[fill=white] (12.000000, 90.000000) +(-45.000000:8.485281pt and 8.485281pt) -- +(45.000000:8.485281pt and 8.485281pt) -- +(135.000000:8.485281pt and 8.485281pt) -- +(225.000000:8.485281pt and 8.485281pt) -- cycle;
\clip (12.000000, 90.000000) +(-45.000000:8.485281pt and 8.485281pt) -- +(45.000000:8.485281pt and 8.485281pt) -- +(135.000000:8.485281pt and 8.485281pt) -- +(225.000000:8.485281pt and 8.485281pt) -- cycle;
\draw (12.000000, 90.000000) node {T};
\end{scope}
\begin{scope}
\draw[fill=white] (12.000000, 75.000000) +(-45.000000:8.485281pt and 8.485281pt) -- +(45.000000:8.485281pt and 8.485281pt) -- +(135.000000:8.485281pt and 8.485281pt) -- +(225.000000:8.485281pt and 8.485281pt) -- cycle;
\clip (12.000000, 75.000000) +(-45.000000:8.485281pt and 8.485281pt) -- +(45.000000:8.485281pt and 8.485281pt) -- +(135.000000:8.485281pt and 8.485281pt) -- +(225.000000:8.485281pt and 8.485281pt) -- cycle;
\draw (12.000000, 75.000000) node {T};
\end{scope}
\begin{scope}
\draw[fill=white] (12.000000, 60.000000) +(-45.000000:8.485281pt and 8.485281pt) -- +(45.000000:8.485281pt and 8.485281pt) -- +(135.000000:8.485281pt and 8.485281pt) -- +(225.000000:8.485281pt and 8.485281pt) -- cycle;
\clip (12.000000, 60.000000) +(-45.000000:8.485281pt and 8.485281pt) -- +(45.000000:8.485281pt and 8.485281pt) -- +(135.000000:8.485281pt and 8.485281pt) -- +(225.000000:8.485281pt and 8.485281pt) -- cycle;
\draw (12.000000, 60.000000) node {T};
\end{scope}
\begin{scope}
\draw[fill=white] (12.000000, 45.000000) +(-45.000000:8.485281pt and 8.485281pt) -- +(45.000000:8.485281pt and 8.485281pt) -- +(135.000000:8.485281pt and 8.485281pt) -- +(225.000000:8.485281pt and 8.485281pt) -- cycle;
\clip (12.000000, 45.000000) +(-45.000000:8.485281pt and 8.485281pt) -- +(45.000000:8.485281pt and 8.485281pt) -- +(135.000000:8.485281pt and 8.485281pt) -- +(225.000000:8.485281pt and 8.485281pt) -- cycle;
\draw (12.000000, 45.000000) node {T};
\end{scope}
\begin{scope}
\draw[fill=white] (12.000000, 30.000000) +(-45.000000:8.485281pt and 8.485281pt) -- +(45.000000:8.485281pt and 8.485281pt) -- +(135.000000:8.485281pt and 8.485281pt) -- +(225.000000:8.485281pt and 8.485281pt) -- cycle;
\clip (12.000000, 30.000000) +(-45.000000:8.485281pt and 8.485281pt) -- +(45.000000:8.485281pt and 8.485281pt) -- +(135.000000:8.485281pt and 8.485281pt) -- +(225.000000:8.485281pt and 8.485281pt) -- cycle;
\draw (12.000000, 30.000000) node {T};
\end{scope}
\begin{scope}
\draw[fill=white] (12.000000, 15.000000) +(-45.000000:8.485281pt and 8.485281pt) -- +(45.000000:8.485281pt and 8.485281pt) -- +(135.000000:8.485281pt and 8.485281pt) -- +(225.000000:8.485281pt and 8.485281pt) -- cycle;
\clip (12.000000, 15.000000) +(-45.000000:8.485281pt and 8.485281pt) -- +(45.000000:8.485281pt and 8.485281pt) -- +(135.000000:8.485281pt and 8.485281pt) -- +(225.000000:8.485281pt and 8.485281pt) -- cycle;
\draw (12.000000, 15.000000) node {T};
\end{scope}
\begin{scope}
\draw[fill=white] (12.000000, -0.000000) +(-45.000000:8.485281pt and 8.485281pt) -- +(45.000000:8.485281pt and 8.485281pt) -- +(135.000000:8.485281pt and 8.485281pt) -- +(225.000000:8.485281pt and 8.485281pt) -- cycle;
\clip (12.000000, -0.000000) +(-45.000000:8.485281pt and 8.485281pt) -- +(45.000000:8.485281pt and 8.485281pt) -- +(135.000000:8.485281pt and 8.485281pt) -- +(225.000000:8.485281pt and 8.485281pt) -- cycle;
\draw (12.000000, -0.000000) node {T};
\end{scope}
\draw (33.000000,225.000000) -- (33.000000,210.000000);
\filldraw (33.000000, 225.000000) circle(1.500000pt);
\begin{scope}
\draw[fill=white] (33.000000, 210.000000) circle(3.000000pt);
\clip (33.000000, 210.000000) circle(3.000000pt);
\draw (30.000000, 210.000000) -- (36.000000, 210.000000);
\draw (33.000000, 207.000000) -- (33.000000, 213.000000);
\end{scope}
\draw (51.000000,210.000000) -- (51.000000,195.000000);
\filldraw (51.000000, 210.000000) circle(1.500000pt);
\begin{scope}
\draw[fill=white] (51.000000, 195.000000) circle(3.000000pt);
\clip (51.000000, 195.000000) circle(3.000000pt);
\draw (48.000000, 195.000000) -- (54.000000, 195.000000);
\draw (51.000000, 192.000000) -- (51.000000, 198.000000);
\end{scope}
\draw (69.000000,195.000000) -- (69.000000,180.000000);
\filldraw (69.000000, 195.000000) circle(1.500000pt);
\begin{scope}
\draw[fill=white] (69.000000, 180.000000) circle(3.000000pt);
\clip (69.000000, 180.000000) circle(3.000000pt);
\draw (66.000000, 180.000000) -- (72.000000, 180.000000);
\draw (69.000000, 177.000000) -- (69.000000, 183.000000);
\end{scope}
\draw (87.000000,180.000000) -- (87.000000,165.000000);
\filldraw (87.000000, 180.000000) circle(1.500000pt);
\begin{scope}
\draw[fill=white] (87.000000, 165.000000) circle(3.000000pt);
\clip (87.000000, 165.000000) circle(3.000000pt);
\draw (84.000000, 165.000000) -- (90.000000, 165.000000);
\draw (87.000000, 162.000000) -- (87.000000, 168.000000);
\end{scope}
\draw (105.000000,165.000000) -- (105.000000,150.000000);
\filldraw (105.000000, 165.000000) circle(1.500000pt);
\begin{scope}
\draw[fill=white] (105.000000, 150.000000) circle(3.000000pt);
\clip (105.000000, 150.000000) circle(3.000000pt);
\draw (102.000000, 150.000000) -- (108.000000, 150.000000);
\draw (105.000000, 147.000000) -- (105.000000, 153.000000);
\end{scope}
\draw (123.000000,150.000000) -- (123.000000,135.000000);
\filldraw (123.000000, 150.000000) circle(1.500000pt);
\begin{scope}
\draw[fill=white] (123.000000, 135.000000) circle(3.000000pt);
\clip (123.000000, 135.000000) circle(3.000000pt);
\draw (120.000000, 135.000000) -- (126.000000, 135.000000);
\draw (123.000000, 132.000000) -- (123.000000, 138.000000);
\end{scope}
\draw (141.000000,135.000000) -- (141.000000,120.000000);
\filldraw (141.000000, 135.000000) circle(1.500000pt);
\begin{scope}
\draw[fill=white] (141.000000, 120.000000) circle(3.000000pt);
\clip (141.000000, 120.000000) circle(3.000000pt);
\draw (138.000000, 120.000000) -- (144.000000, 120.000000);
\draw (141.000000, 117.000000) -- (141.000000, 123.000000);
\end{scope}
\draw (159.000000,120.000000) -- (159.000000,105.000000);
\filldraw (159.000000, 120.000000) circle(1.500000pt);
\begin{scope}
\draw[fill=white] (159.000000, 105.000000) circle(3.000000pt);
\clip (159.000000, 105.000000) circle(3.000000pt);
\draw (156.000000, 105.000000) -- (162.000000, 105.000000);
\draw (159.000000, 102.000000) -- (159.000000, 108.000000);
\end{scope}
\draw (177.000000,105.000000) -- (177.000000,90.000000);
\filldraw (177.000000, 105.000000) circle(1.500000pt);
\begin{scope}
\draw[fill=white] (177.000000, 90.000000) circle(3.000000pt);
\clip (177.000000, 90.000000) circle(3.000000pt);
\draw (174.000000, 90.000000) -- (180.000000, 90.000000);
\draw (177.000000, 87.000000) -- (177.000000, 93.000000);
\end{scope}
\draw (195.000000,90.000000) -- (195.000000,75.000000);
\filldraw (195.000000, 90.000000) circle(1.500000pt);
\begin{scope}
\draw[fill=white] (195.000000, 75.000000) circle(3.000000pt);
\clip (195.000000, 75.000000) circle(3.000000pt);
\draw (192.000000, 75.000000) -- (198.000000, 75.000000);
\draw (195.000000, 72.000000) -- (195.000000, 78.000000);
\end{scope}
\draw (213.000000,75.000000) -- (213.000000,60.000000);
\filldraw (213.000000, 75.000000) circle(1.500000pt);
\begin{scope}
\draw[fill=white] (213.000000, 60.000000) circle(3.000000pt);
\clip (213.000000, 60.000000) circle(3.000000pt);
\draw (210.000000, 60.000000) -- (216.000000, 60.000000);
\draw (213.000000, 57.000000) -- (213.000000, 63.000000);
\end{scope}
\draw (231.000000,60.000000) -- (231.000000,45.000000);
\filldraw (231.000000, 60.000000) circle(1.500000pt);
\begin{scope}
\draw[fill=white] (231.000000, 45.000000) circle(3.000000pt);
\clip (231.000000, 45.000000) circle(3.000000pt);
\draw (228.000000, 45.000000) -- (234.000000, 45.000000);
\draw (231.000000, 42.000000) -- (231.000000, 48.000000);
\end{scope}
\draw (249.000000,45.000000) -- (249.000000,30.000000);
\filldraw (249.000000, 45.000000) circle(1.500000pt);
\begin{scope}
\draw[fill=white] (249.000000, 30.000000) circle(3.000000pt);
\clip (249.000000, 30.000000) circle(3.000000pt);
\draw (246.000000, 30.000000) -- (252.000000, 30.000000);
\draw (249.000000, 27.000000) -- (249.000000, 33.000000);
\end{scope}
\draw (267.000000,30.000000) -- (267.000000,15.000000);
\filldraw (267.000000, 30.000000) circle(1.500000pt);
\begin{scope}
\draw[fill=white] (267.000000, 15.000000) circle(3.000000pt);
\clip (267.000000, 15.000000) circle(3.000000pt);
\draw (264.000000, 15.000000) -- (270.000000, 15.000000);
\draw (267.000000, 12.000000) -- (267.000000, 18.000000);
\end{scope}
\draw (285.000000,15.000000) -- (285.000000,0.000000);
\filldraw (285.000000, 15.000000) circle(1.500000pt);
\begin{scope}
\draw[fill=white] (285.000000, 0.000000) circle(3.000000pt);
\clip (285.000000, 0.000000) circle(3.000000pt);
\draw (282.000000, 0.000000) -- (288.000000, 0.000000);
\draw (285.000000, -3.000000) -- (285.000000, 3.000000);
\end{scope}
\draw[decorate,decoration={brace,amplitude = 4.000000pt},very thick] (3.000000,232.500000) -- (291.000000,232.500000);
\draw (147.000000, 236.500000) node[text width=144pt,above,text centered] {Repeat $d$ times};
\end{tikzpicture}